\documentclass{aa}
\usepackage{color}
\usepackage{colortbl}
\usepackage{multirow}
\usepackage{dcolumn}
\usepackage{amsmath}
\usepackage{amssymb}
\usepackage{graphicx}
\usepackage{epsfig}
\usepackage{changebar}
\usepackage{natbib}
\usepackage{multirow}
\usepackage{subfigure}
\usepackage{txfonts}
\usepackage{relsize}

\usepackage{longtable,lscape}
\usepackage[figuresright]{rotating}
\newcolumntype{d}[1]{D{.}{\cdot}{#1}}
\newcolumntype{.}{D{.}{.}{-1}}

\newcommand{\msun}{M$_\odot$}
\newcommand{\lsun}{L$_\odot$}

\newcommand{\kms}{km~s$^{-1}$}
\newcommand{\Tr}{$T^*_{\rm{R}}$}
\newcommand{\Ta}{$T^*_{\rm{A}}$}
\newcommand{\Tv}{$\int T_{\rm{R}} \rm{d}V$}

\newcommand{\cmthree}{cm$^{-3}$}

\newcommand{\vlsr}{V$_{\rm{LSR}}$}

\begin{document}
   \title{The RMS Survey:}

   \subtitle{$^{13}$CO observations of candidate massive YSOs in the southern hemisphere}

   \author{J.~S.~Urquhart
          \inst{1}
          \and
			 A.~L.~Busfield
			 \inst{1}
			 \and
			 M.~G.~Hoare
			 \inst{1}
			 \and
			 S.~L.~Lumsden
			 \inst{1}
			 \and
			 R.~D.~Oudmaijer
			 \inst{1}
			 \and
			 T.~J.~T.~Moore
			 \inst{2}
			 \and
			 A.~G.~Gibb
			 \inst{3}
			 \and
			 C.~R.~Purcell
			 \inst{4,5}
			 \and
			 M.~G.~Burton
			 \inst{4}
			 \and
			 L.~J.~L.~Marechal
			 \inst{6}
          }

   \offprints{J. S. Urquhart: jsu@ast.leeds.ac.uk}

   \institute{School of Physics and Astrophysics, University of Leeds, Leeds, LS2~9JT, UK 
         \and
             Astrophysics Research Institute, Liverpool John Moores University, Twelve Quays House, Egerton Wharf, Birkenhead, CH41~1LD, UK 
			\and
			Department of Physics and Astronomy, University of British Columbia, 6224 Aricultural Road, Vancouver, BC, V6T 1Z1, Canada 
			\and
				School of Physics and Astronomy, University of New South Wales, Sydney, NSW 2052, Australia  
			\and 
			Jodrell Bank Observatory, University of Manchester, Cheshire, SK11~9DL, UK
			\and
			\'Ecole Normale Sup\'erieure,
D\'epartement de Physique, 24 rue Lhomond, F-75005, Paris, France	
             }

   \date{}

\abstract
   {The Red MSX Source (RMS) survey is an ongoing multi-wavelength observational programme designed to return a large, well-selected sample of massive young stellar objects (MYSOs). We have identified $\sim$2000 MYSOs candidates located within our Galaxy by comparing the colours of MSX and 2MASS point sources to those of known MYSOs. The aim of our follow-up observations is to identify other contaminating objects such as ultra compact (UC) HII regions, evolved stars and planetary nebulae (PNe) and distinguish between genuine MYSOs and nearby low-mass YSOs.}      
   {A critical part of our follow-up programme is to conduct $^{13}$CO molecular line observations in order to determine kinematic distances to all of our MYSO candidates. These distances will be used in combination with far-IR and (sub)millimetre fluxes to determine bolometric luminosities which will allow us to identify and remove nearby low-mass YSOs. In addition these molecular line observations will help in identifying evolved stars which are weak CO emitters.}
   {We have used the  22~m Mopra telescope, the 15~m JCMT and the 20~m Onsala telescope to conduct molecular line observations towards 854 MYSOs candidates located in the 3rd and 4th quadrants. These observations have been made at the $J$=1--0 (Mopra and Onsala) and $J$=2--1 (JCMT) rotational transition frequency of $^{13}$CO molecules and have a spatial resolution of $\sim$20\arcsec--40\arcsec, a sensitivity of $T_{\rm{A}}^*$ $\simeq$ 0.1~K and a velocity resolution of $\sim$0.2~\kms.}
   {We detected $^{13}$CO emission towards a total of 751 of the 854 RMS sources observed ($\sim88$\%). In total 2185 emission components are detected above 3$\sigma$ level (typically $T^*_{\rm{A}} \ge 0.3$~K). Multiple emission profiles are observed towards the majority of these sources -- 455 sources ($\sim$60\%) -- with an average of $\sim$4 molecular clouds along the line of sight. These multiple emission features make it difficult to assign a kinematic velocity to many of our sample. We have used archival CS ($J$=2--1) and maser velocities to resolve the component multiplicity towards 82 sources and have derived a criterion which is used to identify the  most likely component for a further 202 multiple component sources. Combined with the single component detections we have obtained unambiguous kinematic velocities towards 580 sources ($\sim$80\% of the detections). The 171 sources for which we have not been able to determine the kinematic velocity will require additional line data. Using the rotation curve of Brand and Blitz (1993) and their radial velocities we calculate kinematic distances for all components detected.}
	{}

\authorrunning{J. S. Urquhart et al.}
\titlerunning{$^{13}$CO observations of MYSO candidates}
\maketitle
\section{Introduction}
\subsection{Background}

Massive stars ($>$8~\msun, $>$10$^4$~\lsun) are responsible for some of the most spectacular phenomena in astrophysics. They inject huge amounts of energy into the interstellar medium (ISM) in the form of UV radiation, molecular outflows, powerful stellar winds and eventually supernova explosions. They are also responsible for depositing large amounts of enriched material back into the ISM. Through these processes massive stars can have an enormous impact on their local environment and can influence the evolution of their host galaxies. They are also thought to play an important role in regulating star formation through the propagation of strong shocks into their surrounding environment. These shocks may be responsible for triggering subsequent generations of star formation or disrupting conditions necessary for star formation in nearby clouds (\citealt{elmegreen1998}).

Given the profound impact massive stars have, not only on their local environment, but also on a galactic scale, it is vital that we understand the environmental conditions and processes involved in their birth and the earliest stages of their evolution. Their rarity and short evolutionary time scales means that massive stars are generally located much farther away than low-mass star formation sites. Additionally, massive stars are known to form exclusively in clusters making it difficult to identify and attribute derived quantities to individual sources. Furthermore, they reach the main sequence whilst still accreting material, and heavily embedded within their natal molecular cloud, hidden away beneath many magnitudes of visual extinction. For these reasons, until relatively recently, the only catalogue of massive young stellar objects (MYSOs) had been limited to 30 or so serendipitously detected sources (\citealt{Henning1984}) which were mostly nearby. These difficulties have resulted in our understanding of the earliest stages being far less developed than in the case of low mass star formation, which in broad terms is well understood (\citealt{shu1987,shu1993}). 

The situation has improved considerably in recent years with a number of studies using a variety of selection criteria (e.g., \citealt{molinari1996,walsh1997,sridharan2002}) which have identified many new MYSOs. However, all of these studies are based on IRAS colours which bias them towards bright, isolated sources and tend to avoid dense clustered environments and the Galactic mid-plane where the majority of MYSOs are expected to be found -- the scale height of massive stars is $\sim$30\arcmin\  (\citealt{reed2000}). The small numbers of MYSOs identified by these studies, and the way they have been selected, means that they are probably not representative of the general MYSO population and make statistical studies of many aspects of massive star formation difficult. Therefore there is an obvious need for a large, well-selected sample with sufficient numbers of sources in each luminosity bin to allow statistical studies as a function of luminosity. The Red MSX Source (RMS) survey is an ongoing observational programme designed to return just such a sample with complementary data with which to address these issues.

\subsection{The RMS Survey}

MYSOs are mid- and far-infrared bright with luminosities of 10$^4$--10$^5$ L$_\odot$ (\citealt{wynn-williams1982}). They are likely to be somewhat older than the hot molecular core stage and nuclear fusion has probably begun, however ongoing accretion in some way prevents the ionisation of the surroundings to form an HII region (see \citealt{hoare2007}). They are almost always associated with massive bipolar molecular outflows (e.g., \citealt{wu2004}), another indication that major accretion is likely to be ongoing. MYSOs are also known to possess ionised stellar winds that are weak thermal radio sources ($\sim$1~mJy at 1 kpc; \citealt{hoare2002}). These objects can therefore be roughly parameterized as near and/or mid-infrared bright,  have luminosities consistent with young O and early B-type stars and are relatively radio continuum quiet with some exceptions (e.g., S:106IR; \citealt{drew1993}).

\citet{lumsden2002} compared the colours of sources from the MSX and 2MASS point source catalogues to those of known MYSOs to develop a colour selection criteria from which we identified approximately 2000 MYSO candidates spread throughout the Galaxy ($|b|<$5\degr). Sources toward the Galactic centre were excluded ($|l|<10$\degr) due to confusion and difficulties in calculating kinematic distances. A problem with a colour selected sample is that the shape of the spectral energy distribution from an optically thick cloud is insensitive to the type of heating source. Therefore there are several other types of embedded, or dust enshrouded objects, that have similar colours to MYSOs and contaminate our sample, such as ultra compact (UC) HII regions, evolved stars and a small number of planetary nebulae (PNe). 

The RMS survey is a multi-wavelength programme of follow-up observations designed to distinguish between genuine MYSOs and these other embedded or dusty objects (\citealt{hoare2005}) and to compile a database of complementary multi-wavelength data with which to study their properties.\footnote{http://ast.leeds.ac.uk/RMS} These include
high resolution cm continuum observations to identify UCHII regions and PNe (\citealt{urquhart2007}), mid-infrared imaging to identify genuine point sources, obtain accurate astrometry and avoid excluding MYSOs located near UCHII regions (e.g., \citealt{mottram2006}), near-infrared spectroscopy (e.g., \citealt{clarke2006}) to distinguish between MYSOs and evolved stars. 

Another crucial part of our campaign is to obtain kinematic distances and luminosities towards our whole sample. These allow us to distinguish between nearby low and intermediate mass YSOs from the more massive YSOs. We have therefore conducted a programme of molecular line observations using transitions of the $^{13}$CO molecule towards all RMS sources not previously observed or for which good data was not available (see Sect.~\ref{sect:source_selection} for details). Additionally, these molecular line observations will allow us to eliminate a large proportion of evolved stars since they are not generally associated with strong CO emission ($^{12}$CO ($J$=1--0 and $J$=2--1) typically less than 1~K; \citealt{loup1993}). In this paper we present the results of our molecular line observations towards 854 RMS sources located in the southern Galactic plane (180\degr\ $< l <$ 350\degr). In Sect.~\ref{sect:observations} we outline our source selection, observational and data reduction procedures. We present our results and statistical analysis in Sect.~\ref{sect:results}. In Sect.~\ref{sect:summary} we present a summary of the results and highlight our main findings.  

\section{Observations and data reduction}
\label{sect:observations}

The $^{13}$CO transition was chosen for our observations as it is generally found to be only moderately optically thick and is thus better able to probe the bulk parameters of gas along the line of sight than $^{12}$CO. Moreover, using $^{13}$CO avoids many of the problems often encountered when observing with $^{12}$CO, which is nearly always optically thick and can often result in complex structures with multiple components and/or heavily self-absorbed spectral profiles, particularly towards the Galactic centre. The CS transition was considered as its high critical density (excitation threshold $\sim$10$^4$--$10^5$~\cmthree) makes it an excellent tracer of high density molecular gas. However, CS is much less abundant than $^{13}$CO and as a consequence  requires significantly longer on-source integration time to obtain an equivalent signal to noise. Additionally, given the lower abundance of CS there is a the danger that many of the more distant sources might be missed entirely. Given its higher abundance and low excitation threshold $^{13}$CO is generally much stronger than CS and less problematic than $^{12}$CO (c.f. \citealt{jackson2006,wu2001}).

\subsection{Source selection}
\label{sect:source_selection}

Of our $\sim$2000 colour selected MYSO candidates (hereafter RMS catalogue) 963 are located within the 3rd and 4th quadrants. The large majority of our southern RMS sources were previously unknown in the literature. However, approximately a quarter are associated with bright IRAS point sources, and as such, might have been previously observed as part of another programme. In order to avoid re-observing these sources we conducted an extensive literature search using SIMBAD\footnote{http://simbad.u-strasbg.fr/simbad/sim-fid.}. This search  identified a number of sources for which archival data was already available, the majority of these were found in two large observational programmes: a CS survey by \citet{bronfman1996} of colour selected IRAS sources, and a CO survey by \citet{wouterloot1989} which concentrated mainly on IRAS sources located in the outer Galaxy.

\citet{bronfman1996} conducted a programme of CS ($J$=2--1) transition observations towards 1427 IRAS point sources proposed by \citet{wood1989} as sites of massive star formation. We have 259 sources in common with sources observed by \citet{bronfman1996} of which they report 55 as non-detections. The low number of sources in common is because the \citet{wood1989} criteria preferentially selects UCHII regions rather than the earlier evolutionary phases of massive star formation. The majority of these were excluded from our observations, but, because  CS is generally much harder to detect than $^{13}$CO we re-observed all of their non-detections (with the exception of G206.7804-01.9395 for which there was good CO data available). In addition to the their non-detections we re-observed $\sim$25\% of their detections to allow for consistency checks and comparison of the two tracers. 

\citet{wouterloot1989} used the $^{12}$CO~($J$=1--0) transition to observe 1302 IRAS selected sources located within 10\degr\ of the Galactic plane and between 85\degr\ $< l <$ 280\degr. We find a total of 76 sources in common between their observations and our southern MYSO candidates. Some of these were excluded but a number were found to possess complex profiles, or were found to be saturated, which led to a large number being re-observed. Using data in the literature we were able to reduce the number of observations needed by over a hundred to 856.

\begin{sidewaysfigure*}
\begin{center}
\includegraphics[height=0.95\textwidth,angle=270, trim=90 0 90 0]{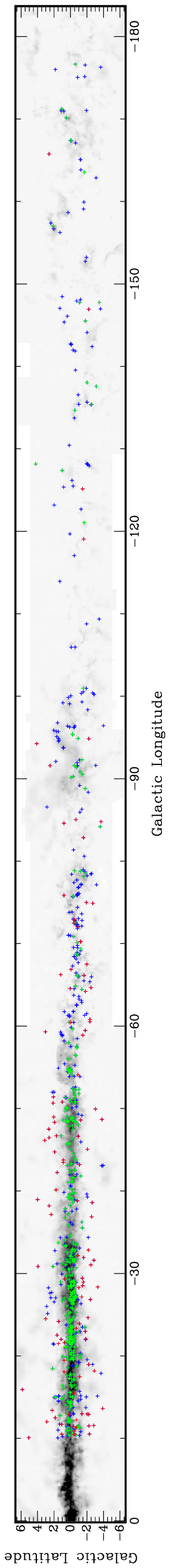}\\
\includegraphics[height=0.935\textwidth, angle=270, trim=0 10 0 0]{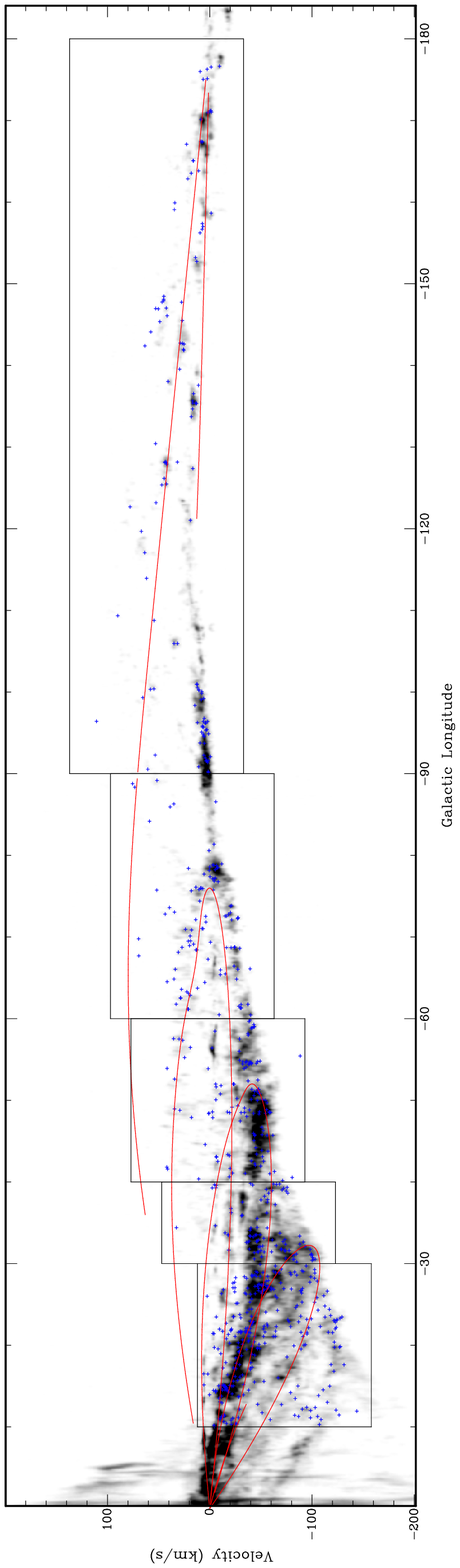}

\caption{Top panel: Galactic distribution of all southern hemisphere RMS sources located in the 3rd and 4th quadrants observed in $^{13}$CO and presented in this paper. The colours correspond to whether an observation resulted in a non-detection (red), a single component was detected (blue) or multiple components are detected (green). The integrated $^{12}$CO emission map of \citet{dame2001} (grey scale) shows the Galactic distribution of molecular material. Bottom panel: Galactic longitude-velocity plot showing the velocities of observed $^{13}$CO components as a function of Galactic longitude. The distribution of molecular material is shown in grey scale (again from \citealt{dame2001}) for comparison. The location of the spiral arms taken from model by \citet{tayor1993} and updated by \citet{cordes2004} are over-plotted in red. The regions covered by these observations are outlined in black (see text and  Table~2 for details).}

\label{fig:rms_distribution}
\end{center}
\end{sidewaysfigure*}
 
\subsection{Observation procedures}

We made $^{13}$CO observations towards 856 RMS sources from a total of 963 located in the southern Galactic plane.  The vast majority of which were observed using the 22 m Mopra telescope located near Coonabarabran, New South Wales, Australia\footnote{Mopra is operated by the Australia Telescope National Facility, CSIRO and the University of New South Wales during the time of the observations.}. The regions of the plane not accessible from Mopra were observed using the 15~m James Clerk Maxwell Telescope (JCMT) located on Mauna Kea, Hawaii\footnote{The James Clerk Maxwell Telescope is operated by The Joint Astronomy Centre on behalf of the Particle Physics and Astronomy Research Council of the United Kingdom, the Netherlands Organisation for Scientific Research, and the National Research Council of Canada.} and the 20~m Onsala telescope  located approximately 50~km south of Goteborg\footnote{Onsala is operated by the Swedish National Facility for Radio Astronomy, Onsala Space Observatory at Chalmers University of Technology.}. The $^{13}$CO ($J$=1--0) transition was used for the Mopra and Onsala observations, however, the instruments available on the JCMT are unable to operate at this frequency and therefore the higher excitation $^{13}$CO ($J$=2--1) transition was used. The use of these two different transitions should not affect our findings since YSOs are expected to have similar intensities at these frequencies (\citealt{little1994}). The number of sources observed by each telescope and the details of each telescopes' observational set-up are summarised in Table~\ref{tbl:CO_radio_parameters}.

All of these observations were performed in position-switching mode, with typical on-source integration times of $\sim$10~minutes for both the Mopra and Onsala observations, and $\sim$6~minutes for the JCMT observations (however, these integration times were reduced for stronger sources). The on-source integration time was split into a number of separate scans consisting of 1 minute of on- and off-source integration. Reference positions were offset from source positions by 1 degree in a direction perpendicular to the Galactic plane. These were chosen to avoid contamination of source emission from emission in the reference position at a similar velocity. In some cases, particularly towards the Galactic centre, several positions needed to be tried before a suitable reference position was found.

\begin{table}[!tbp]
\begin{center}
\caption{Observational parameters for the $^{13}$CO observations.}
\label{tbl:CO_radio_parameters}
\begin{minipage}{\linewidth}
\begin{tabular}{lccc}
\hline
\hline
Parameter& Mopra & Onsala & JCMT\\
\hline
\# of sources observed & 818 & 9 & 27\\
Rest frequency (GHz) & 110.201& 110.201& 220.3986 \\
Total bandwidth (MHz)& 64 & 80 & 267 \\
Vel. bandpass (\kms) & 174 & 224 & 360 \\ 
Number of channels   & 1024 & 1600 & 1713 \\
Vel. resolution (km s$^{-1}$) & 0.17 & 0.14 & 0.21\\
Beam size (\arcsec) & 33 & 35 & 21 \\
Date of observations & 2002-2005 & March 2003&2003-2004\\
Integration time (mins) & 10 & 10 & 6 \\
Telescope efficiency ($\eta_{\rm{fss}}$) & 0.55 & 0.43 & 0.69\\
\hline
\end{tabular}\\
\\
\end{minipage}
\end{center}
\end{table}

Mopra was used to observe 818 RMS sources during three periods between 2002 and 2005. The receiver can be tuned to either single or double side-band mode. The incoming signal is separated into
two channels using a polarisation splitter, each of which can be tuned separately allowing two channels to be observed simultaneously. However, the large overheads involved in periodically retuning the second IF between the $^{13}$CO and SiO maser frequency (\mbox{86.2 GHz}), which was used for pointing corrections, meant that for the observations made before 2004 the second IF was set to the maser frequency and was not retuned back to the $^{13}$CO frequency between pointing observations. An upgrade of the tuning system in 2004 significantly improved the tuning capabilities of the receiver and made it feasible to use both polarisations when observing our sources which significantly improved the signal-to-noise and lowered integration times. 

Of the remaining southern sources, 27 were observed with the JCMT during 2003 and 2004 as part of a long-term bad weather backup programme and nine were observed with the Onsala telescope during a run in March 2003. 

Typical system temperatures obtained for both the Mopra and Onsala observations are between 250--450~K depending on weather conditions and telescope elevation, with values typically below 300~K. The higher frequency observations conducted with the JCMT have system temperatures between 400--500~K. Telescope pointing was regularly checked ($\sim$2--3~hours) and found to be better than 3\arcsec\ for the JCMT and 10\arcsec\ for the Mopra and Onsala telescopes. To correct the measured antenna temperatures ($T^*_{\rm{A}}$) for atmospheric absorption, ohmic losses and rearward spillover and scattering, a measurement was made of an ambient load (assumed to be at 290~K) following the method of \citet{kutner1981}; this was done approximately every two hours.

In the upper panel of Fig.~\ref{fig:rms_distribution} we present a plot of the distribution of molecular material as a function of Galactic longitude and latitude as traced by the \citet{dame2001} $^{12}$CO survey. The coloured crosses indicate the RMS positions observed as part of the $^{13}$CO observations presented here; the different colours indicate whether the observation resulted in a non-detection, detection of a single or multiple components by colouring the symbols red, blue and green respectively. The RMS sources towards which $^{13}$CO emission has been detected correlate extremely well with the distribution of the $^{12}$CO. In the lower panel of Fig.~\ref{fig:rms_distribution} we present a plot of the distribution of $^{12}$CO as a function of Galactic longitude and V$_{\rm{LSR}}$ (again taken from \citealt{dame2001}). We have over-plotted the velocities of all detected components. Additionally, we have overlaid the longitude-velocity of the updated spiral arm model of \citet{tayor1993} by \citet{cordes2004} in red to illustrate how their velocities vary as a function of longitude in our Galaxy.

The velocities over which molecular clouds can be found in the southern Galactic plane range from approximately $-150$~\kms\ to +100~\kms. Since the total Galactic velocity range ($\sim$250~\kms) is larger than the total velocity bandpass available to our Mopra and Onsala observations it was necessary to change the central velocity as a function of Galactic longitude in order to optimise our velocity coverage. We used the Galactic longitude-velocity distribution maps of \citet{dame2001} to centre the available velocity bandpass as a function of longitude to cover as much of the Galaxy as possible. The central velocities used at various Galactic longitudes are tabulated in Table~\ref{tbl:central_vlsr} and the region covered by each are outlined in black on the lower panel of Fig.~\ref{fig:rms_distribution}. Comparing the regions covered by our observations with that of the distribution of $^{12}$CO shows that, although our velocity bandpass was limited, we have obtained an almost complete coverage of the southern part of the Galaxy.

\begin{table}[!tbp]
\begin{center}
\caption{Systemic spectrometer velocities used at different Galactic longitudes to optimise available velocity bandpass. These were chosen from visual inspection of the longitude-velocity maps of \citet{dame2001}.}
\label{tbl:central_vlsr}
\begin{minipage}{\linewidth}
\begin{tabular}{lccc}
\hline
\hline
Gal. longitude& Centred velocity (\kms)\\
\hline
180--270& 50 \\
271--300& 20 \\
301--320& -10 \\ 
321--330& -40\\
331--350& -75\\
\hline
\end{tabular}\\
\\
\end{minipage}
\end{center}
\end{table}

\subsection{Data reduction}

\begin{figure}
\begin{center}
\includegraphics[width=0.9\linewidth]{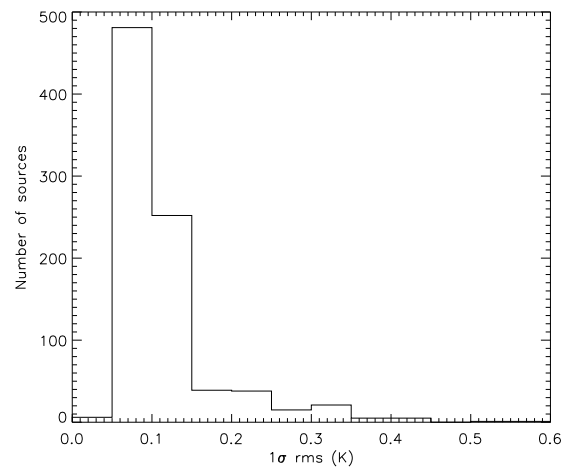}

\caption{Histograms of 1$\sigma$ rms obtained from 6--10 minute observations using a bin size of 0.05 K. Over 90\% of the observations have a sensitivity of better than 0.15~K with typical values of $\sim$0.1~K.}

\label{fig:rms_hist}
\end{center}
\end{figure}

\begin{figure*}[!t]
\begin{center}

\includegraphics[width=0.4\linewidth]{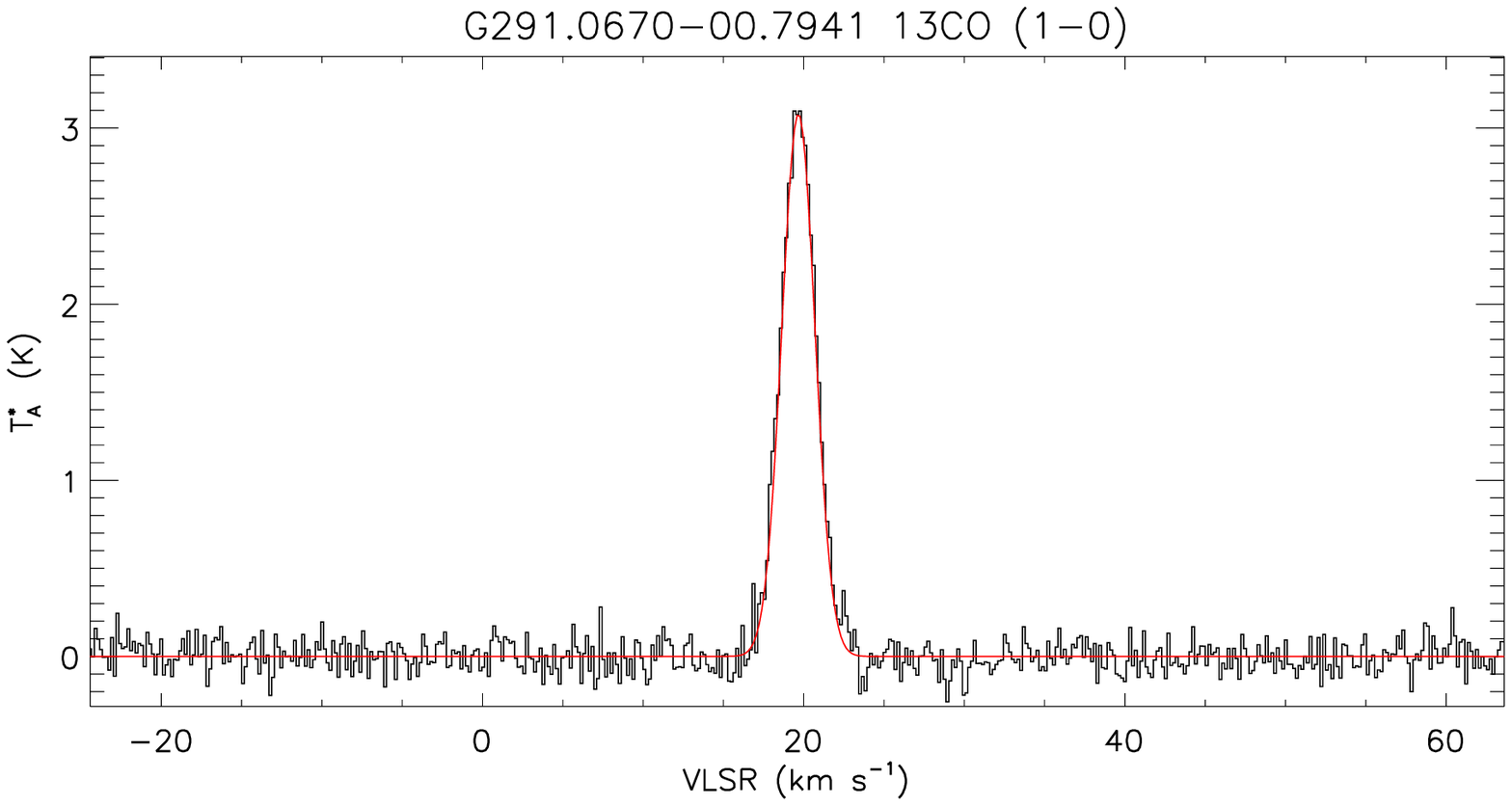} 
\includegraphics[width=0.4\linewidth]{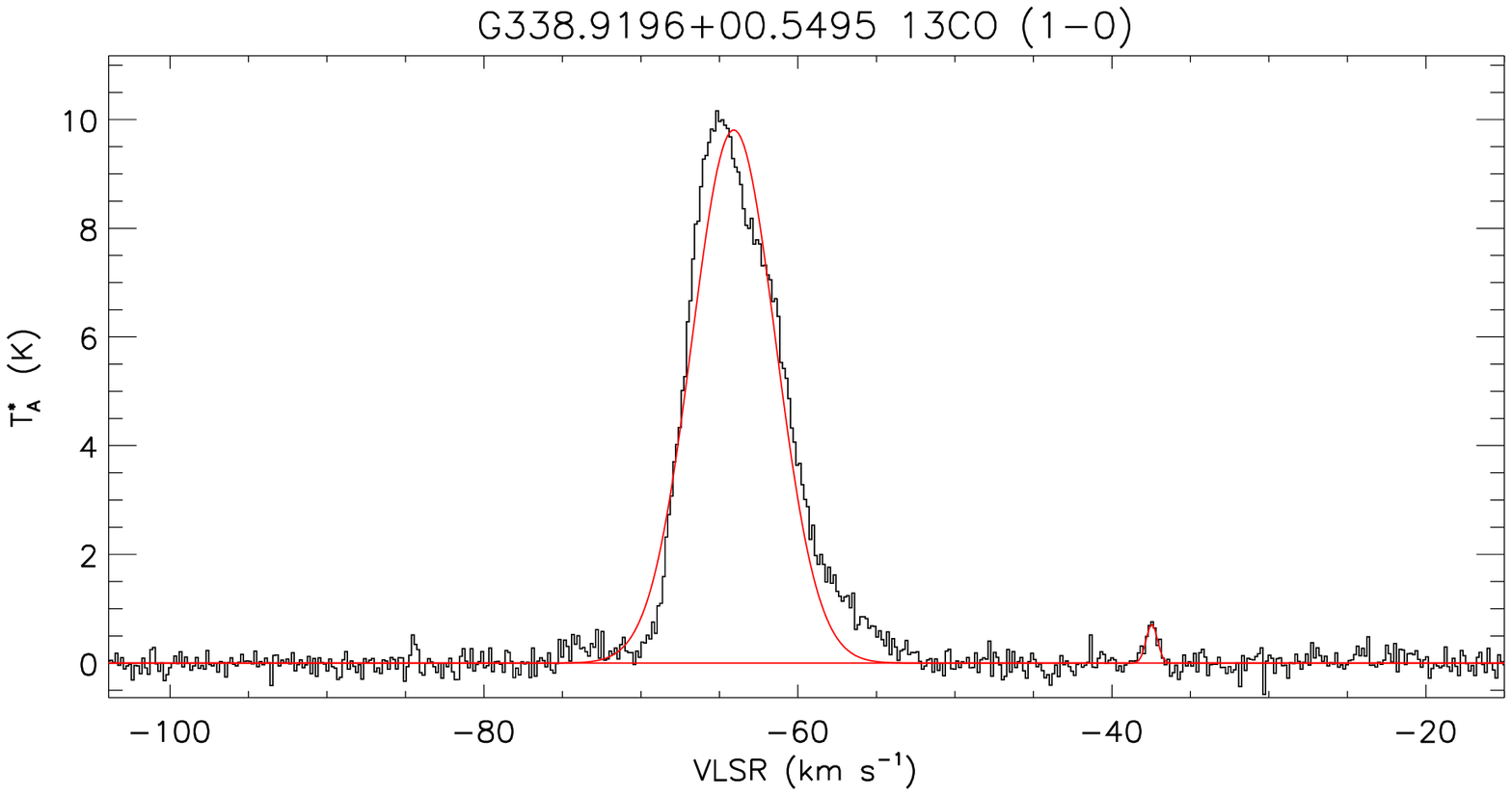} 
\includegraphics[width=0.4\linewidth]{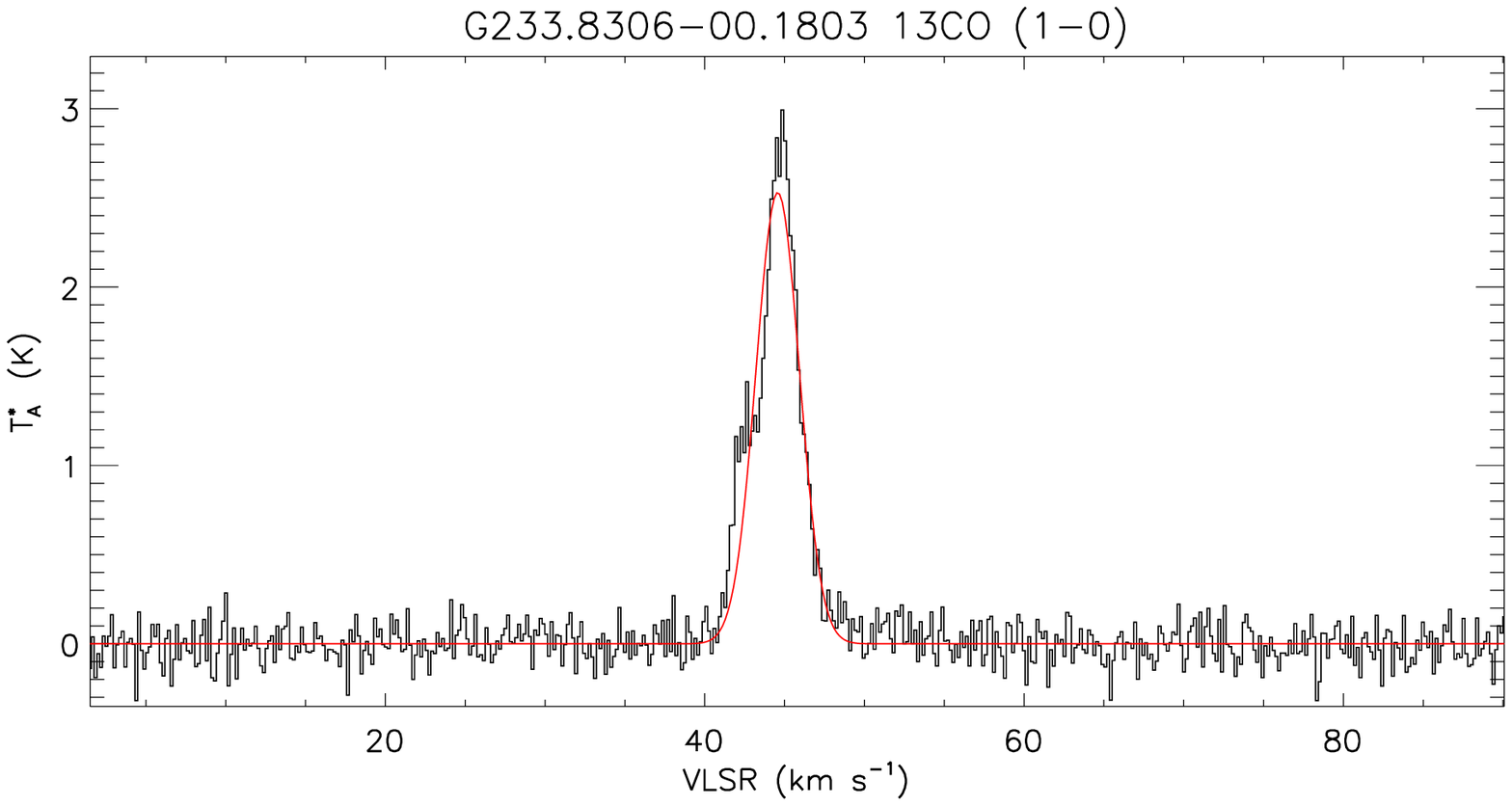} 
\includegraphics[width=0.4\linewidth]{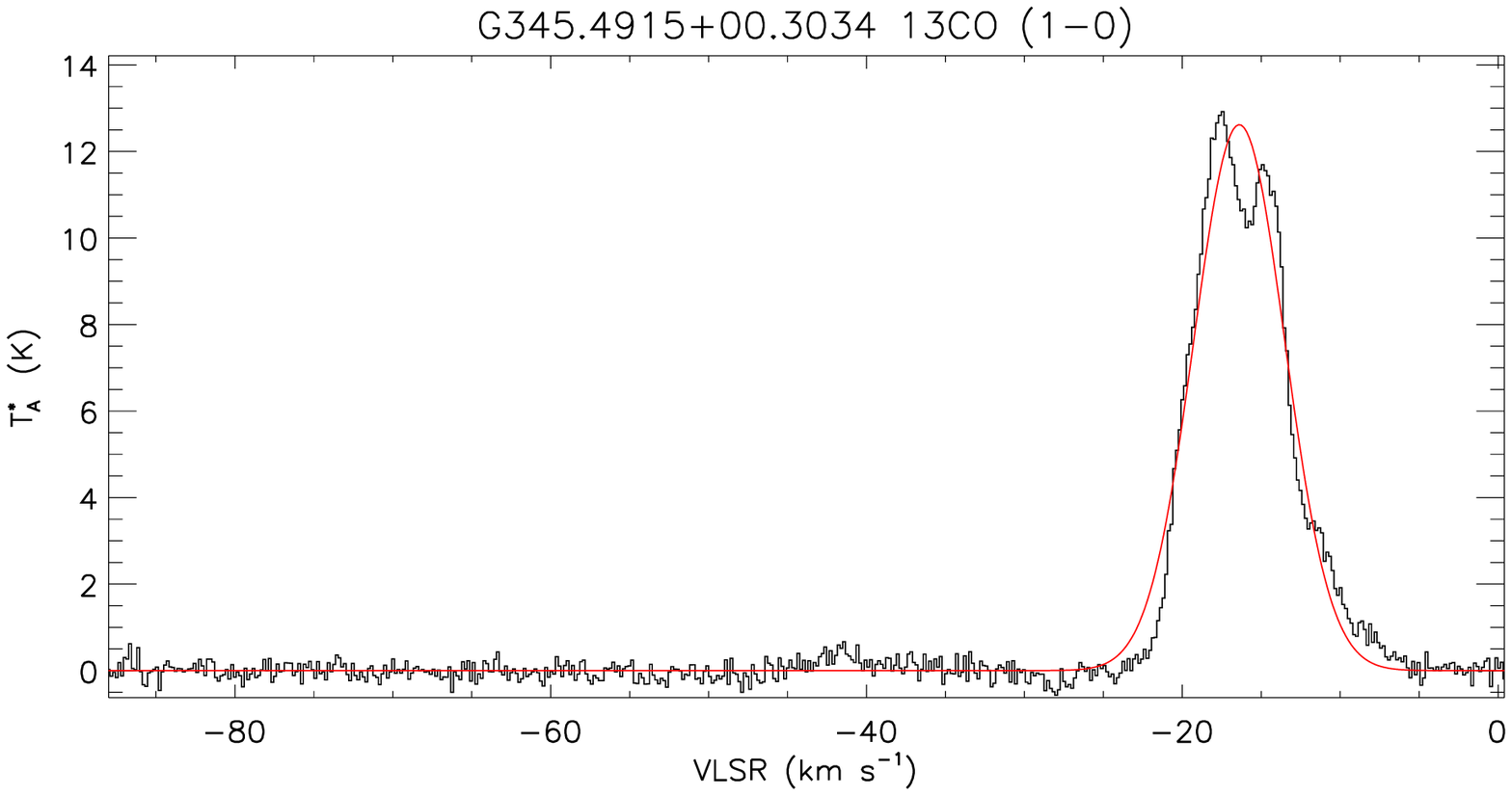} 
\includegraphics[width=0.4\linewidth]{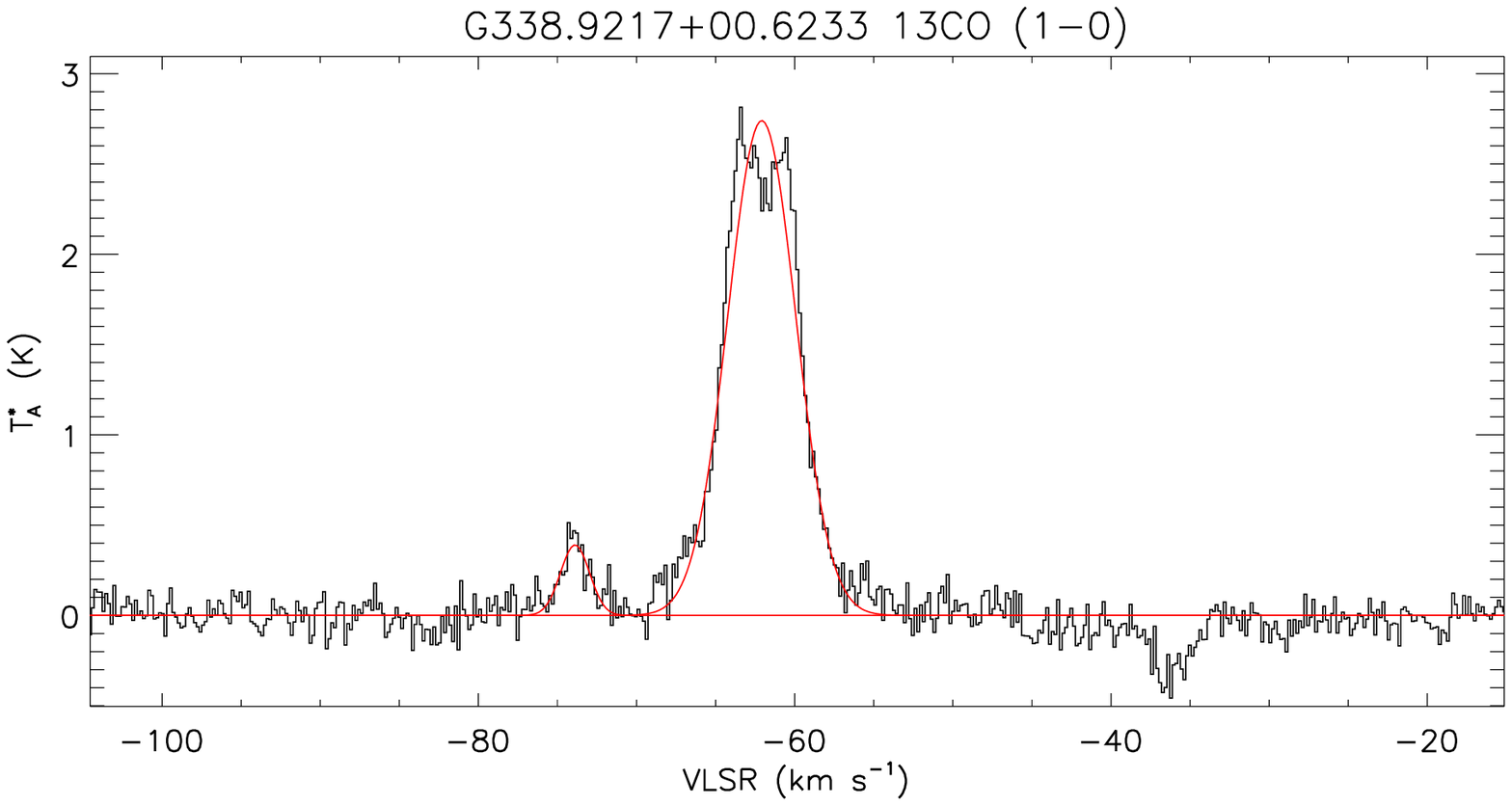} 
\includegraphics[width=0.4\linewidth]{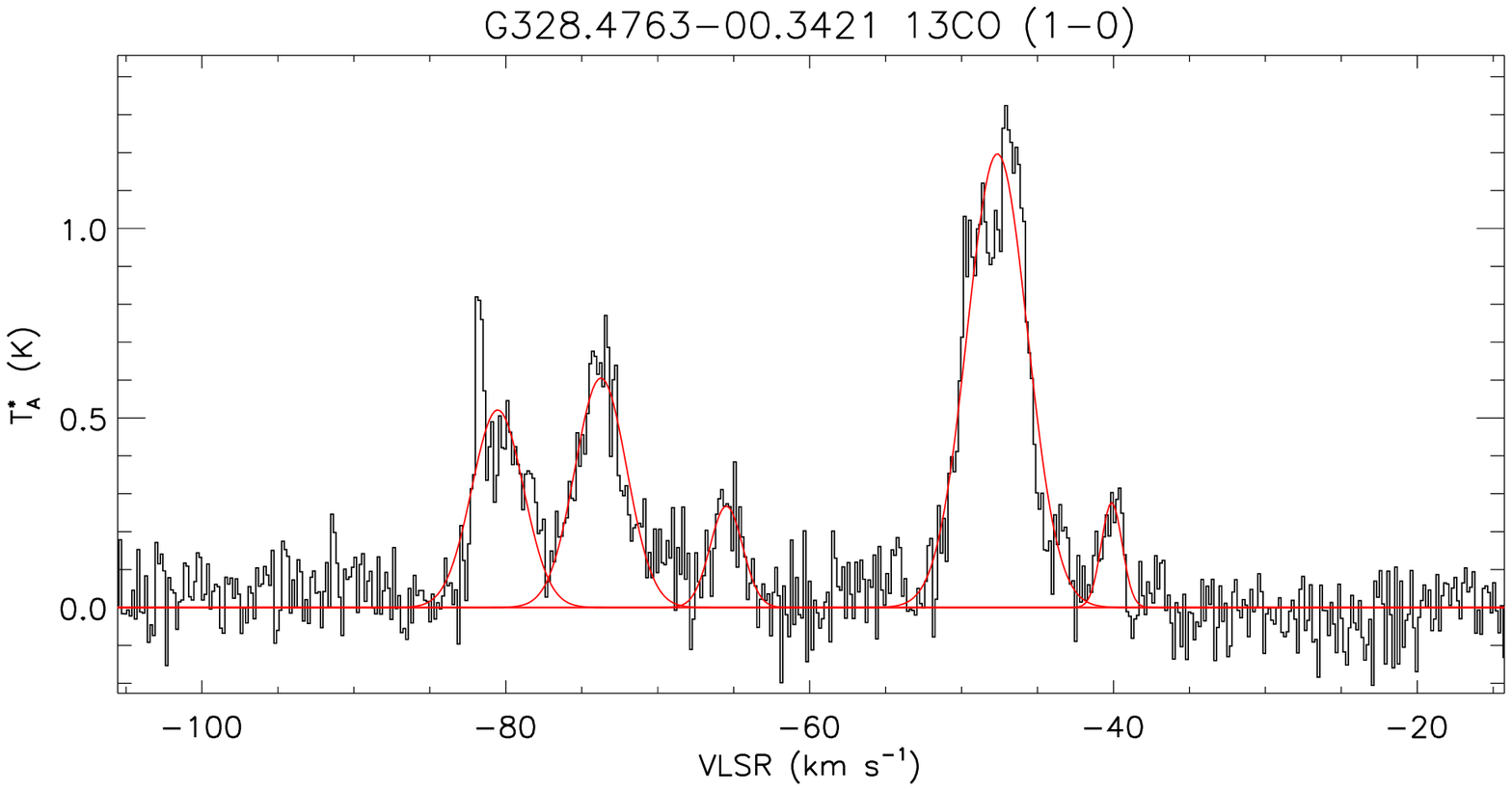} 
\includegraphics[width=0.4\linewidth]{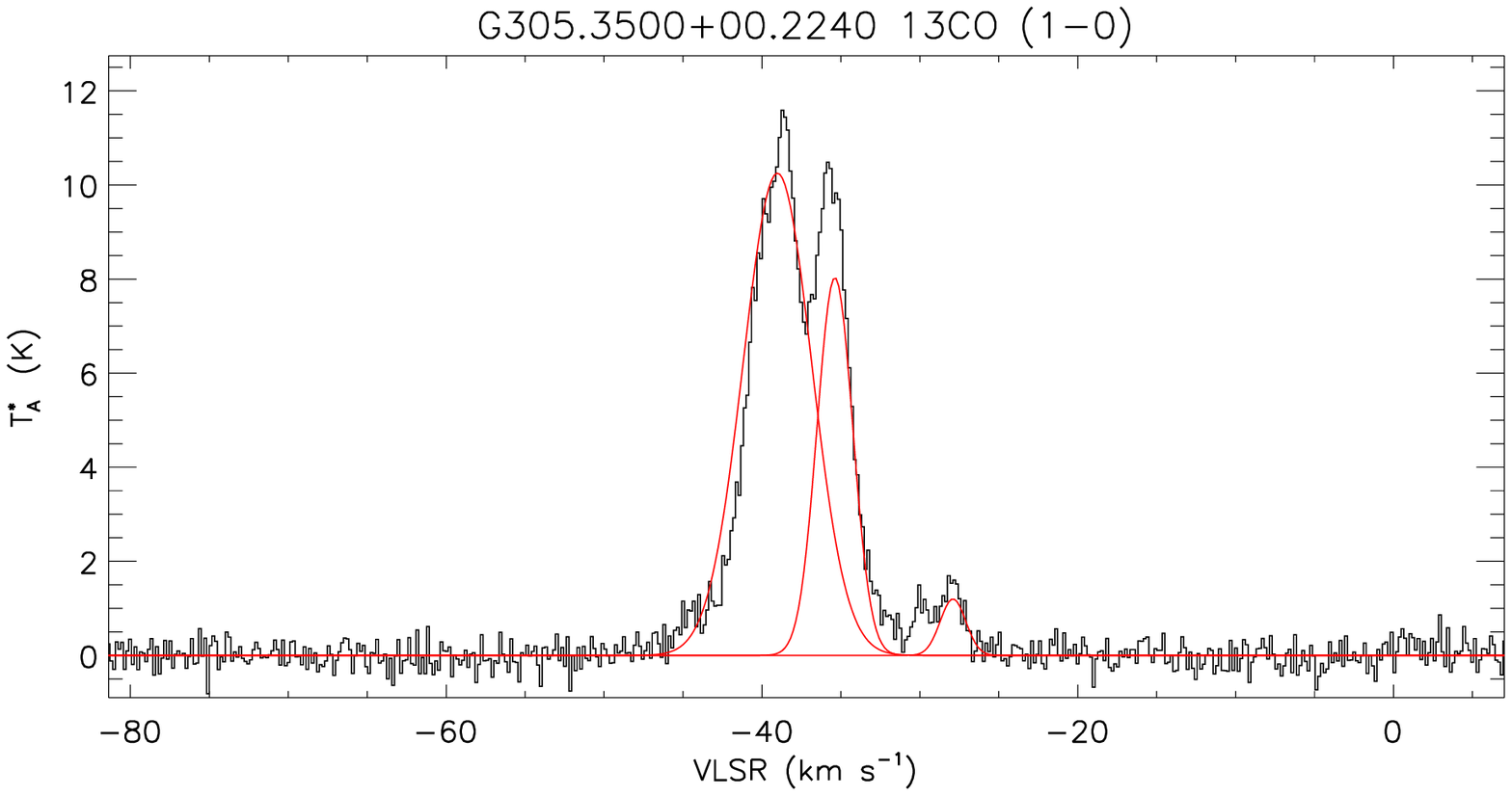} 
\includegraphics[width=0.4\linewidth]{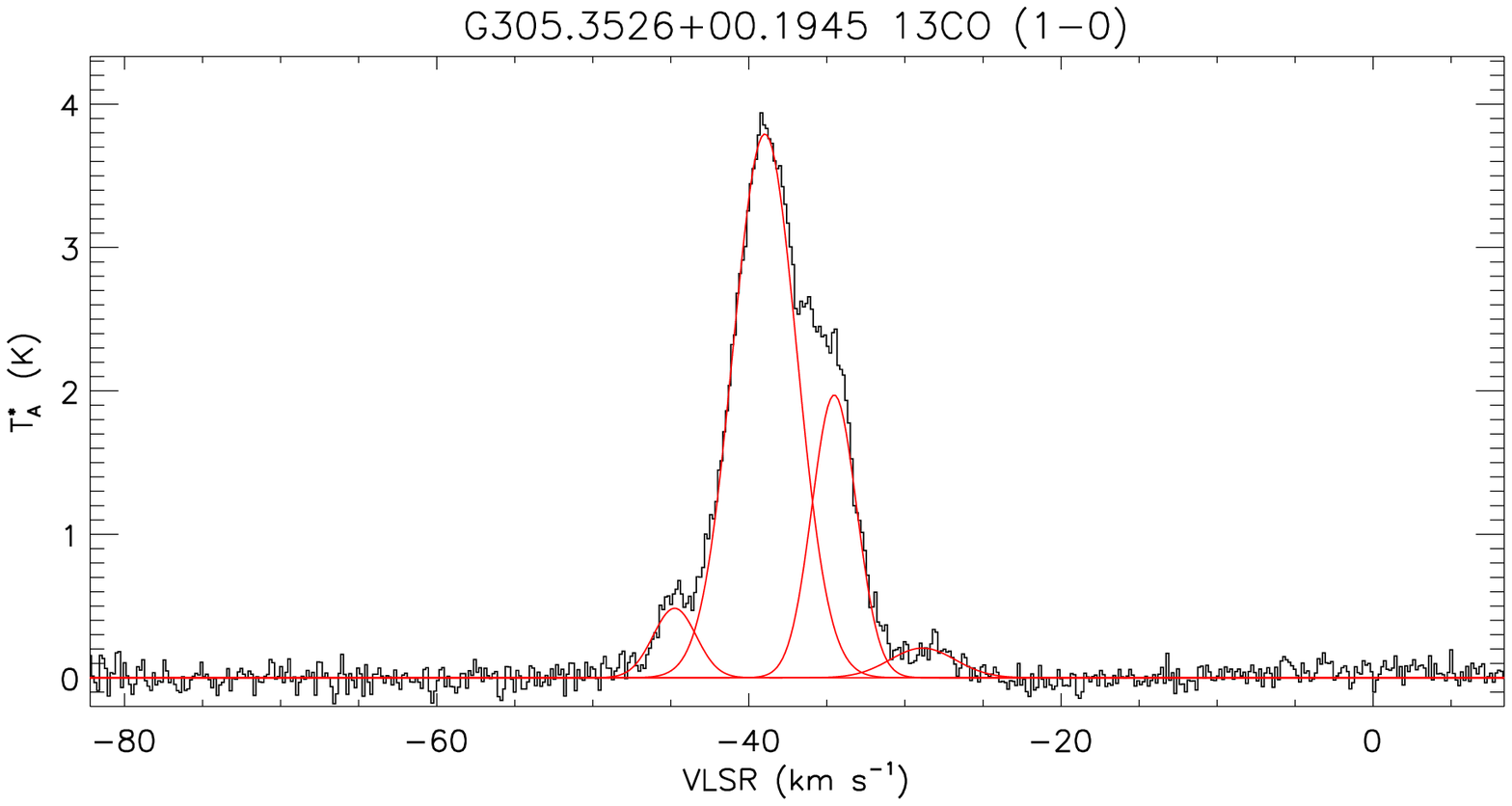} 

\includegraphics[width=0.4\linewidth]{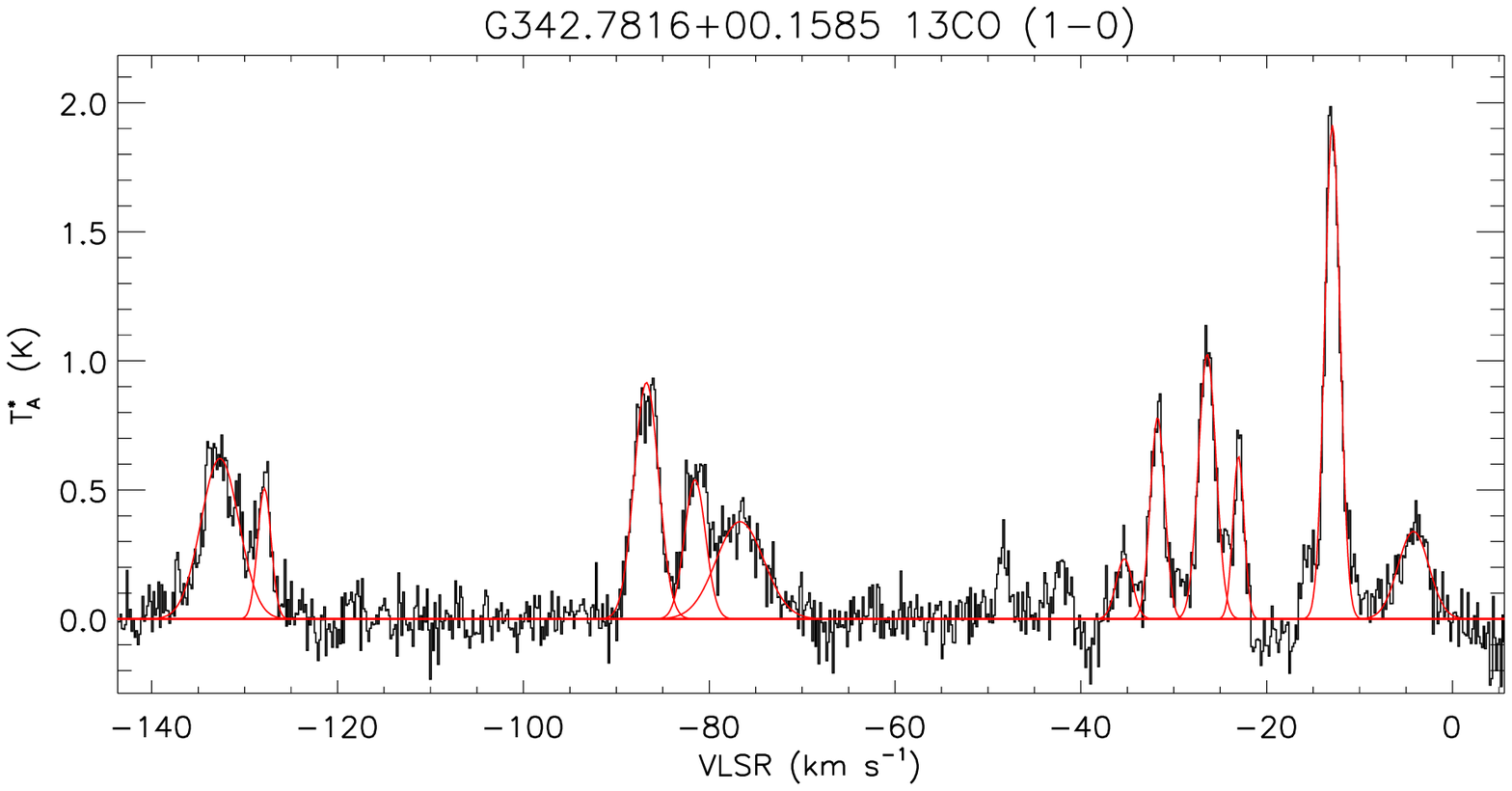}
\includegraphics[width=0.4\linewidth]{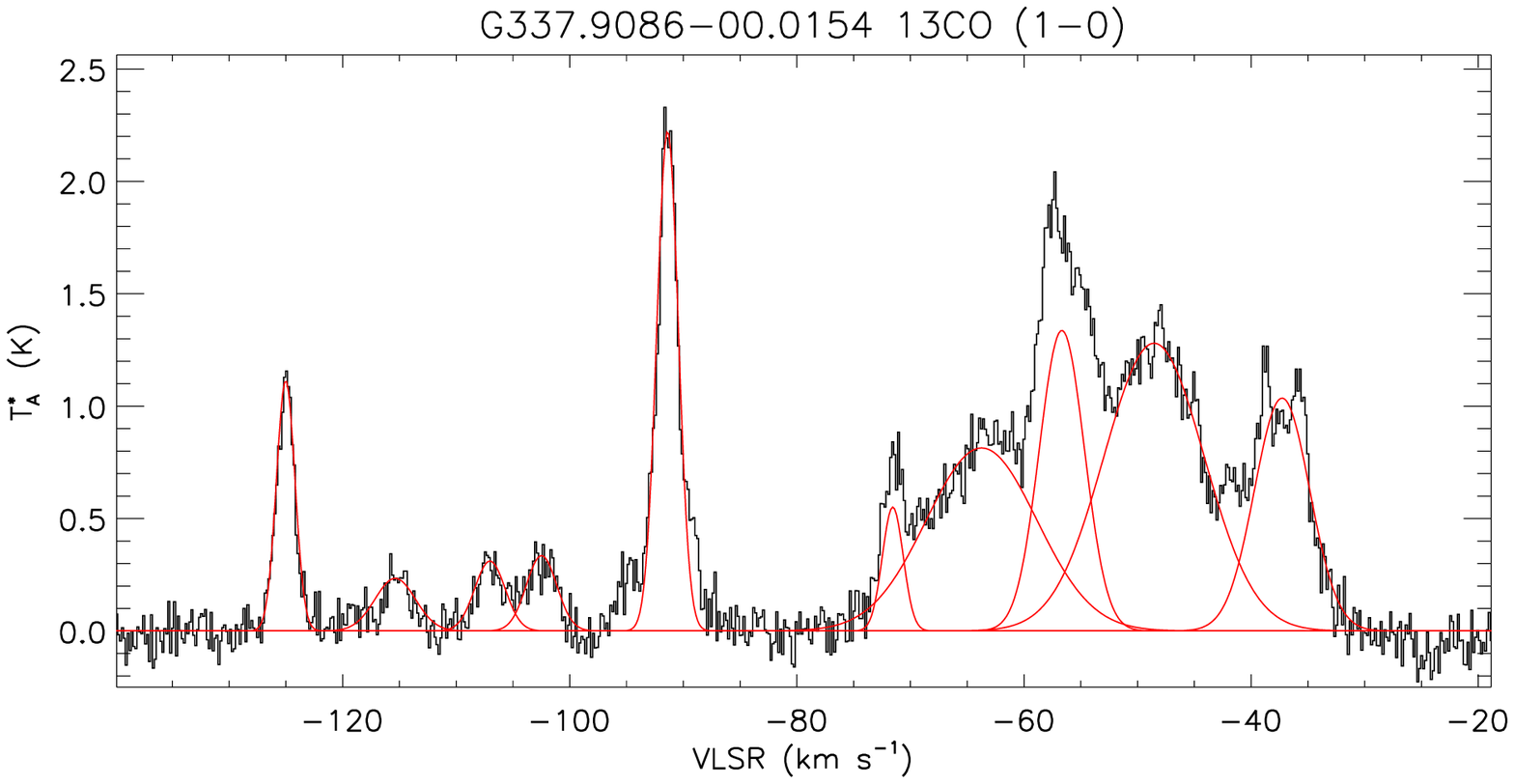}
\includegraphics[width=0.4\linewidth]{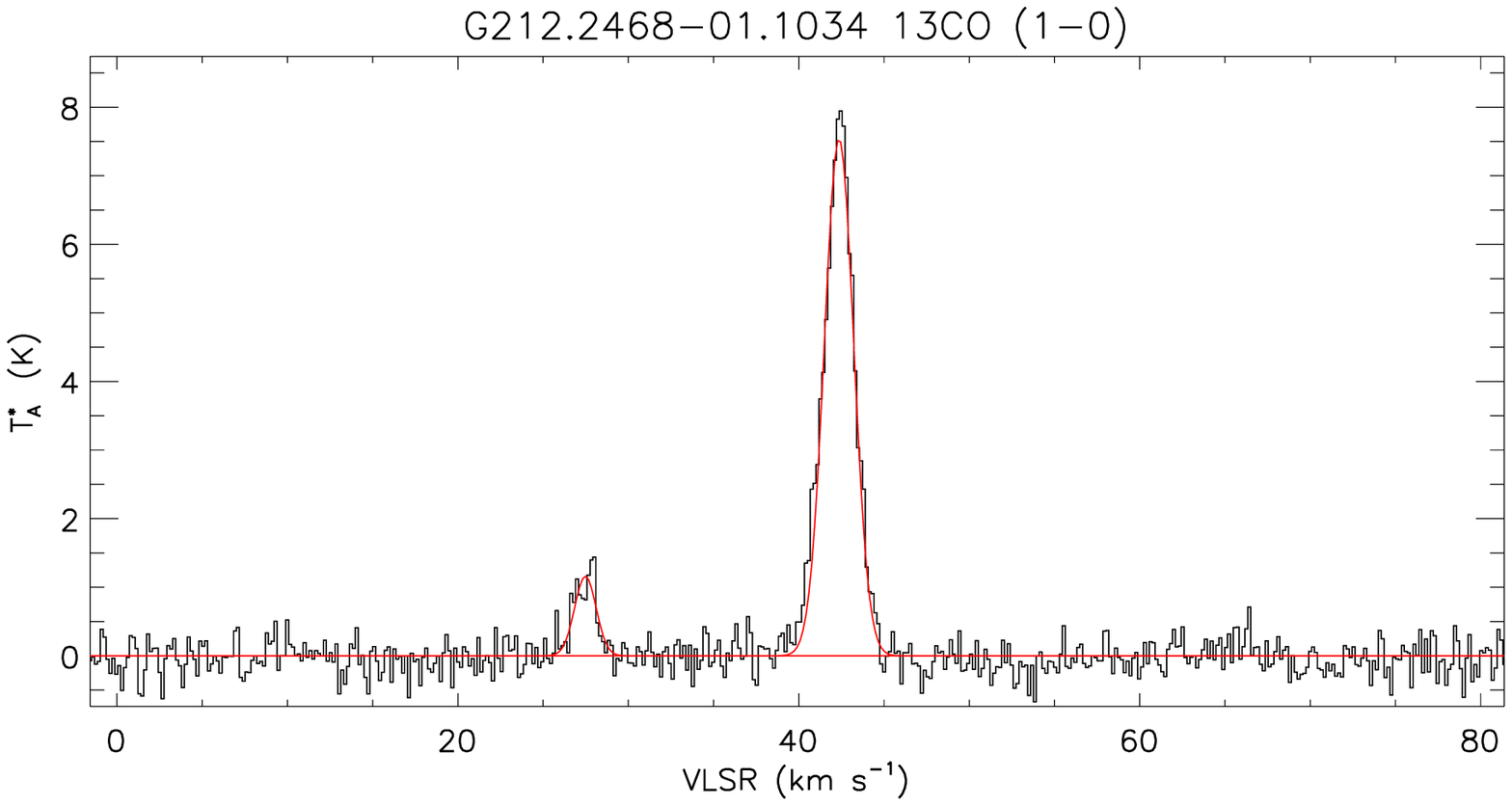}
\includegraphics[width=0.4\linewidth]{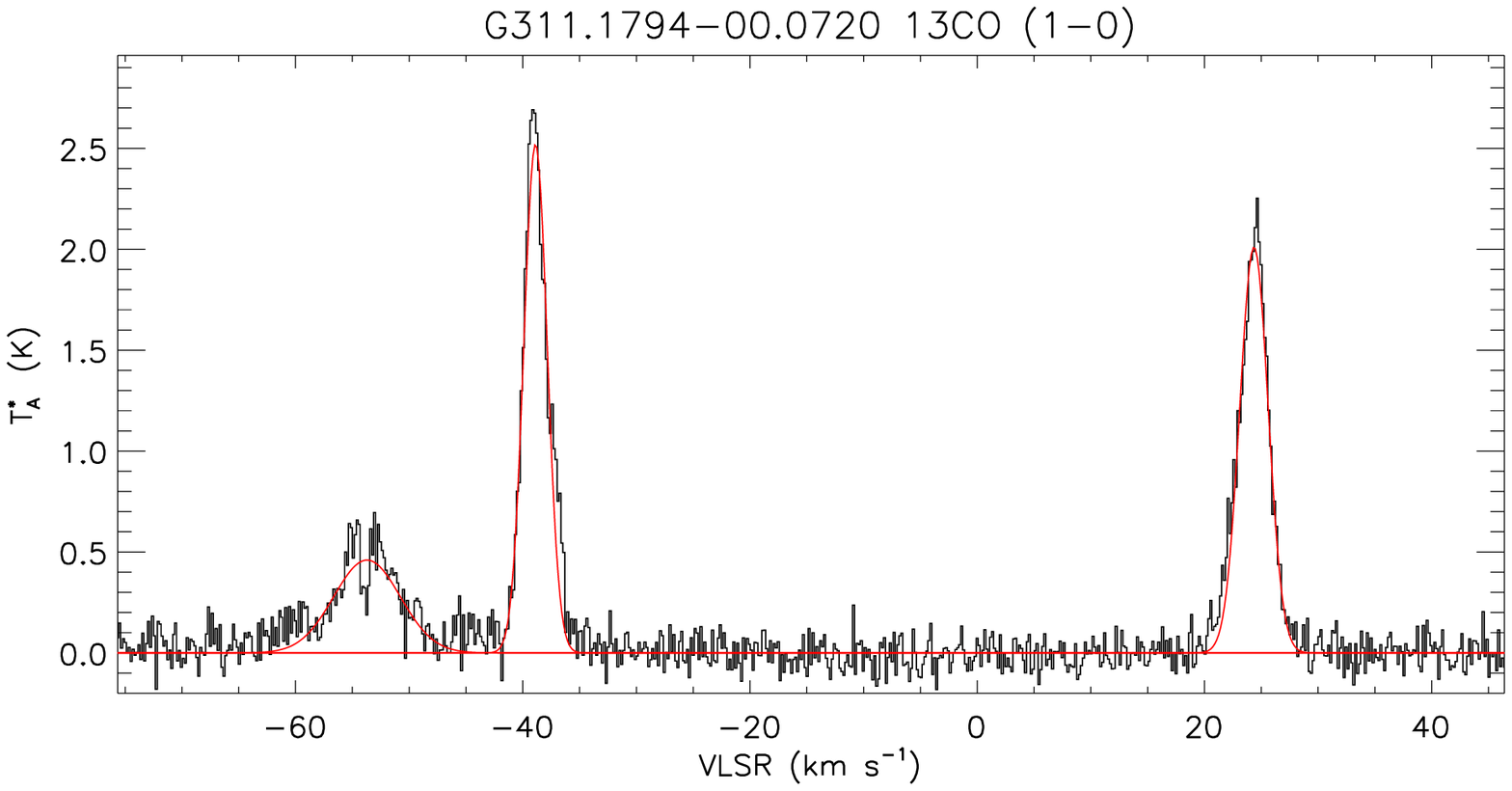}

\caption{Example spectra observed towards RMS sources. The spectra presented in the top four rows have been chosen to illustrate the different kinds of emission profiles exhibited in the data and to clarify the classification scheme used for the non-Gaussian profiles. Moving left to right from the top left the profiles are classified as follows: Gaussian, red wing, blue shoulder, self-absorbed, flat-top, red asymmetic flat-top, blended and blended feature with a main and shoulder component. The bottom two rows show examples of multiple components seen towards the majority of our sample.}

\label{fig:example_spectra}
\end{center}
\end{figure*}

The basic spectral line data reduction and processing was performed using standard procedures and the software package specific to the individual telescopes used: \textsc{DFM} (Data
From Mopra)\footnote{A tcl/tk graphical interface written for
\textsc{SPC} (Spectral Line Reduction Package) by C. Purcell.} for
Mopra data, the \textsc{XS}\footnote{Written by P. Bergman, Onsala
Space Observatory.} package for Onsala data and
\textsc{CLASS}
(Continuum and Line Analysis Single-dish Software)\footnote{Part of the \textsc{GILDAS} (Grenoble
Image and Line Data Analysis Software) working group software.} for JCMT
data. As previously mentioned the observation of each source resulting in between six and ten separate integrations. Sky emission was subtracted from these  individual scans which were inspected to remove poor data. The remaining scans were averaged together to produce a single spectrum for each source and a low-order polynomial was subtracted from the baseline. The resulting 1$\sigma$ sensitivity for each spectrum was typically $T^*_{\rm{A}} \simeq 0.1$~K per channel (see Fig.~\ref{fig:rms_hist}).

The spectral line parameters were obtained by fitting Gaussian profiles to each component present in a given spectrum using \textsc{XS}. Where necessary, higher-order polynomials were fitted and subtracted from the baselines before the Gaussians were fitted. In the majority of cases a Gaussian profile provided a good fit to the data, however, in a significant number of cases, line components displayed significant deviation from a Gaussian profile to warrant an additional comment. We have adopted the classification scheme used by  \citet{wouterloot1989} for their outer Galaxy $^{12}$CO observations of colour selected IRAS  sources. They distinguish seven different non-Gaussian profiles which are:  flat top, asymmetric blue or red top, blue or red asymmetry, self-absorption, blue and/or red wing, blue or red shoulder, self-absorption. 
However, assigning these can be a very subjective exercise and care must be taken when drawing conclusions from any analysis performed. For example, in some cases the distinction between some of the profiles can be quite arbitrary such as between  asymmetric top and shoulder profiles. Moreover, some emission features can exhibit several different kinds of profile simultaneously. In Fig.~\ref{fig:example_spectra} we present some examples of the emission profiles observed towards RMS sources.

In addition to the profiles just mentioned we encountered a number of cases where two or more separate profiles in close proximity overlap with each other to form complex features, or where the shoulder component was particularly pronounced and a single Gaussian produced a poor fit to the data. In these cases it is not clear whether these represent the superposition  of a number of clouds along the line of sight or supersonic motions within a particular cloud. However, their close proximity in velocity and resulting kinematic distances (see Sect.~\ref{sect:kinematic_distance}) would suggest that if they are not associated with the same cloud, they are probably part of the same molecular complex. In some cases it is difficult, if not impossible, to distinguish between a profile that appears to be self-absorbed and two separate, but overlapping components. We identify sources as blended features if either source overlaps with the other above the half-power level. 

Finally the data were calibrated to the corrected receiver temperature scale $T_{\rm{R}}^*$, which accounts for loses due to forward spillover and scattering, by dividing the antenna temperatures by the telescope efficiency where  $T_{\rm{R}}^*=T^*_{\rm{A}}/\eta_{\rm{fss}}$, and $\eta_{\rm{fss}}$ is the forward spillover and scattering efficiency (see Table~\ref{tbl:CO_radio_parameters} for individual telescope efficiencies).
Since the main objective of these observations is to obtain kinematic velocities, rather than accurate line intensities, little absolute calibration was performed. We therefore estimate the calibration uncertainties could be as much as a factor of two.

In a small number of cases some of our spectra have been contaminated by emission present in the reference position which was not evident in the individual integrations and only became apparent when they were averaged together. The data obtained for two sources (i.e., G336.0778+00.0361 and G338.4921+00.0683) were corrupted and will not be presented here. 

\begin{figure}[!t]
\begin{center}

\includegraphics[width=0.9\linewidth]{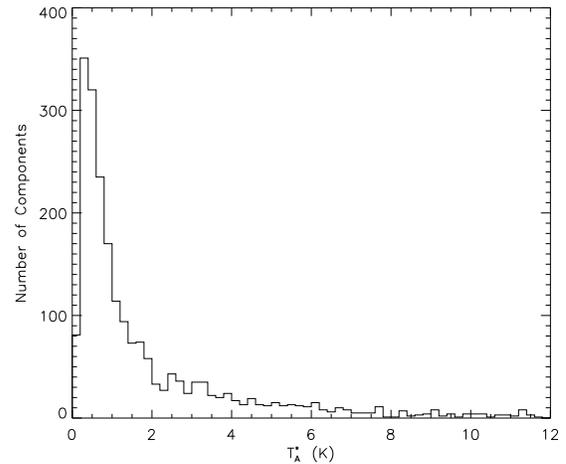}

\caption{Histogram of the distribution of \Ta\ measured towards RMS sources. The hard cutoff under 0.25~K on the temperature plot is due to the sensitivity limit. The bin size is 0.2~K.}

\label{fig:amp_hist}
\end{center}
\end{figure}

\begin{figure}[!t]
\begin{center}

\includegraphics[width=0.9\linewidth]{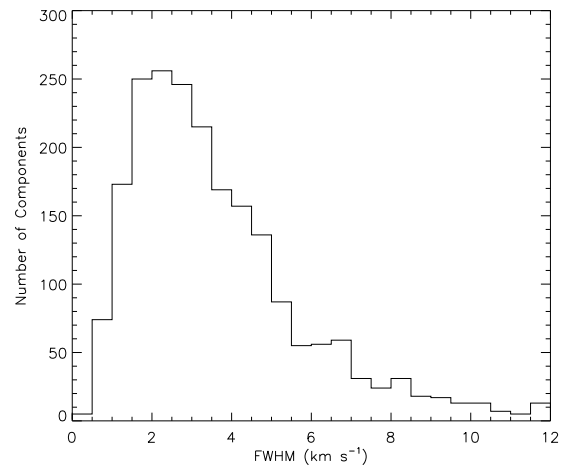}

\caption{Histogram showing the distribution of FWHM observed towards RMS sources. The distribution peaks at $\sim$2.5~\kms. Bin size used is 0.5~\kms.}

\label{fig:fwhm_hist}
\end{center}
\end{figure}
\begin{figure}[!t]
\begin{center}

\includegraphics[width=0.9\linewidth]{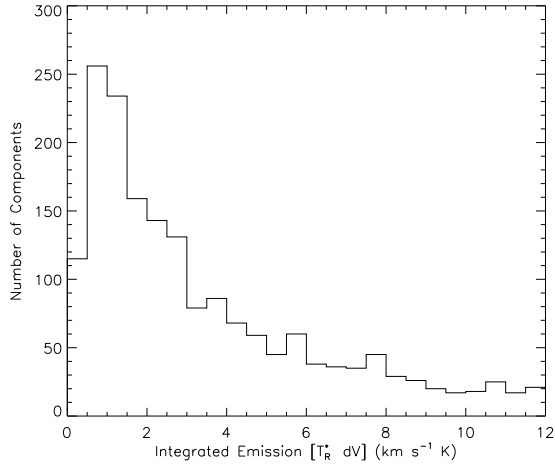}

\caption{Histogram of integrated intensity ($\int T_{\rm{R}}^*~{\rm{d}}V$) of all components detected towards RMS sources.}

\label{fig:int_data}
\end{center}
\end{figure}

\section{Results and analysis}
\label{sect:results}

\subsection{Detection statistics and distributions}

\begin{figure*}
\begin{center}
\includegraphics[width=0.33\linewidth, trim=20 0 20 0]{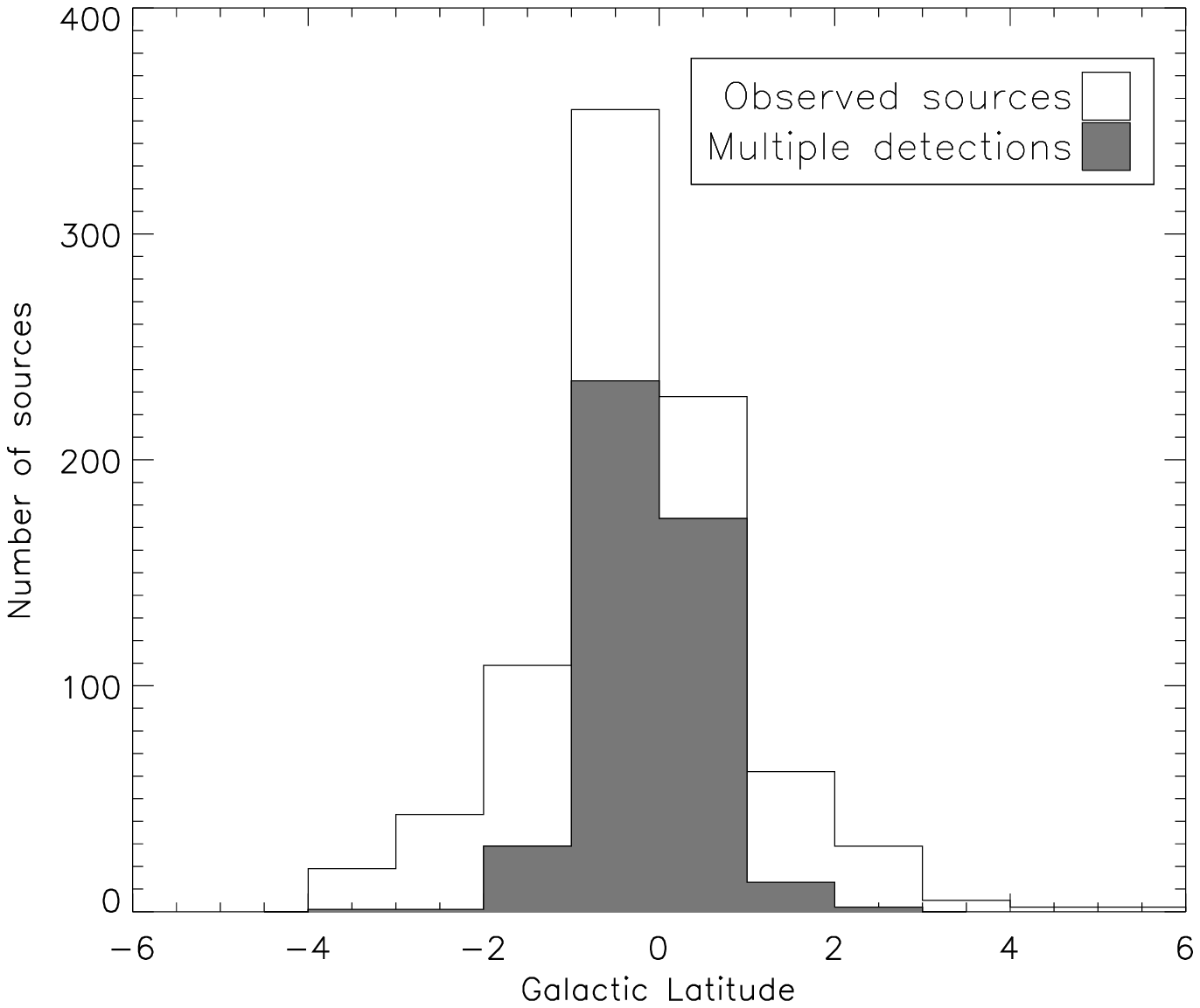}
\includegraphics[width=0.33\linewidth, trim=20 0 20 0]{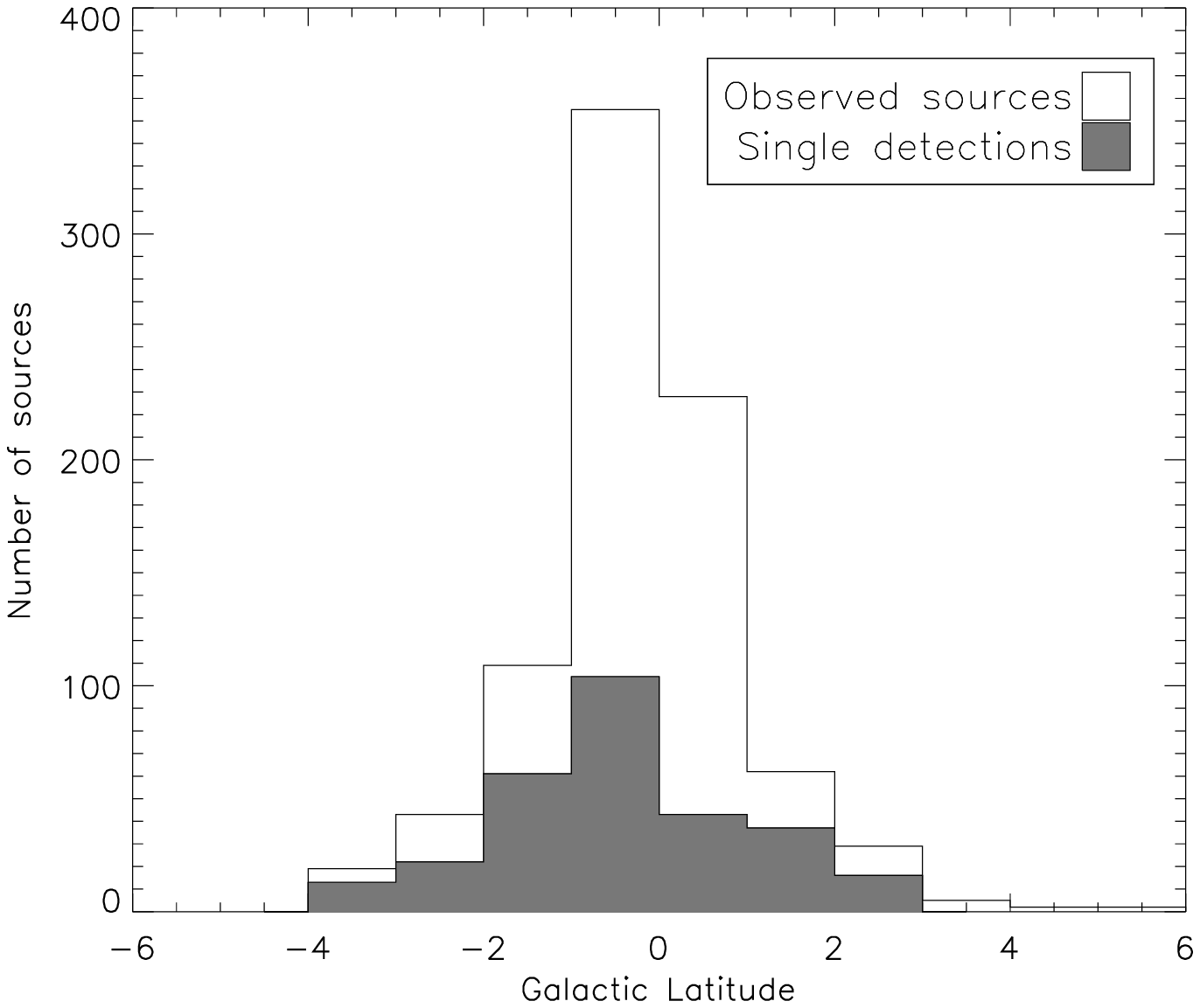}
\includegraphics[width=0.33\linewidth, trim=20 0 20 0]{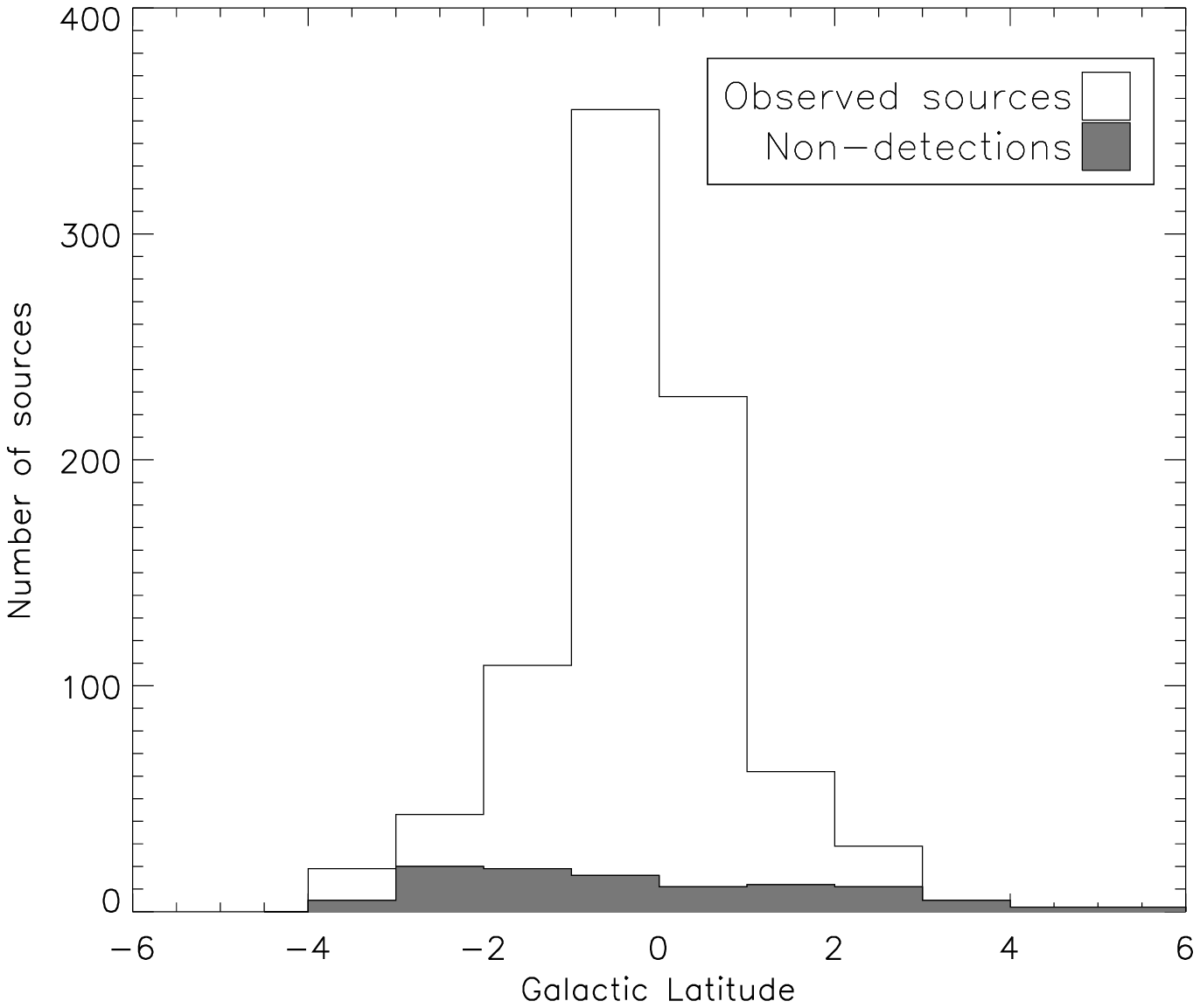}

\caption{Histograms of the Galactic latitude distribution of observed RMS sources. The unfilled histogram in each plot consists of all sources observed and reported in this paper. Histograms of the latitude distributions  of the multiple component detections, single component detections and non-detections are over-plotted (grey histogram) on the left, middle and right panels respectively. A bin size of 1 degree has been used for all of these histograms.}

\label{fig:lat_dist_histo}
\end{center}
\end{figure*} 

We detected $^{13}$CO emission towards 751 of the 854 RMS sources observed ($\sim$88\%) with just over a 100 non-detections. Multiple emission profiles are observed towards the majority of these sources -- 455 sources ($\sim$60\%) -- with an average of $\sim$4 components found in each spectrum. A total of 2185 emission components are detected above 3$\sigma$ (typically $T^*_{\rm{A}} \ge 0.3$~K). In  Figs.~\ref{fig:amp_hist} and \ref{fig:fwhm_hist} we present histograms of the peak temperatures (\Ta) and FWHM of all detected components respectively. The hard cutoff at low \Ta\ seen in Fig.~\ref{fig:amp_hist} is due to the observational detection limit. The average temperature is $\sim$2 K and FWHM is $\sim$4 \kms, however, both distributions are highly skewed to large values and consequently more typical (i.e., median) values  are $\sim$0.7 K and 2.5 \kms\ respectively. In Fig.~\ref{fig:int_data} we present the distribution of integrated intensities ($\int T_{\rm{R}}~{\rm{d}}V=1.06~T^*_{\rm{R}}~dv$, where ${\rm{d}}v$ is the FWHM of the line) of all of the detected components.

We present the parameters obtained from Gaussian fits to the data and derived values in Table~\ref{tbl:source_parameters} and present plots of the fitted spectra in Fig.~11 (only available in the online version). The table format is as follows: the RMS names, positions and observations 1$\sigma$ rms in columns 1--4; component number and parameters extracted from Gaussian fits  (i.e., \vlsr, \Tr\ and FWHM) and profile type in columns 5--9; component integrated intensity in column 10; in columns 11--13 we present the Galactic centre radius and kinematic distances determined from the \citet{brand1993} rotation model (see Sect.~\ref{sect:kinematic_distance} for details). 
 
The spatial distributions of the non-detections, single and multiple component detections can be seen in the Galactic longitude-latitude plot presented in the upper panel of Fig.~\ref{fig:rms_distribution}. In this plot the multiple component detections are clearly concentrated not only towards the Galactic centre but also have a very narrow latitude range. The distribution of the single peak detections are more evenly spread in Galactic longitude but still show a narrow latitude range. The non-detections appear to be evenly distibuted in both longitude and latitude. The different latitude distributions can be clearly seen in the histograms presented in Fig.~\ref{fig:lat_dist_histo}. The distribution of the multiple component sources, which constitute the majority of the detections, is strongly peaked around the Galactic mid-plane with nearly all of these sources confined within two degrees of the plane, and the majority within one degree. The single component detections are also strongly peaked, but have a much broader latitude distribution. The non-detections appears to be evenly distributed in latitude; these sources are most likely evolved stars which have similar colours to embedded YSOs but have very little, if any, associated molecular material. On closer examination of the distribution of non-detections there appears to be a minimum at $|b|\simeq 0$\degr\ possibly indicating an unfortunate line of sight alignment between a small number of evolved stars and molecular clouds towards the Galactic mid-plane. However, these will be identified through our near infrared spectroscopic observations.

\subsection{Identification of multiple components}

\begin{figure}[!t]
\begin{center}

\includegraphics[width=0.9\linewidth]{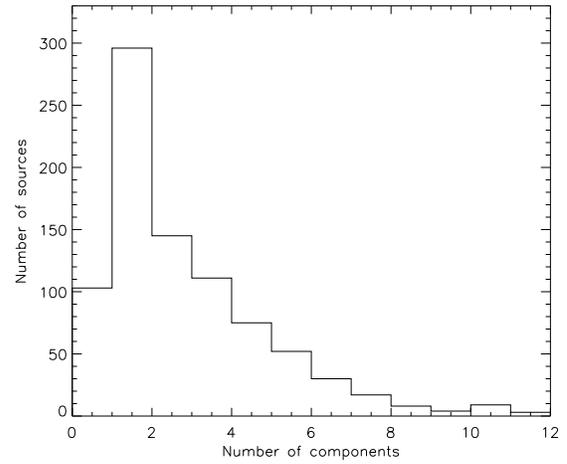}

\caption{Histogram of the number of components detected towards RMS sources reported in this paper.}

\label{fig:comp_hist}
\end{center}
\end{figure}

Although using the $^{13}$CO transition has proved successful in detecting emission towards potential MYSOs one significant drawback is the large number of components detected towards the majority of our sample. A histogram of the number of components detected towards each of RMS source is presented in Fig.~\ref{fig:comp_hist}. As previously mentioned we re-observed a number of sources towards which \citet{bronfman1996} has reported detecting CS emission. In sources towards which only a single CS and $^{13}$CO component have been detected we find their velocities agree to better than 1~\kms. Using this velocity correlation  we are able to identify the most likely component associated with an RMS source where multiple components are detected in 18 cases using CS ($J$=2--1) results found in the literature (i.e., \citealt{bronfman1996,fontani2005}). Additionally, in a number of cases the associated component can easily be identified as they are far stronger than any other component seen in the spectrum. However, identifying the correct component is not always so obvious, and before we can assign kinematic distances to our MYSOs candidates we need to find a way to resolve these multi-component sources. In bottom panels of Fig.~\ref{fig:example_spectra} we present two examples of multiple component detections. In the left hand panel it is easy to identify the component associated with the RMS source, however, in the panel on the right two of the components are approximately equal in strength making it difficult to identify the one associated with the RMS source.

\begin{figure}
\begin{center}

\includegraphics[width=0.90\linewidth]{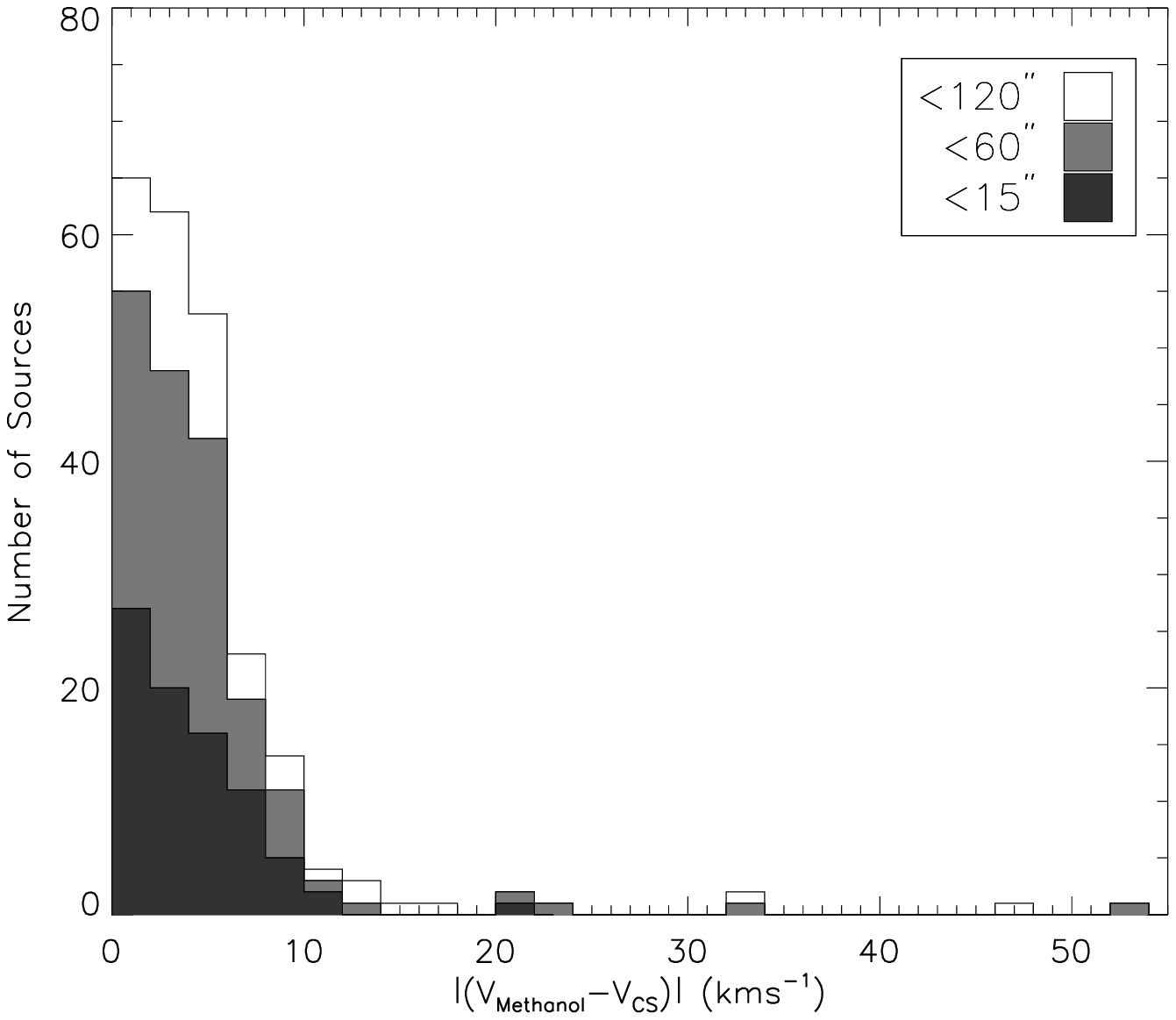}
\includegraphics[width=0.90\linewidth]{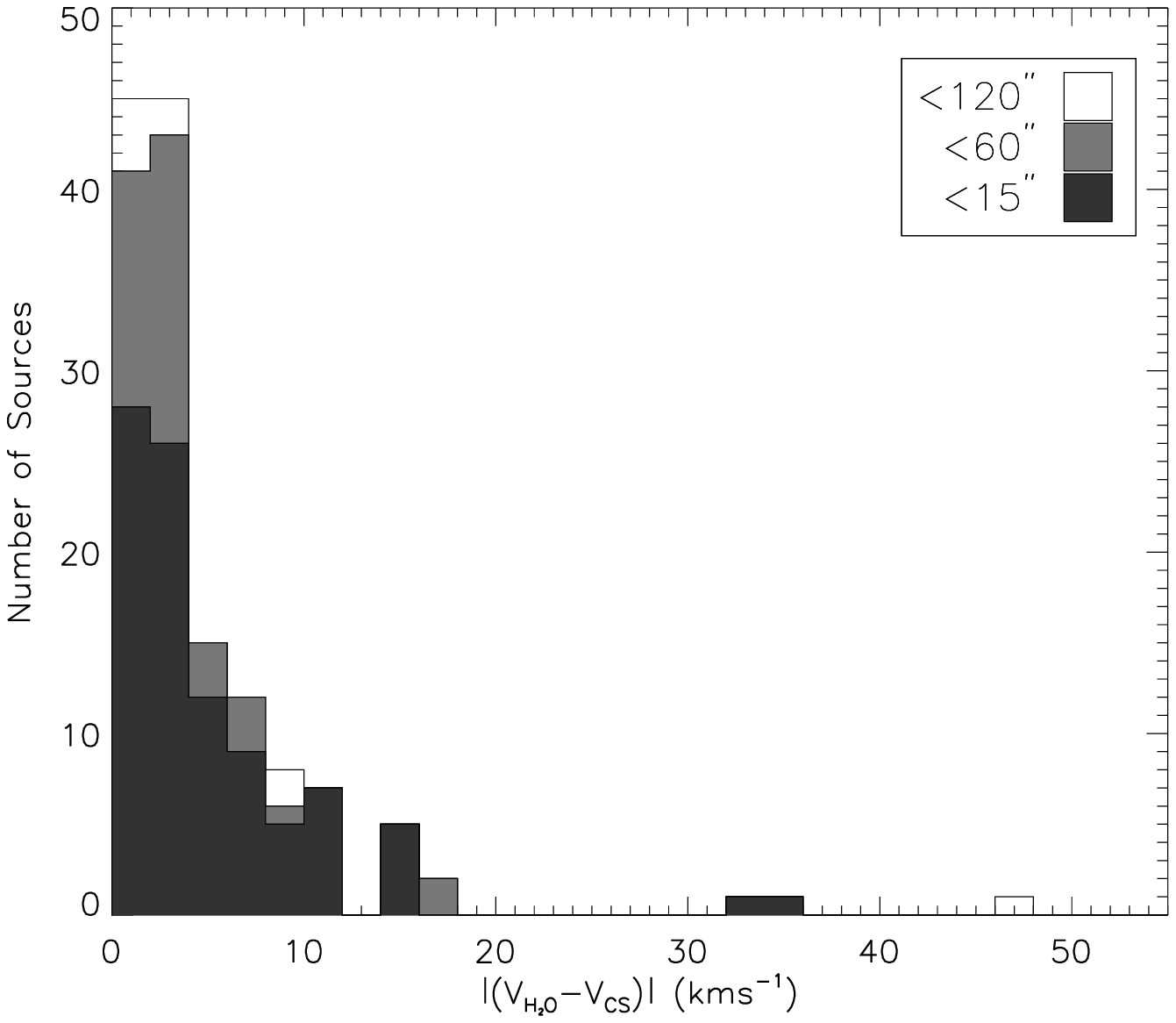}

\caption{Histogram illustrating the difference between the velocities of methanol (upper panel) and water masers (lower panel) and the velocities of their host molecular clouds as traced by CS (see text for details). The bin size is 2 \kms\ and the total number of sources used in each plot are 233 and 148 for methanol and water masers respectively. The unfilled histogram includes all maser-CS matches within a 120\arcsec\ search radius, over-plotted in light and dark grey include all matches with in a 60\arcsec\ and 15\arcsec\ radius respectively. The velocity range of the water-CS sources have been truncated to facilitate comparison between the to distributions shown, however, only a few sources have velocity differences of greater than the range plotted.}

\label{fig:maser_velocities}
\end{center}
\end{figure}

Water and methanol masers are often associated with massive star forming regions and their velocities may help identify the component of interest towards some of these multiple component detections. Maser emission can consist of many components spread over a wide range of velocities, typically 10--15~\kms\ for methanol and up to 70 \kms\ for water masers (\citealt{sridharan2002}). Despite these large velocity ranges and number of components, the velocity of the brightest component correlates extremely well with the velocity of the molecular clouds with which they are associated. In order to derive a velocity criterion with which to associate masers with a particular molecular component we cross-matched methanol masers from the \citet{pestalozzi2005} catalogue and water masers from the Arcetri catalogue (\citealt{valdettaro2001}) with CS detections reported by \citet{bronfman1996}. In order to see how the velocities of the molecular clouds and masers varied as a function of angular separation between them we used three search radii (120\arcsec, 60\arcsec\ and 15\arcsec) to match masers and CS sources. We found 233 methanol masers and 148 water masers in common with the CS sources within the 120\arcsec\ search radius. In Fig.~\ref{fig:maser_velocities} we present histograms of the differences in the peak methanol and water maser velocities compared to the CS velocities as a function of angular separation.

These histograms reveal a fairly robust correlation, even at rather large angular separations, between the observed velocities of CS emission which gives the velocity of the molecular cloud and the velocities of nearby masers. These  plots show that both distributions level off at approximately the same $\Delta V$ ($\sim$15~\kms) with standard deviations of $\sim$2.5~\kms\ and $\sim$3.5~\kms\ for methanol and water masers respectively for all of the angular distributions plotted (source with velocity differences greater than 200~\kms\ have been excluded). It is therefore possible to identify the most likely CO component in cases where the difference in velocity between any two CO components is large and the velocity difference between a particular CO component and the maser velocities is smaller than three times the standard deviation (3$\sigma$ $\simeq$ 7.5 \kms\ and 10.5 \kms\ for methanol and water masers respectively). 

We have used these 3$\sigma$ values to identify the most likely star forming cloud where multiple components have been detected and either a methanol or water maser has been found within 120\arcsec. We found 78 methanol masers and 12 water masers associated with 79 RMS sources and are able use this association criterion to resolve the multiple component ambiguity towards 64 sources. Using a combination of the archival maser data and CS data we are able to identify the correct component associated with an RMS source in 82 of the 455 multiple component cases. Although using these other tracers of massive star formation has proved useful in reducing the number of multiple component sources by $\sim$20\%, the vast majority  remain unresolved.  We notice that in cases where we have been able to associate a component with an RMS source it is nearly always the component with the largest integrated intensity ($\int T_{\rm{R}}~{\rm{d}}V$) that is identified by maser or CS velocity. In fact the masers and CS lines are almost exclusively associated with the strongest components (in 78 of the 82 cases solved) and only in a few cases (4) is the emission associated with slightly weaker components, but never less than half that of the strongest component. Therefore we can use this information to identify the most likely component associated with the MYSO by picking the strongest component as long as it is more than twice as strong as any other component present in the data.

By applying this criterion to our data we are able to resolve the component ambiguity towards a further 202 sources, bringing the total number resolved to 284 or 63\% of the multi-component detections. Together with the single component detections we have been able to determine kinematic velocities towards $\sim$80\% of our detections. For sources where we are able to resolve the component multiplicity we indicate the most likely component associated with an asterisk in the component number column of Table~\ref{tbl:source_parameters}. Additional observational data will be required to resolve the remaining 171 sources towards which multi-components are detected. A number of these may be resolved using data from the Methanol Multibeam (MMB) survey)\footnote{http://www.jb.man.ac.uk/research/methanol} which is currently underway and plans to survey the whole Galactic plane for $|b| < 2\degr$. For the remaining sources we plan to resolve their component ambiguities with a programme of CS, water masers and CO mapping observations.

\subsection{Kinematic distances}
\label{sect:kinematic_distance}

\begin{figure}
\begin{center}
\includegraphics[width=0.9\linewidth]{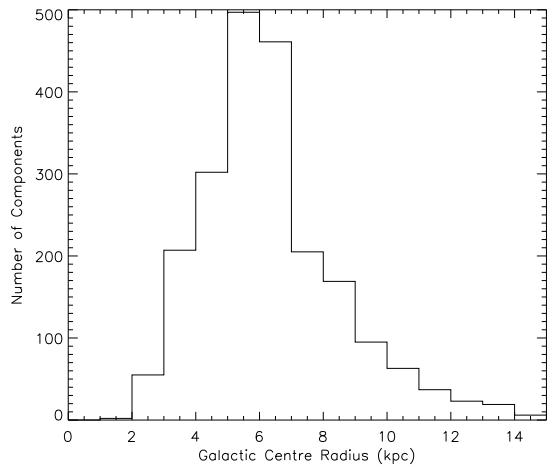}

\caption{Histogram showing the distribution of clouds with respected to their distance from the Galactic centre. The distribution peaks at between 5 and 6 kpc consistent with the location of the 5 kpc molecular ring which surrounds the Galactic centre. The bin size used is 1~kpc.}

\label{fig:rgc_hist}
\end{center}
\end{figure}

We have used the rotation curve of \citet{brand1993} to calculate kinematic distances and  Galactocentric radii to all detected components. We have assumed the distance to the Galactic centre is 8.5~kpc and a solar velocity of 220~\kms. In Fig.~\ref{fig:rgc_hist} we present a histogram of the Galactic radii of all detected components, the distribution peaks between 5 and 6~kpc, consistent with the location of the thick ring of material which surrounds the centre of the Galaxy. The vast majority of sources in our catalogue are located within the solar circle ($\sim$80\%) and therefore their \vlsr\ results in two possible distances equally spaced on either side of the tangent point; these are referred to as the near and far distances. This is known as the kinematic distance ambiguity problem and makes determining accurate distances to objects located within the solar circle difficult. The only sources within the solar circle that do not suffer from this ambiguity are sources located at the tangent point. 

The Galactocentric radii and kinematic distances determined from the rotation model for all sources located between 192\degr\ $< l <$ 350\degr\ are presented in columns 11--13 of Table~\ref{tbl:source_parameters}. We have not presented distances for sources located near the Galactic anti-centre since they are very uncertain. Sources with components outside the solar circle have only a single distance as do sources located at the tangent points. Errors in the distances have been determined by shifting the component \vlsr\ by $\pm$10~\kms\ to account for peculiar motions; these result in typical errors of $\sim$1~kpc. If the near and far distances for a particular component are within 1~kpc of each other we place it at the tangent point. In total 119 components were found to be located at or near the tangent point leaving 1959 components with ambiguous distances.

There are a number of methods discussed in the literature that can be applied to solve the distance ambiguity such as: H$_2$CO (formaldehyde) absorption (e.g., \citealt{downes1980,araya2001,araya2002}) or radio recombination lines in conjunction with HI absorption toward HII regions (e.g., \citealt{kolpak2003}), and HI self-absorption and molecular line emission towards molecular clouds and MYSOs (e.g., \citealt{jackson2002,busfield2006}). The HI self-absorption method has already been successfully used to resolve distance ambiguities towards a subset of the RMS sample (see \citealt{busfield2006} for details). We will use the HI methods to resolve the kinematic distance ambiguity towards the remaining sources in a subsequent paper. 

\section{Summary}
\label{sect:summary}
We present the results of a programme of molecular line observations of towards 854 candidate massive young stellar objects (MYSOs) located in the 3rd and 4th quadrant of the Galaxy. These have been made using the Mopra and Onsala telescopes and the JCMT using the $J$=1--0 and $J$=2--1 rotational transitions of the $^{13}$CO molecules. We detected emission towards 751 MYSOs with multiple components being seen in 60\% of cases. The multiple detections are  concentrated within 2\degr\ of the Galactic mid-plane and strongly peaked within 1\degr. The single component detections are also strongly peaked but have a slightly broader distribution in Galactic latitude $|b|$ $\pm$3\degr. The non-detections are evenly distributed over a latitude range of $\pm$4\degr, consistent with them being identified as evolved stars.

We have used archival methanol and water maser, and CS ($J$=2--1) molecular line velocities to resolve the component ambiguity towards 82 MYSOs candidates where multiple components have been detected. Using the results of our maser and CS associations we have derived a criterion for selecting the most likely component for sources where no useful tracer is present. In total we have been able to identify the component associated with our MYSO candidates in 284 cases of the 455 multiple detections. Combined with the single component detections we have obtained unambiguous kinematic velocities towards 580 sources ($\sim$80\% of the detections). The 171 sources for which we have not been able to determine the kinematic velocity will require additional line data.

These observations of southern sources form part of a larger programme of follow-up observations of near and mid-infrared colour selected sample of MYSO candidates. The results of observations made towards sources in the northern Galactic plane will be presented in a subsequent paper.

\begin{acknowledgements}

The authors would like to thank the Director and staff of the Paul Wild Observatory, the JCMT and Onsala for their hospitality and assistance during our observing runs. We would also like to thank Thomas Dame for providing the integrated $^{12}$CO maps used in Fig.~1. JSU and CRP are supported by PPARC postdoctoral fellowship grants; CRP was also supported by a UNSW Scholarship during the observing. This research would not have been possible without the SIMBAD astronomical database service operated at CDS, Strasbourg, France and the NASA Astrophysics Data System Bibliographic Services. 

\end{acknowledgements}

\bibliography{scuba2.bib}

\bibliographystyle{aa}


\clearpage
\pagestyle{empty}
\begin{onecolumn}
\begin{landscape}

\footnotesize
\begin{longtable}{lcccl....l.ccc}
\caption{\label{tbl:source_parameters} Sample table of source parameters.}\\
\hline\hline
\multirow{2}{24mm}{Field Name$^a$} 	&	RA  		&Dec  		& rms & Id.$^b$	&	\multicolumn{1}{c}{\vlsr} 	& \multicolumn{1}{c}{\Tr}	& \multicolumn{1}{c}{FWHM} 	& \multicolumn{1}{c}{\Tv}&Profile type & \multicolumn{1}{c}{RGC} &Near	&Far & Notes \\ 
&	(J2000) 	& (J2000) 	& (K) & 		& \multicolumn{1}{c}{\kms} &\multicolumn{1}{c}{(T)} 	&\multicolumn{1}{c}{(\kms)} 	&  \multicolumn{1}{c}{(\kms\ K)}&				&\multicolumn{1}{c}{(kpc)}& \multicolumn{1}{c}{(kpc)}& \multicolumn{1}{c}{(kpc)} &  \\  
\hline
\endfirsthead
\caption{continued}\\
\hline \hline
\multirow{2}{24mm}{Field Name$^a$} 	&	RA  		&Dec  		& rms & Id.$^b$	&	\multicolumn{1}{c}{\vlsr} 	& \multicolumn{1}{c}{\Tr}	& \multicolumn{1}{c}{FWHM} 	& \multicolumn{1}{c}{\Tv}&Profile type & \multicolumn{1}{c}{RGC} &Near	&Far & Notes \\ 
&	(J2000) 	& (J2000) 	& (K) & 		& \multicolumn{1}{c}{\kms} &\multicolumn{1}{c}{(T)} 	&\multicolumn{1}{c}{(\kms)} 	&  \multicolumn{1}{c}{(\kms\ K)}&				& \multicolumn{1}{c}{(kpc)}& \multicolumn{1}{c}{(kpc)}& \multicolumn{1}{c}{(kpc)} &  \\  
\hline
\endhead
\hline \multicolumn{14}{l}{$^a$ Observations made with the JCMT and Onsala telescope are indicated by $\dag$ and $\ddag$ respectively.}\\ 
\multicolumn{14}{l}{$^b$ Where multiple components have been detected we identify the most likely component associated with the RMS source by an $\star$.}\\ 
\multicolumn{14}{l}{Notes: (1) strongest component; (2) methanol maser; (3) water maser; (4) and (5) CS detection by \citet{bronfman1996} and \citet{fontani2005}.}\\ 
\endfoot
\hline
\endlastfoot
G300.1615$-$00.0877 & 12:27:08.16 & $-$62:49:43.68 & 0.07 & 1 &$-$39.4 & 2.9 & 2.4 &    7.4 &  & 7.1 &
\multicolumn{2}{c}{4.1 (0.72)} &  \\ 
G300.3412$-$00.2190 & 12:28:36.00 & $-$62:58:33.24 & 0.10 & 1  $^\star$  &$-$38.5 & 6.7 & 2.0 &   14.2 & blue shoulder & 7.1 &
\multicolumn{2}{c}{4.2 (0.84)} & 1 \\ 
 &  &  &  &2 &34.6 & 1.3 & 4.2 &    5.8 &  & 10.6 &
\multicolumn{2}{c}{12.0 (0.99)} &  \\ 
G300.5320$-$01.7926 & 12:29:04.32 & $-$64:33:36.72 & 0.07 & 1 &$-$0.9 & 0.3 & 1.0 &    0.3 &  & 8.6 &
\multicolumn{2}{c}{8.8 (0.91)} &  \\ 
G300.7027+03.0398 & 12:33:44.88 & $-$59:45:19.44 & 0.08 &\multicolumn{1}{c}{$\cdots$} & \multicolumn{1}{c}{$\cdots$} &\multicolumn{1}{c}{$\cdots$} &\multicolumn{1}{c}{$\cdots$} &\multicolumn{1}{c}{$\cdots$} &\multicolumn{1}{c}{$\cdots$} &\multicolumn{1}{c}{$\cdots$} &\multicolumn{1}{c}{$\cdots$} &\multicolumn{1}{c}{$\cdots$}   \\ 
G300.7221+01.2007 & 12:32:50.88 & $-$61:35:28.68 & 0.11 & 1 &$-$43.0 & 9.6 & 2.2 &   22.4 & blue shoulder & 7.0 &
\multicolumn{2}{c}{4.1 (0.19)} &  \\ 
G300.9785+01.1459 & 12:34:58.08 & $-$61:39:48.24 & 0.10 & 1 &$-$43.7 & 6.4 & 3.3 &   22.4 & red asymmetry top & 6.9 &
\multicolumn{2}{c}{4.2 (0.19)} &  \\ 
G301.0130+01.1153 & 12:35:14.64 & $-$61:41:45.96 & 0.10 & 1 &$-$43.1 & 6.1 & 3.3 &   21.3 & red shoulder & 7.0 &
\multicolumn{2}{c}{4.2 (0.36)} &  \\ 
G301.1067$-$01.4919 & 12:34:35.52 & $-$64:18:14.76 & 0.09 &\multicolumn{1}{c}{$\cdots$} & \multicolumn{1}{c}{$\cdots$} &\multicolumn{1}{c}{$\cdots$} &\multicolumn{1}{c}{$\cdots$} &\multicolumn{1}{c}{$\cdots$} &\multicolumn{1}{c}{$\cdots$} &\multicolumn{1}{c}{$\cdots$} &\multicolumn{1}{c}{$\cdots$} &\multicolumn{1}{c}{$\cdots$}   \\ 
G301.1175+00.9627 & 12:36:02.64 & $-$61:51:17.28 & 0.20 & 1 &$-$41.6 & 6.7 & 4.7 &   33.4 & red shoulder & 7.0 &
\multicolumn{2}{c}{4.2 (0.66)} &  \\ 
G301.1726+01.0034 & 12:36:31.92 & $-$61:49:02.64 & 0.10 & 1 &$-$40.6 & 5.4 & 3.8 &   21.8 &  & 7.0 &
\multicolumn{2}{c}{4.3 (0.78)} &  \\ 
G301.5309$-$00.8231 & 12:38:48.24 & $-$63:39:38.16 & 0.07 & 1 &$-$5.1 & 3.4 & 0.7 &    2.5 &  & 8.4 &
0.2 (0.93) &8.7 (0.91)&  \\ 
G301.8147+00.7808 & 12:41:53.76 & $-$62:04:13.08 & 0.07 & 1 &$-$37.4 & 9.3 & 2.1 &   20.7 & red wing & 7.1 &
\multicolumn{2}{c}{4.4 (1.11)} &  \\ 
G302.0213+00.2542 & 12:43:31.20 & $-$62:36:12.96 & 0.08 & 1  $^\star$  &$-$46.1 & 4.0 & 2.9 &   12.3 &  & 6.9 &
\multicolumn{2}{c}{4.3 (0.19)} & 1 \\ 
 &  &  &  &2 &$-$37.7 & 3.1 & 1.2 &    3.9 & blue shoulder & 7.1 &
\multicolumn{2}{c}{4.5 (1.11)} &  \\ 
G302.4546$-$00.7401 & 12:47:08.64 & $-$63:36:28.80 & 0.07 &1 &30.4 & 0.5 & 3.1 &    1.6 & blended & 10.4 &
\multicolumn{2}{c}{12.1 (0.97)} &  \\ 
 &  &  &  & 2  $^\star$  &35.3 & 1.3 & 3.2 &    4.4 & blended & 10.7 &
\multicolumn{2}{c}{12.6 (1.00)} & 1 \\ 
G302.4867$-$00.0308 & 12:47:31.68 & $-$62:53:57.48 & 0.10 & 1  $^\star$  &$-$37.1 & 4.4 & 3.1 &   14.5 & blended (main component) & 7.1 &
\multicolumn{2}{c}{4.5 (1.19)} & 4 \\ 
 &  &  &  &2 &$-$34.2 & 2.6 & 2.1 &    5.8 & blended (red shoulder) & 7.2 &
3.6 (1.17) &5.6 (1.34)&  \\ 
 &  &  &  &3 &$-$29.1 & 0.9 & 4.0 &    3.8 &  & 7.4 &
2.7 (1.50) &6.5 (1.56)&  \\ 
G302.5005$-$00.7701 & 12:47:33.12 & $-$63:38:19.32 & 0.10 & 1 &30.0 & 2.6 & 3.2 &    8.8 & blue shoulder & 10.4 &
\multicolumn{2}{c}{12.1 (0.96)} &  \\ 
G302.6604$-$00.7908 & 12:48:59.52 & $-$63:39:40.68 & 0.10 &1 &23.1 & 0.7 & 3.4 &    2.5 & blended & 9.9 &
\multicolumn{2}{c}{11.4 (0.92)} &  \\ 
 &  &  &  & 2  $^\star$  &26.4 & 1.3 & 2.1 &    2.9 & blended & 10.1 &
\multicolumn{2}{c}{11.8 (0.94)} & 1 \\ 
G303.5353$-$00.5982 & 12:56:50.40 & $-$63:27:49.32 & 0.10 &1 &18.9 & 3.3 & 2.3 &    8.0 &  & 9.7 &
\multicolumn{2}{c}{11.3 (0.90)} &  \\ 
 &  &  &  &2 &31.6 & 0.9 & 4.1 &    3.9 &  & 10.5 &
\multicolumn{2}{c}{12.5 (0.98)} &  \\ 
G303.5990$-$00.6524 & 12:57:25.20 & $-$63:30:59.40 & 0.08 &1 &$-$2.3 & 1.6 & 0.4 &    0.7 &  & 8.5 &
\multicolumn{2}{c}{9.4 (0.87)} &  \\ 
 &  &  &  & 2  $^\star$  &19.1 & 1.8 & 2.5 &    4.8 &  & 9.7 &
\multicolumn{2}{c}{11.3 (0.90)} & 1 \\ 
G303.9973+00.2800 & 13:00:41.52 & $-$62:34:21.72 & 0.10 &1 &$-$45.1 & 1.1 & 2.3 &    2.7 & blended & 6.9 &
\multicolumn{2}{c}{4.6 (0.87)} &  \\ 
 &  &  &  &2 &-42.3 & 2.2 & 2.4 &    5.6 & blended & 6.9 &
\multicolumn{2}{c}{4.7 (1.06)} &  \\ 
 &  &  &  &3 &-28.9 & 0.4 & 4.4 &    1.9 &  & 7.4 &
2.5 (1.47) &7.0 (1.47)&  \\ 
 &  &  &  &4 &30.7 & 2.4 & 2.5 &    6.4 &  & 10.5 &
\multicolumn{2}{c}{12.5 (0.97)} &  \\ 
G304.5719+00.7163 & 13:05:28.08 & $-$62:06:38.88 & 0.10 &\multicolumn{1}{c}{$\cdots$} & \multicolumn{1}{c}{$\cdots$} &\multicolumn{1}{c}{$\cdots$} &\multicolumn{1}{c}{$\cdots$} &\multicolumn{1}{c}{$\cdots$} &\multicolumn{1}{c}{$\cdots$} &\multicolumn{1}{c}{$\cdots$} &\multicolumn{1}{c}{$\cdots$} &\multicolumn{1}{c}{$\cdots$}   \\ 
G304.7700$-$00.5193 & 13:07:49.92 & $-$63:19:57.72 & 0.11 & 1 &24.6 & 1.7 & 1.9 &    3.4 &  & 10.0 &
\multicolumn{2}{c}{12.1 (0.94)} &  \\ 
G305.1097+01.4684 & 13:09:36.24 & $-$61:19:35.76 & 0.09 & 1 &$-$29.2 & 1.0 & 0.9 &    1.0 &  & 7.4 &
2.4 (1.21) &7.4 (1.21)&  \\ 
G305.1997+00.0216 & 13:11:16.80 & $-$62:45:46.08 & 0.28 &1 &$-$49.2 & 4.2 & 1.2 &    5.3 &  & 6.7 &
\multicolumn{2}{c}{4.7 (0.72)} &  \\ 
 &  &  &  &2 &$-$45.5 & 5.6 & 3.5 &   20.8 &  & 6.8 &
\multicolumn{2}{c}{4.8 (1.01)} &  \\ 
 &  &  &  & 3  $^\star$  &-37.2 & 23.3 & 3.6 &   88.9 & blended (main component) & 7.1 &
3.4 (1.29) &6.4 (1.41)& 2 \\ 
 &  &  &  &4 &$-$33.4 & 8.3 & 6.4 &   56.3 & blended (red shoulder) & 7.2 &
2.9 (1.52) &6.9 (1.56)&  \\ 
G305.2242+00.2028 & 13:11:22.08 & $-$62:34:48.72 & 0.24 &1 &$-$42.9 & 5.4 & 3.7 &   21.2 &  & 6.9 &
\multicolumn{2}{c}{4.9 (1.16)} &  \\  
G305.2535+00.2412 & 13:11:35.76 & $-$62:32:22.92 & 0.08 &1 &$-$46.9 & 0.7 & 2.4 &    1.8 &  & 6.8 &
\multicolumn{2}{c}{4.8 (0.92)} &  \\ 
 &  &  &  & 2  $^\star$  &$-$40.9 & 3.7 & 3.0 &   11.8 & blended (main component) & 7.0 &
4.3 (1.05) &5.5 (1.26)& 1 \\ 
 &  &  &  &3 &$-$37.3 & 1.6 & 3.3 &    5.6 & blended (red shoulder) & 7.1 &
3.4 (1.29) &6.4 (1.41)&  \\ 
 &  &  &  &4 &$-$32.4 & 0.7 & 4.1 &    3.0 &  & 7.3 &
2.8 (1.59) &7.0 (1.60)&  \\ 
G305.2694$-$00.0072 & 13:11:54.24 & $-$62:47:10.32 & 0.13 & 1 &$-$32.3 & 16.3 & 5.0 &   86.4 & red shoulder & 7.3 &
2.8 (1.59) &7.1 (1.60)&  \\ 
G305.3500+00.2240 & 13:12:26.64 & $-$62:32:57.12 & 0.27 & 1  $^\star$  &$-$39.0 & 18.6 & 5.1 &  100.6 & blended & 7.0 &
3.7 (1.19) &6.1 (1.35)& 1 \\ 
 &  &  &  &2 &$-$35.4 & 14.6 & 2.7 &   41.8 & blended & 7.2 &
3.1 (1.42) &6.7 (1.49)&  \\ 
 &  &  &  &3 &$-$27.9 & 2.2 & 1.9 &    4.4 &  & 7.4 &
2.3 (1.11) &7.6 (1.11)&  \\ 
G305.3611+00.1494 & 13:12:35.52 & $-$62:37:21.00 & 0.20 & 1  $^\star$  &$-$39.7 & 7.1 & 5.4 &   40.6 & red shoulder & 7.0 &
3.9 (1.15) &6.0 (1.32)& 2 \\ 
\end{longtable}
\end{landscape}
\end{onecolumn}

\clearpage
\setcounter{figure}{10}
\begin{figure*}
\begin{center}
\caption{$^{13}$CO spectra for all RMS sources towards which emission is detected.}
\includegraphics[width=0.33\linewidth]{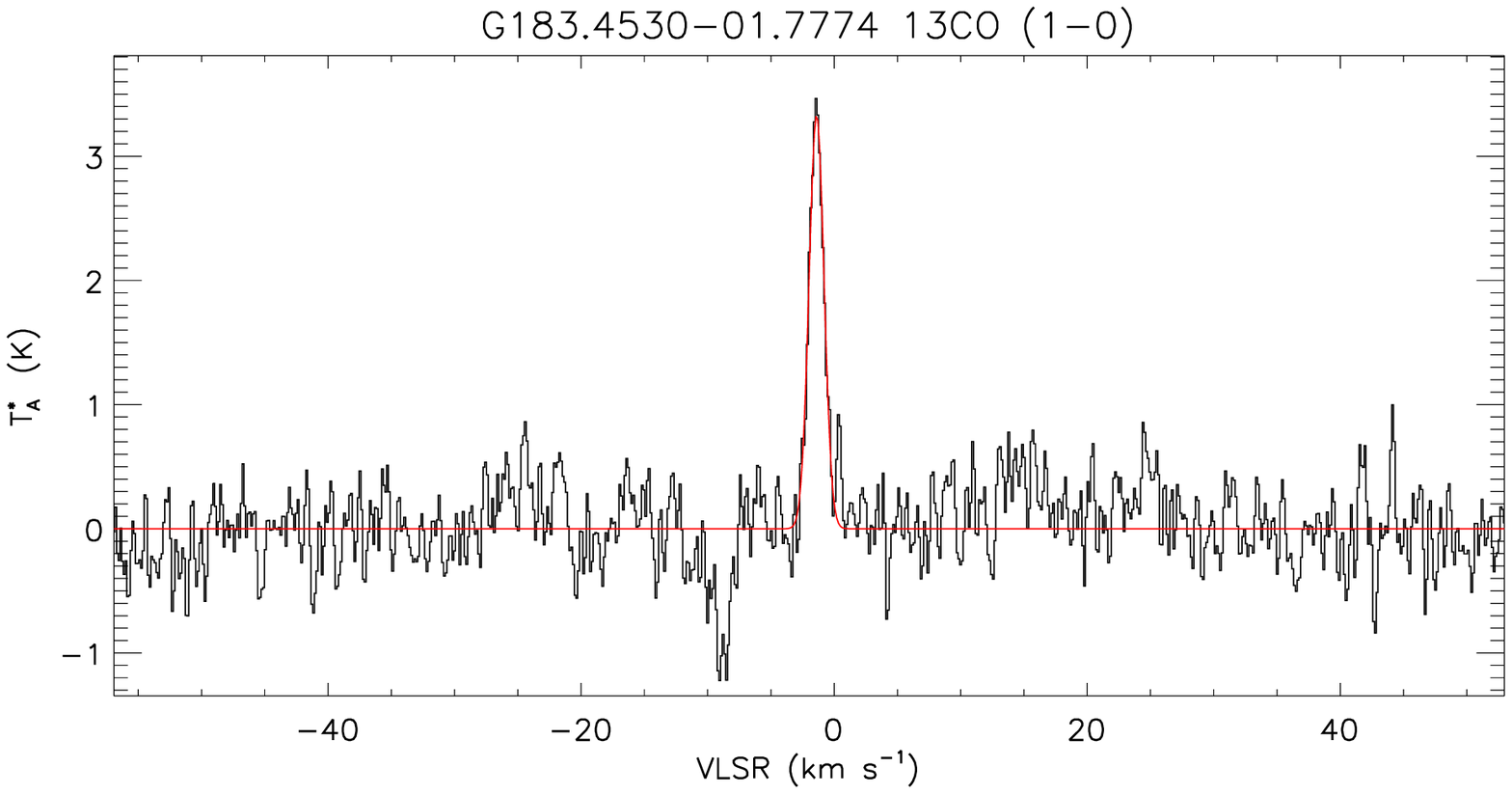} 
\includegraphics[width=0.33\linewidth]{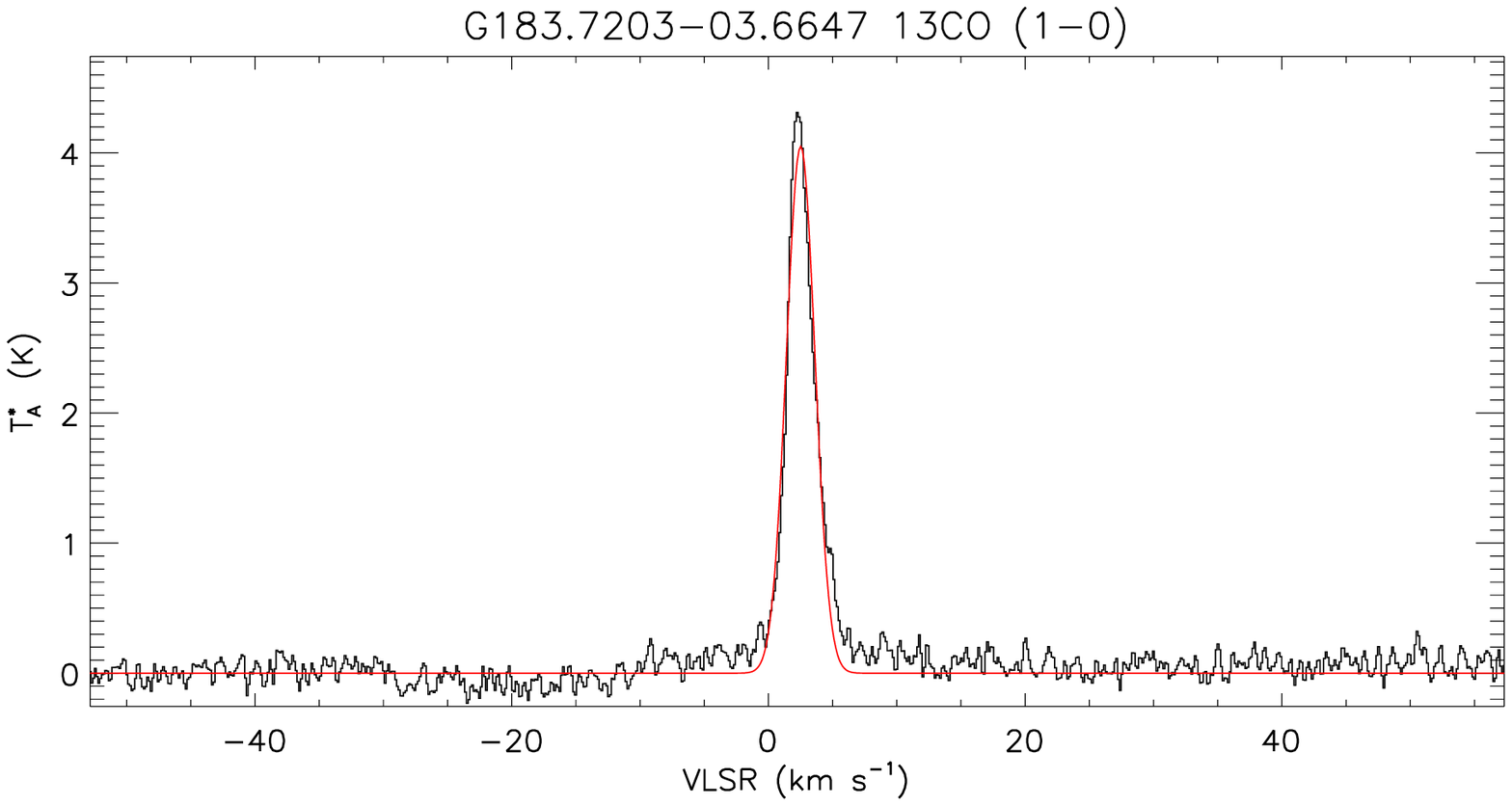} 
\includegraphics[width=0.33\linewidth]{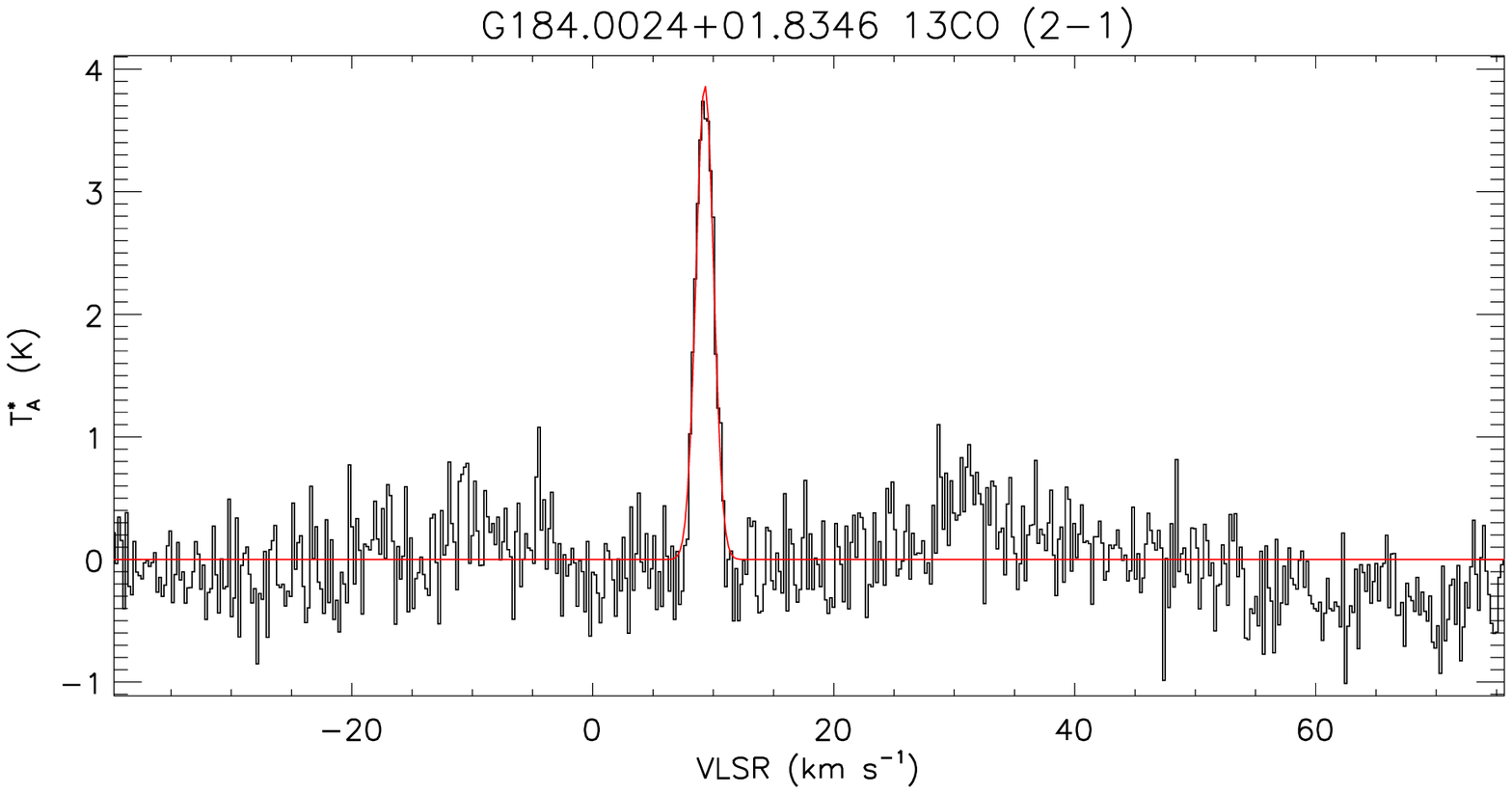} \\ 
\includegraphics[width=0.33\linewidth]{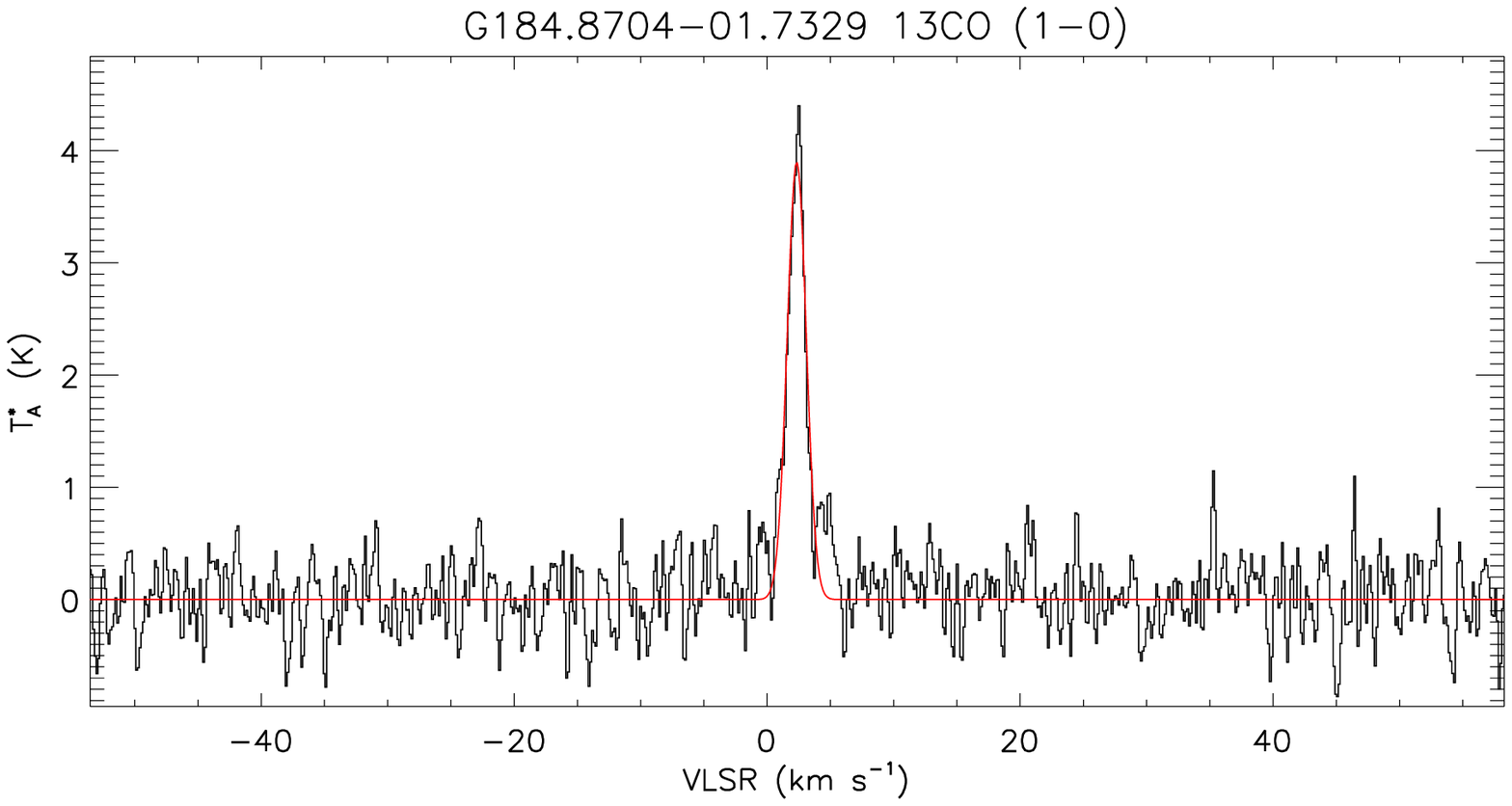} 
\includegraphics[width=0.33\linewidth]{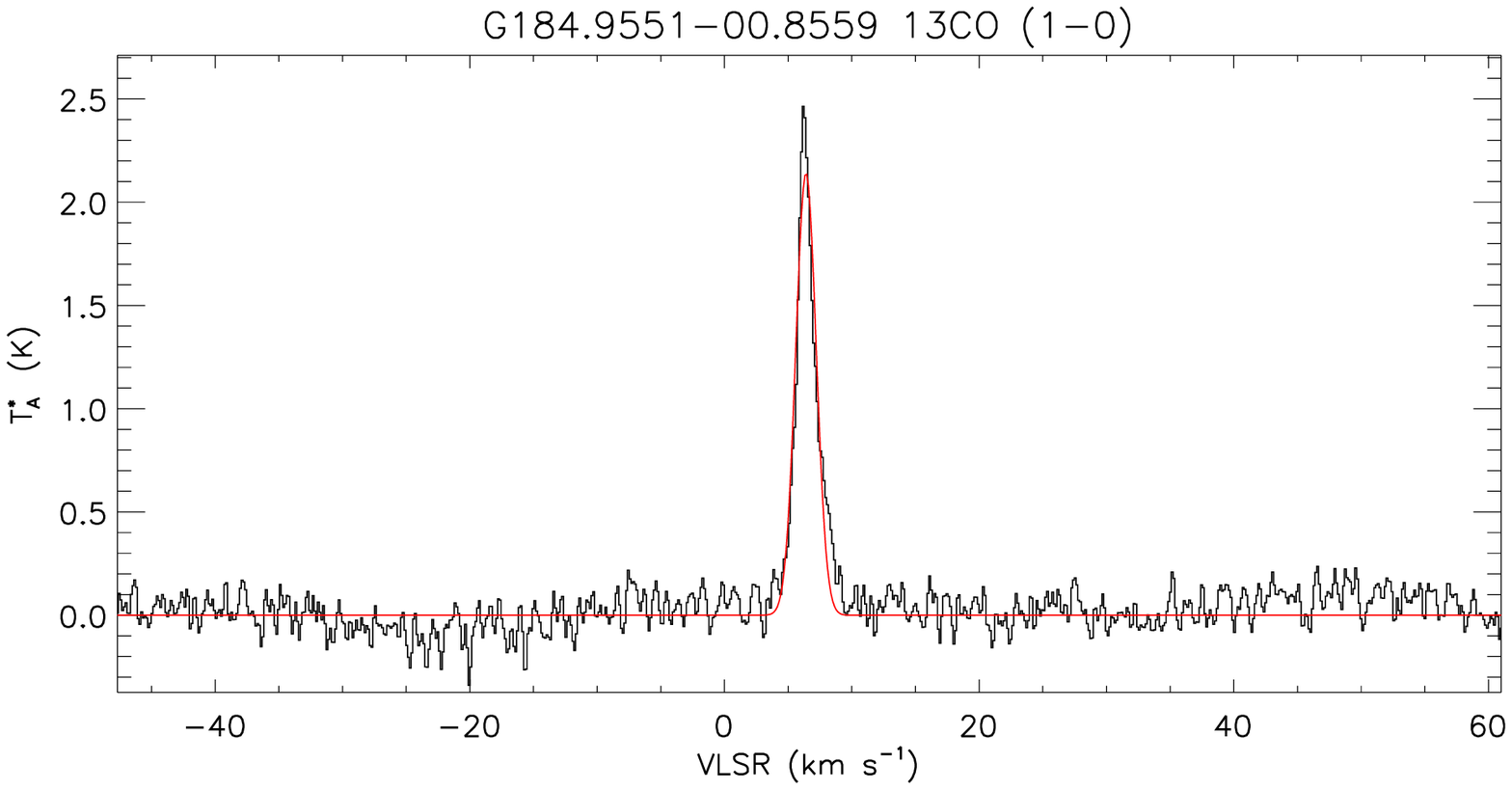} 
\includegraphics[width=0.33\linewidth]{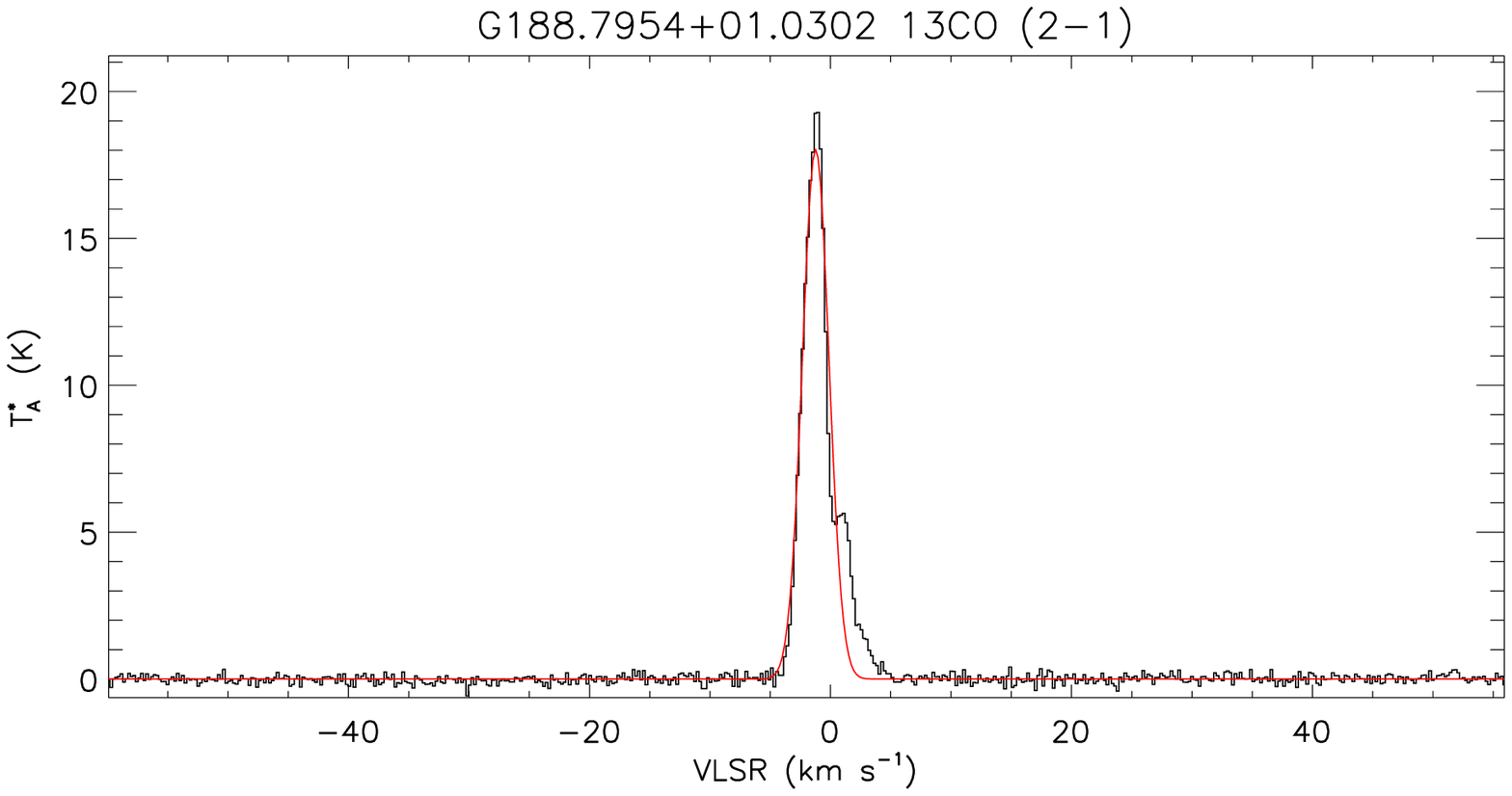} \\ 
\includegraphics[width=0.33\linewidth]{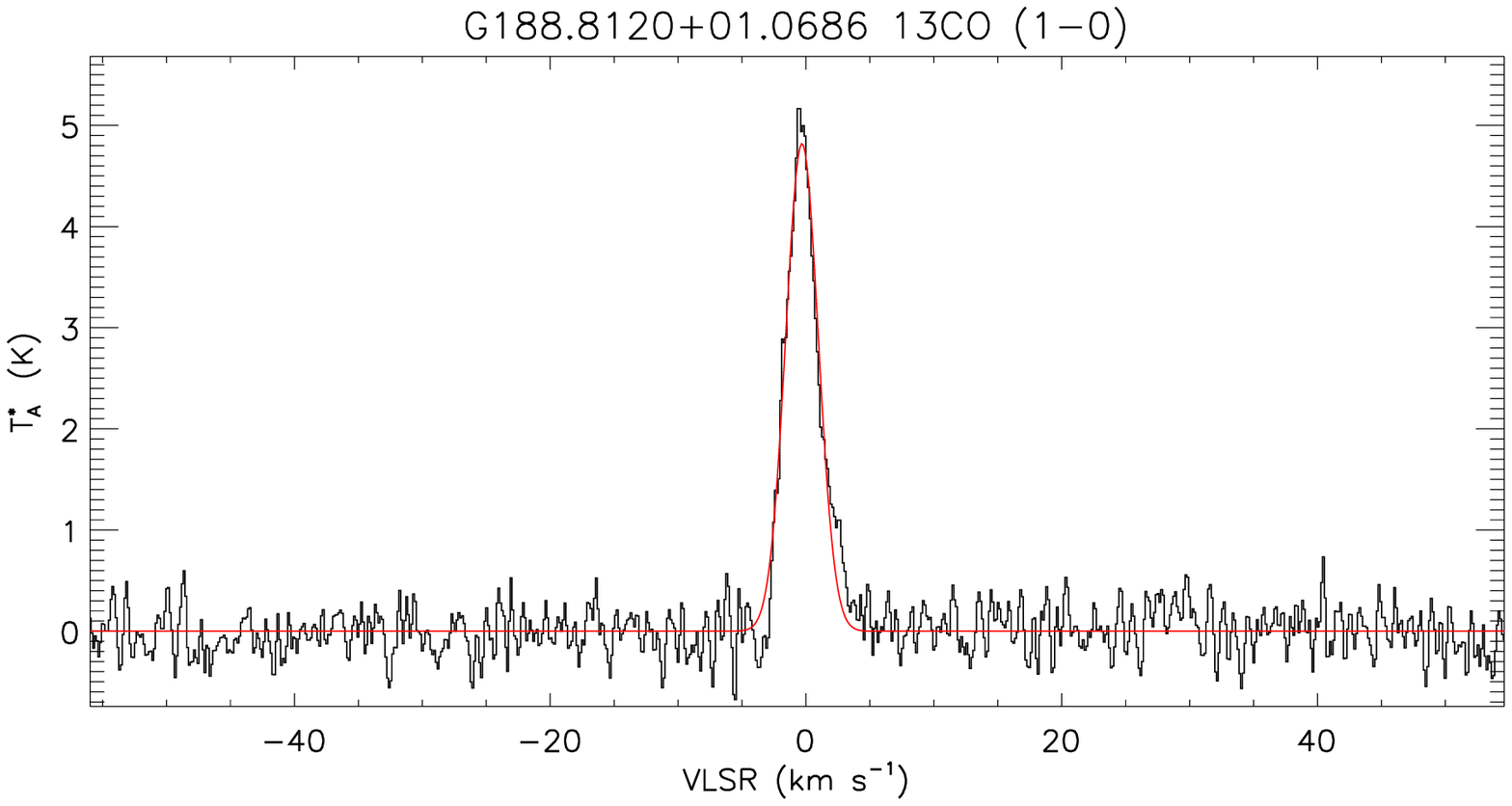} 
\includegraphics[width=0.33\linewidth]{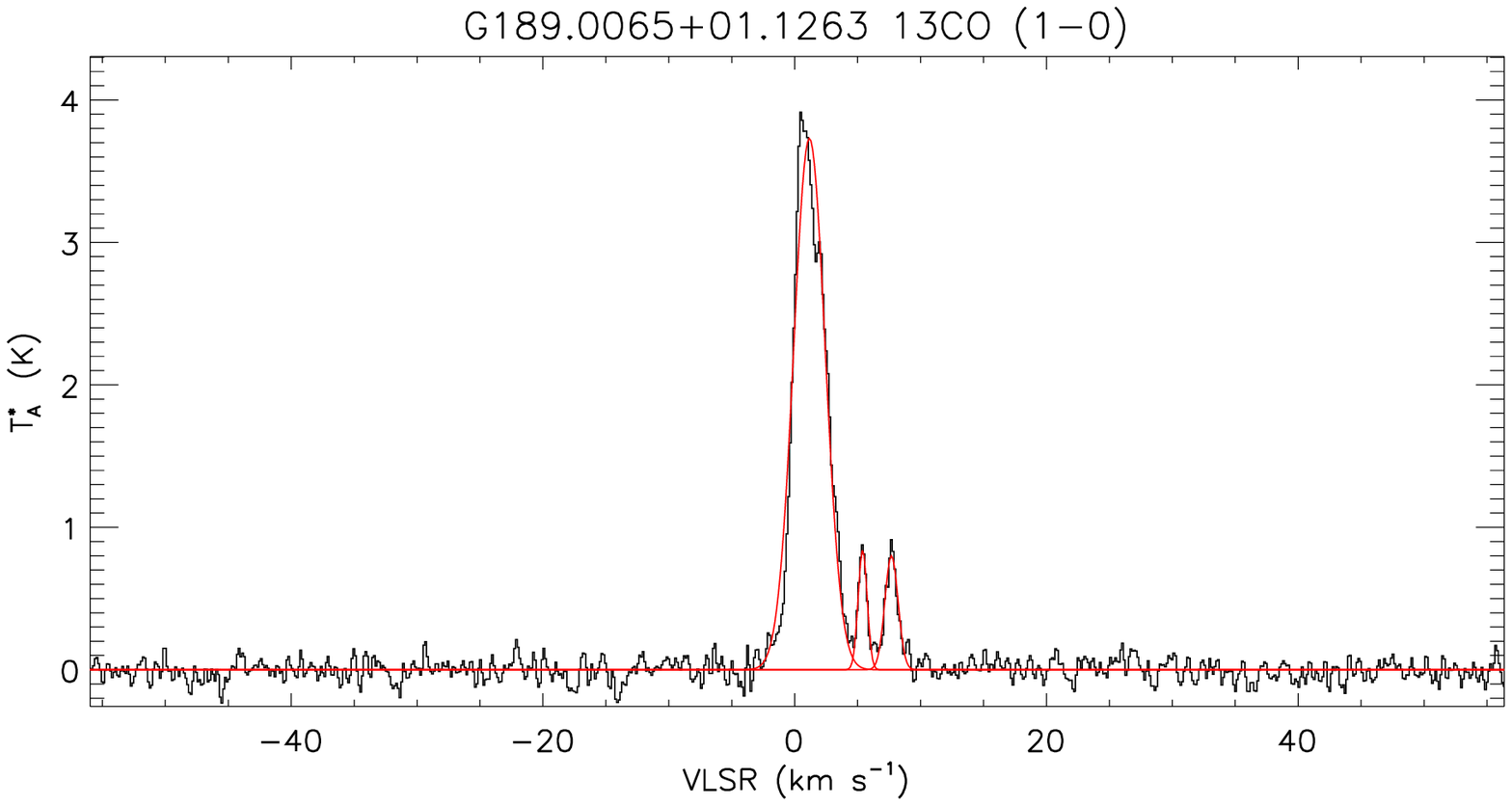} 
\includegraphics[width=0.33\linewidth]{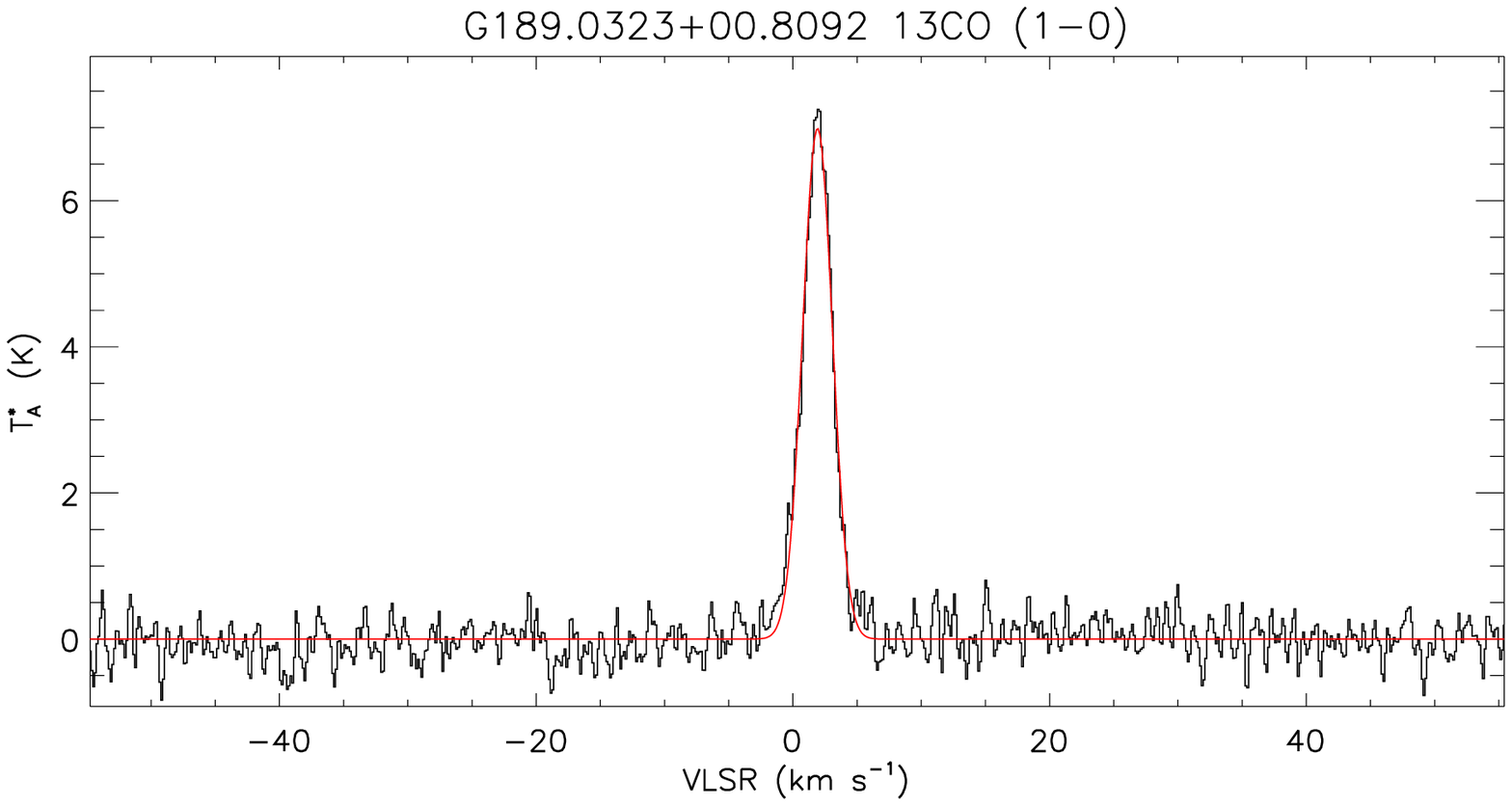} \\ 
\includegraphics[width=0.33\linewidth]{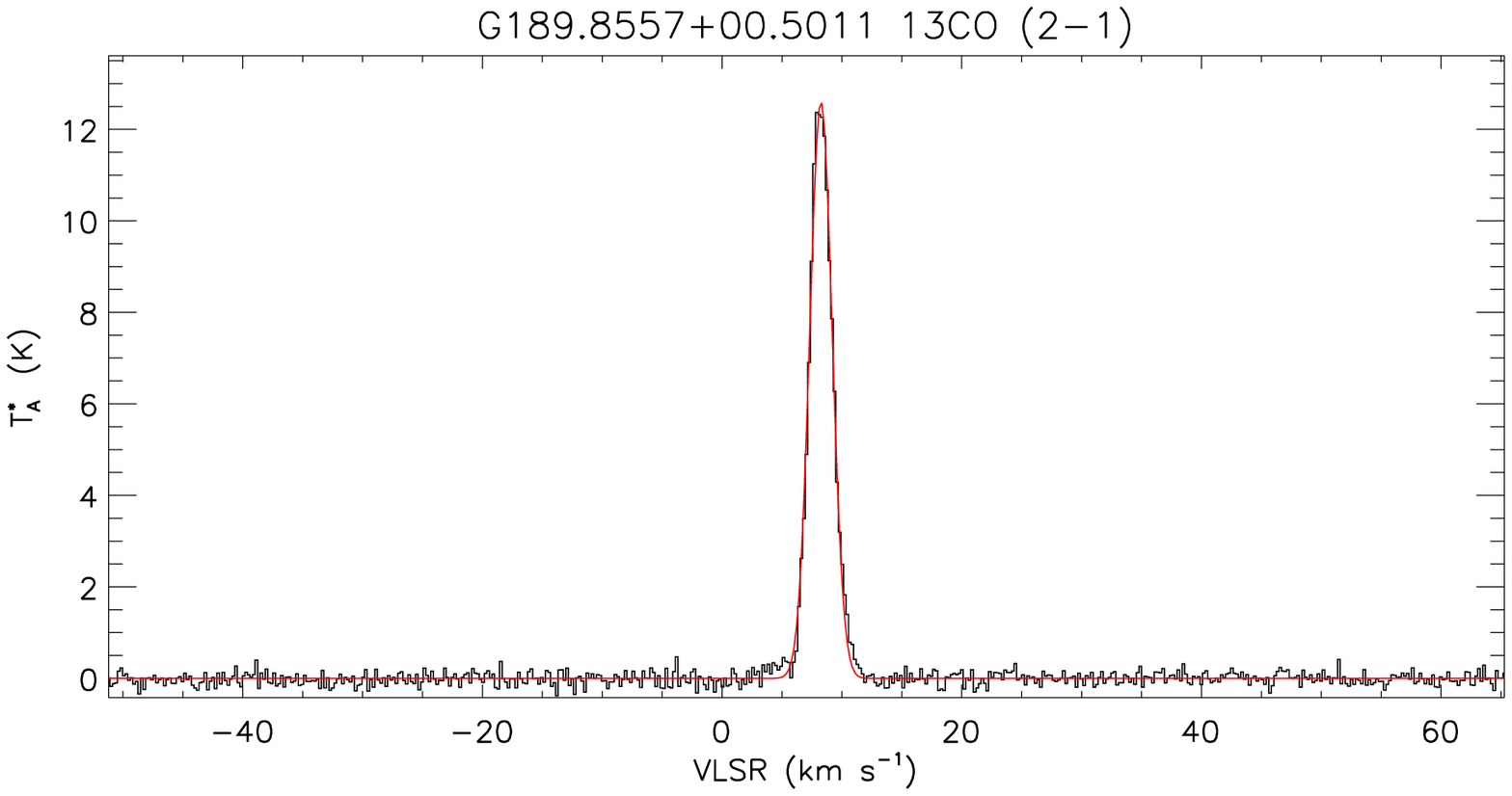} 
\includegraphics[width=0.33\linewidth]{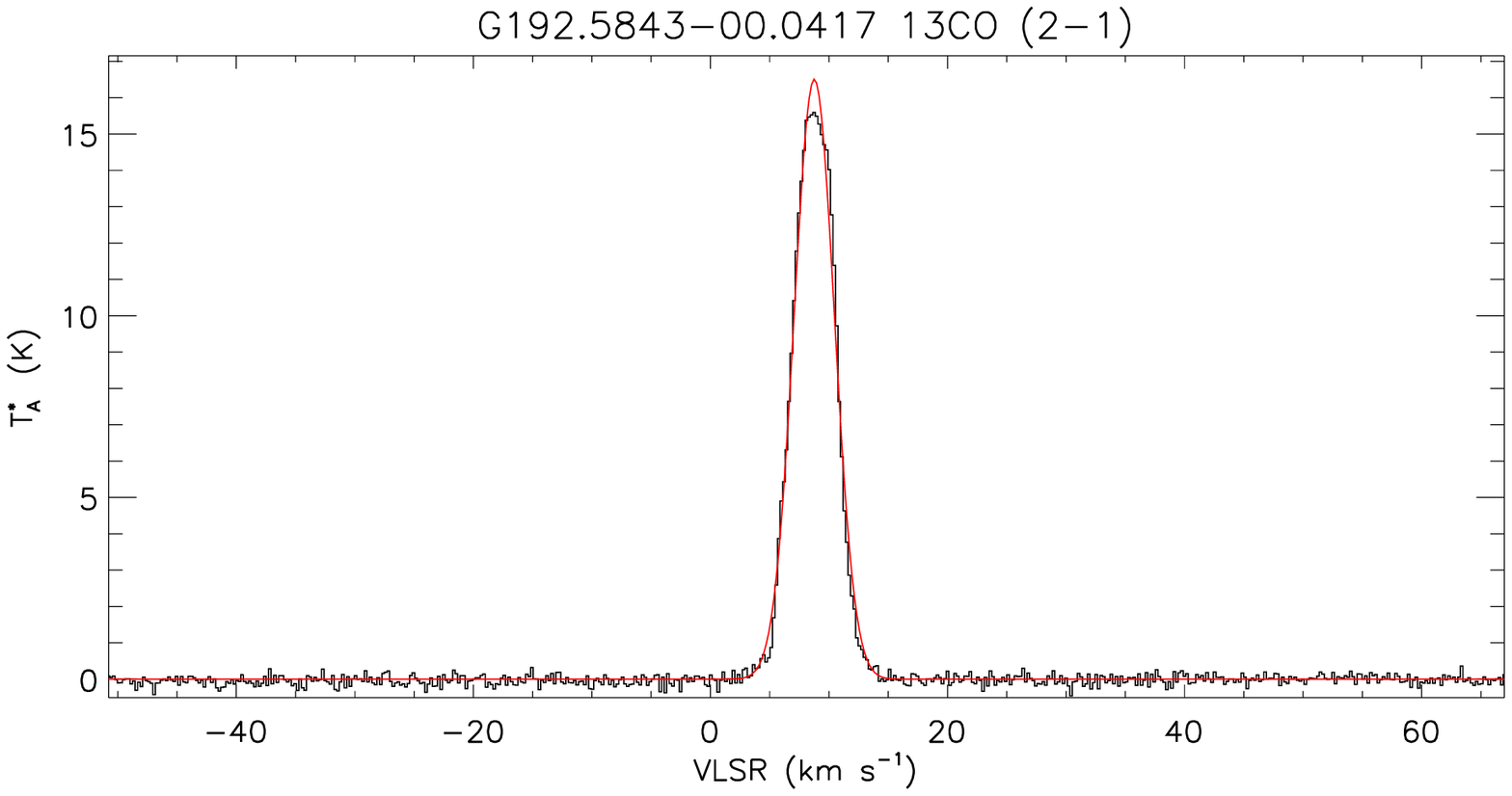} 
\includegraphics[width=0.33\linewidth]{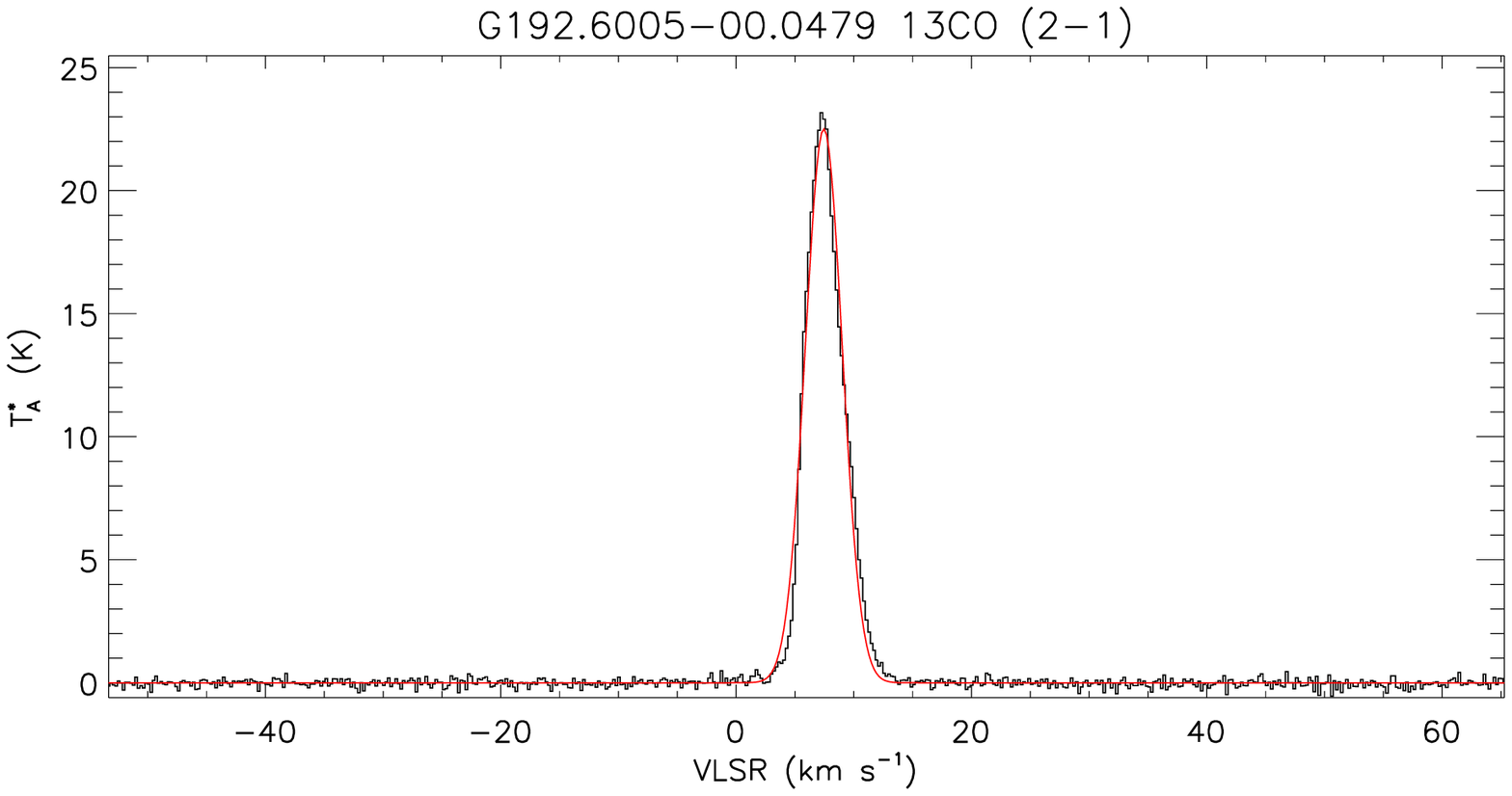} \\ 
\includegraphics[width=0.33\linewidth]{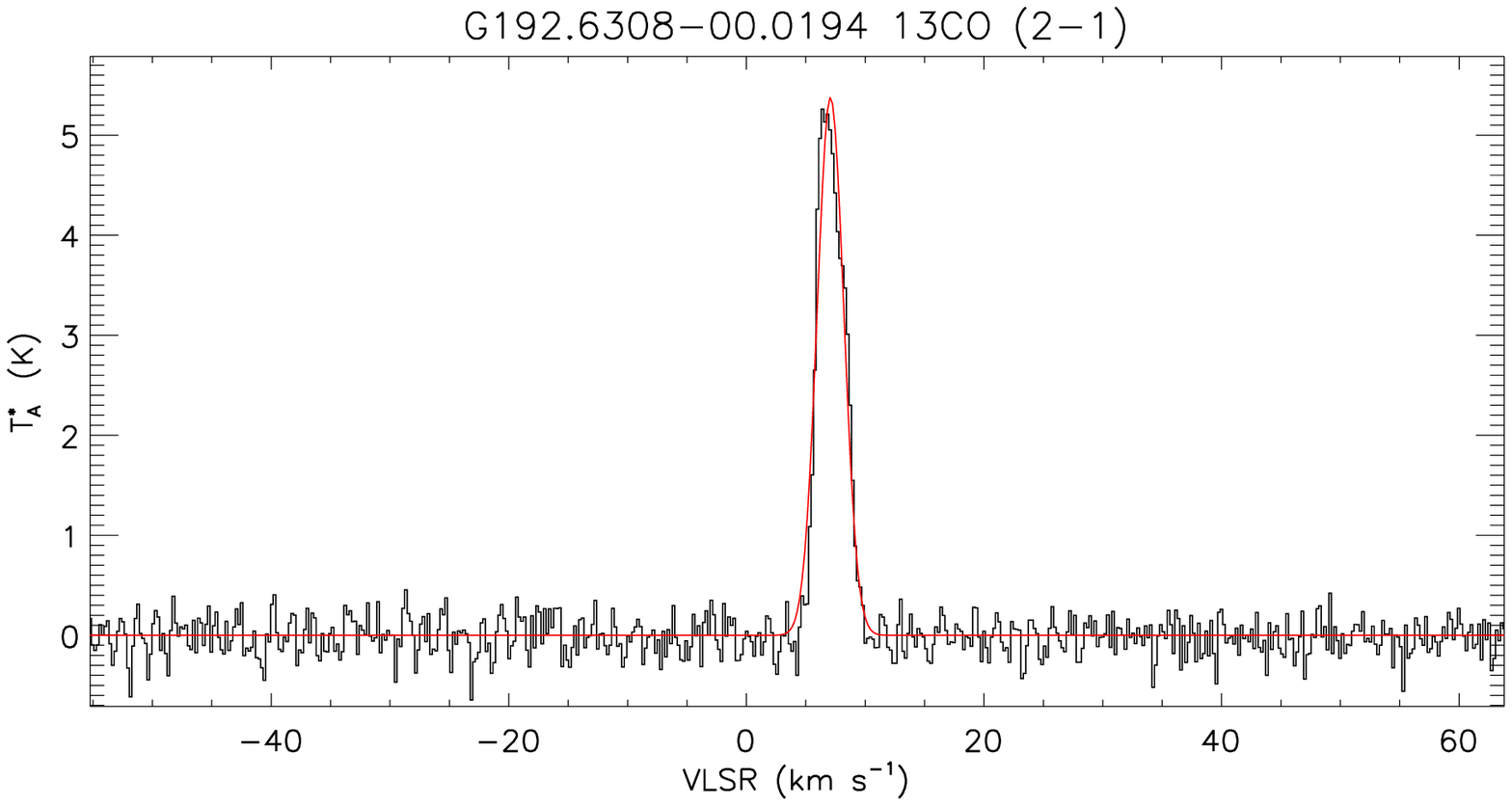} 
\includegraphics[width=0.33\linewidth]{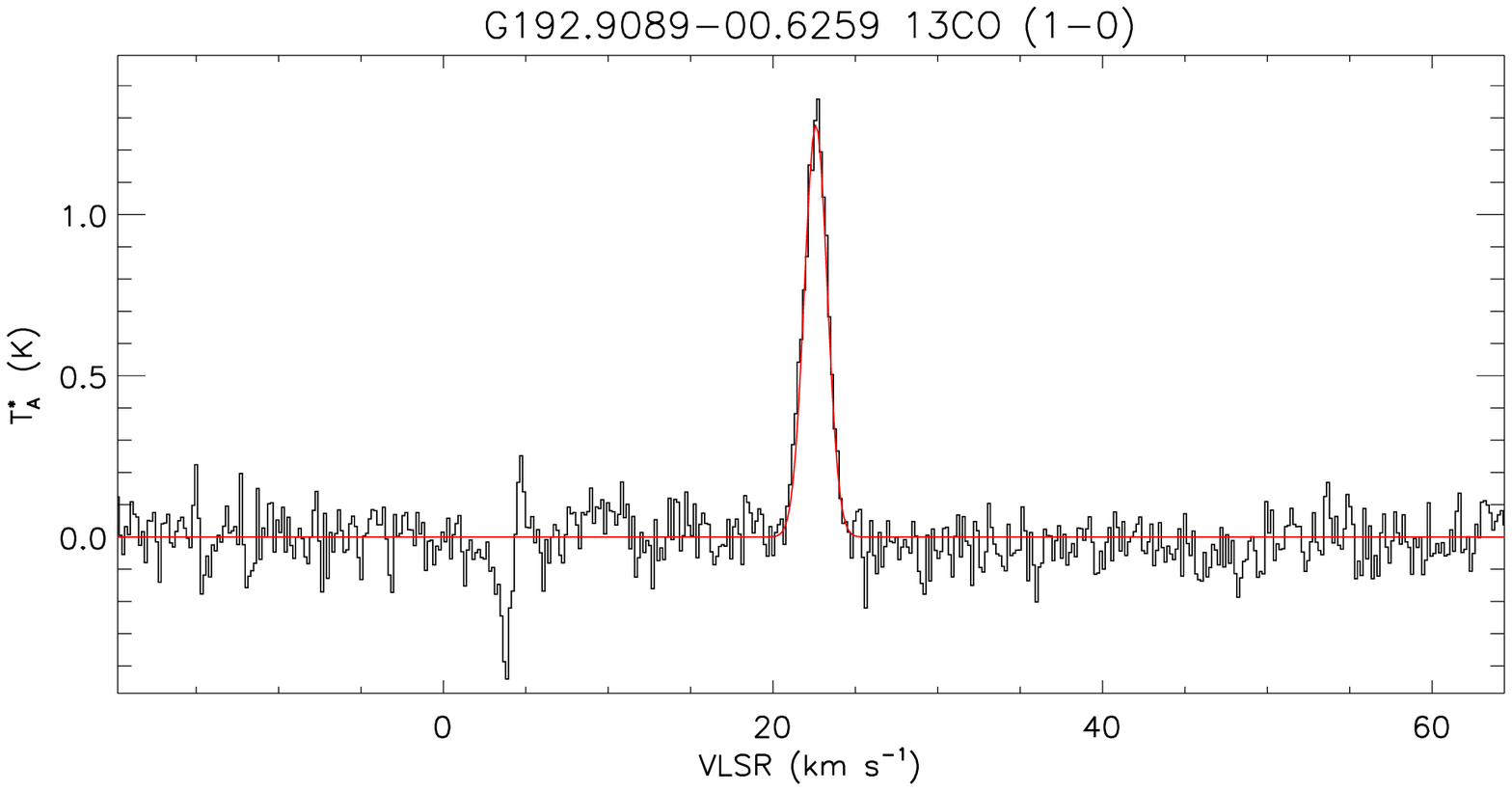} 
\includegraphics[width=0.33\linewidth]{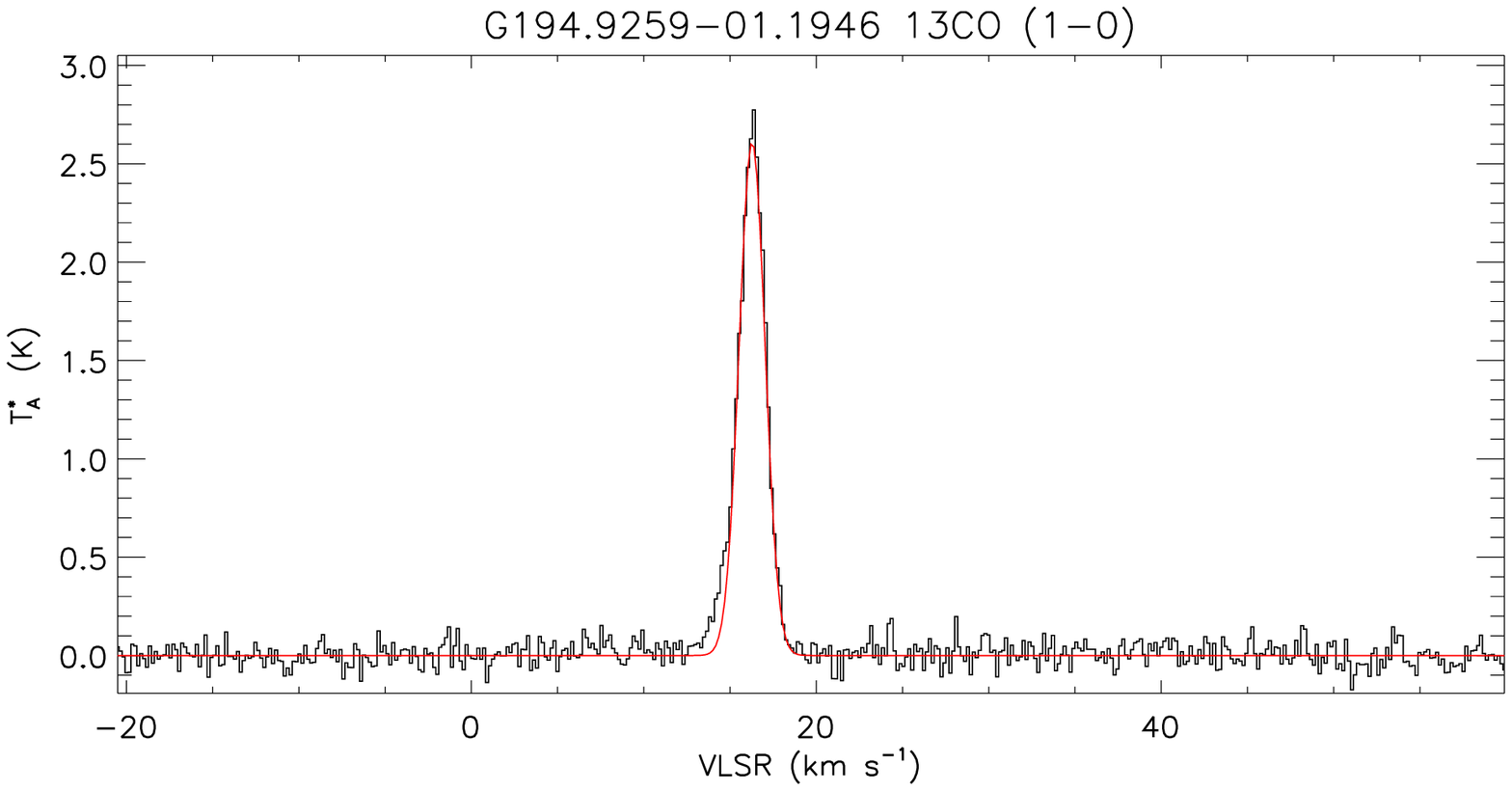} \\ 
\includegraphics[width=0.33\linewidth]{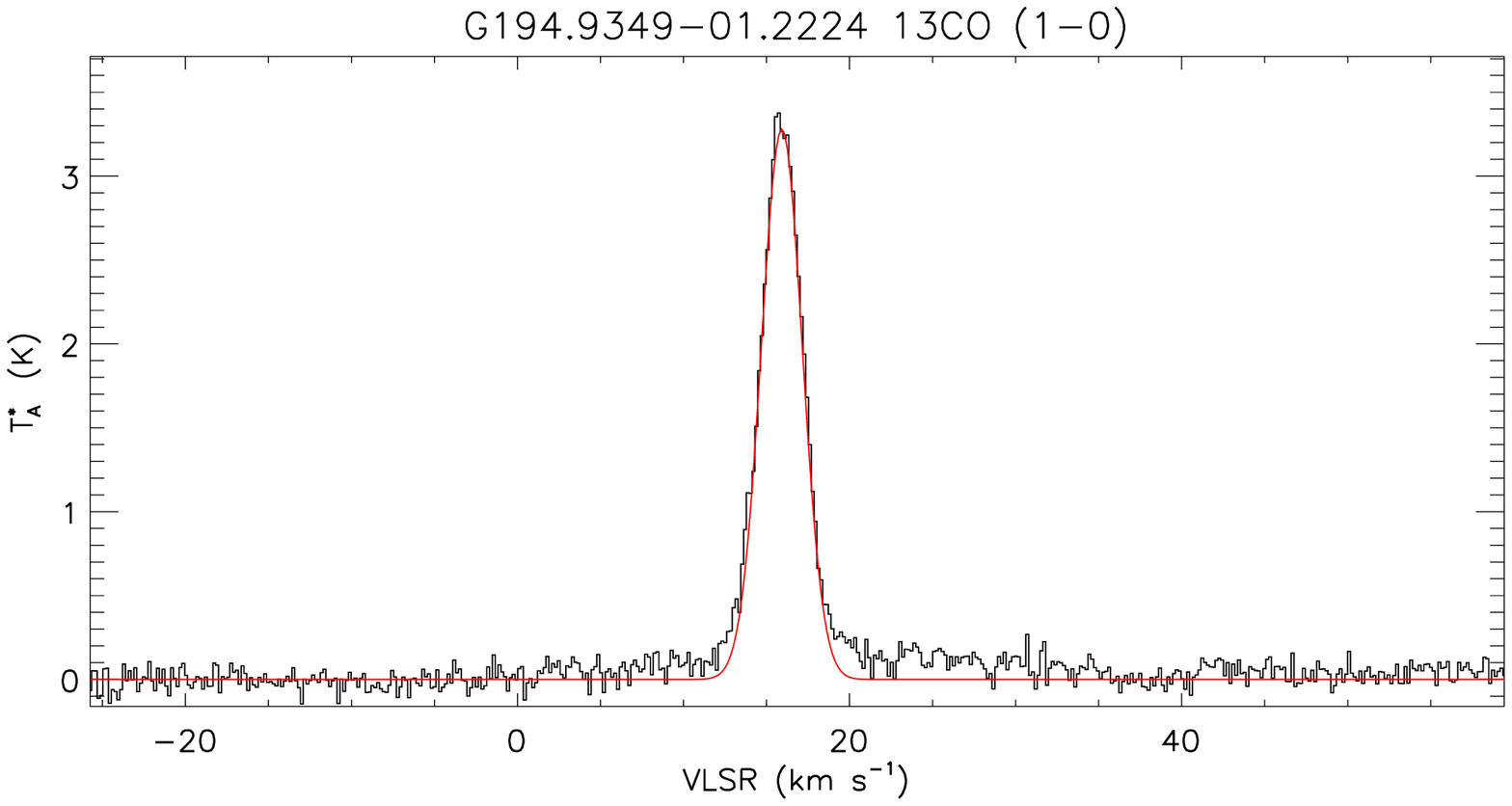} 
\includegraphics[width=0.33\linewidth]{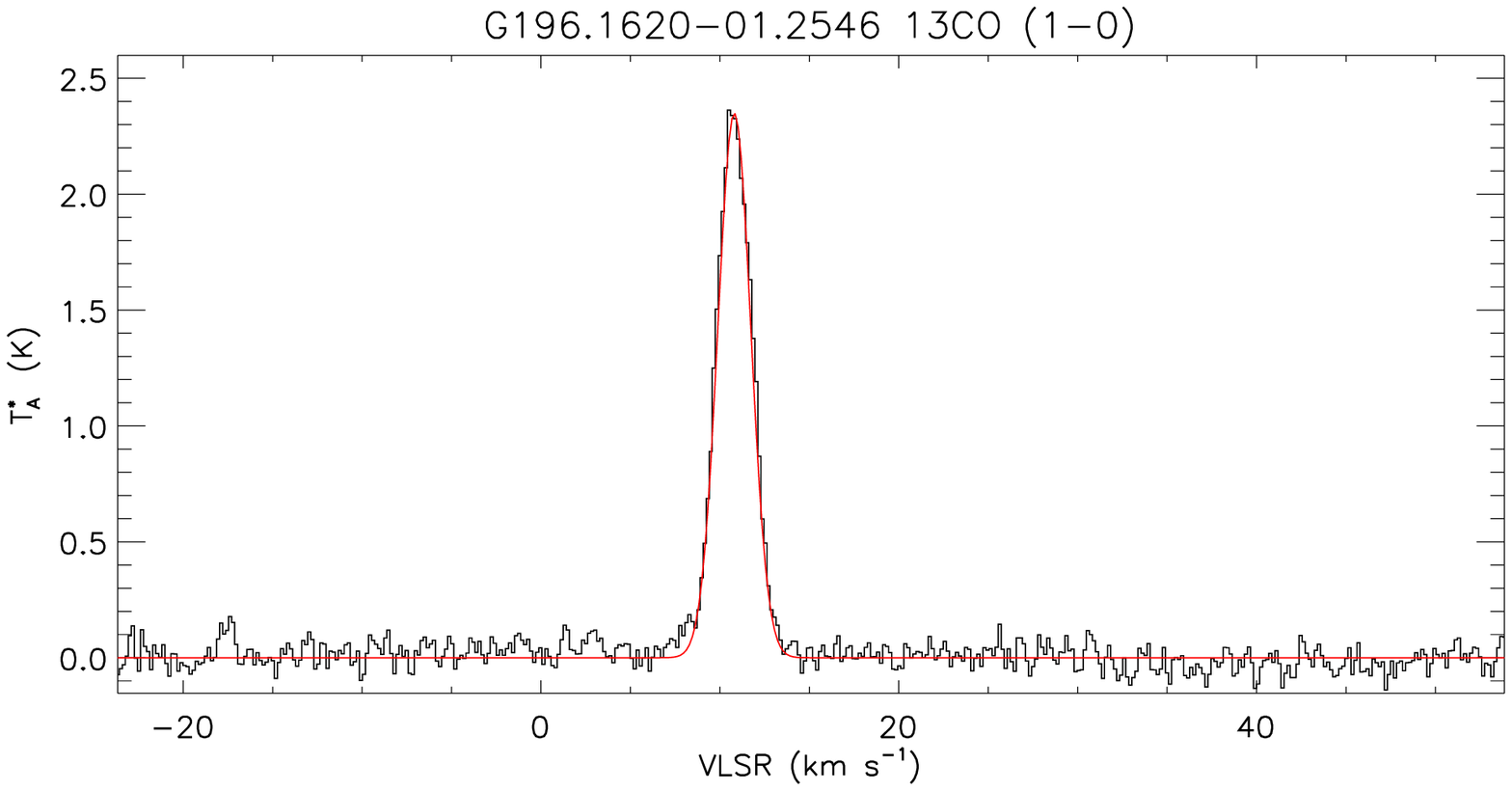} 
\includegraphics[width=0.33\linewidth]{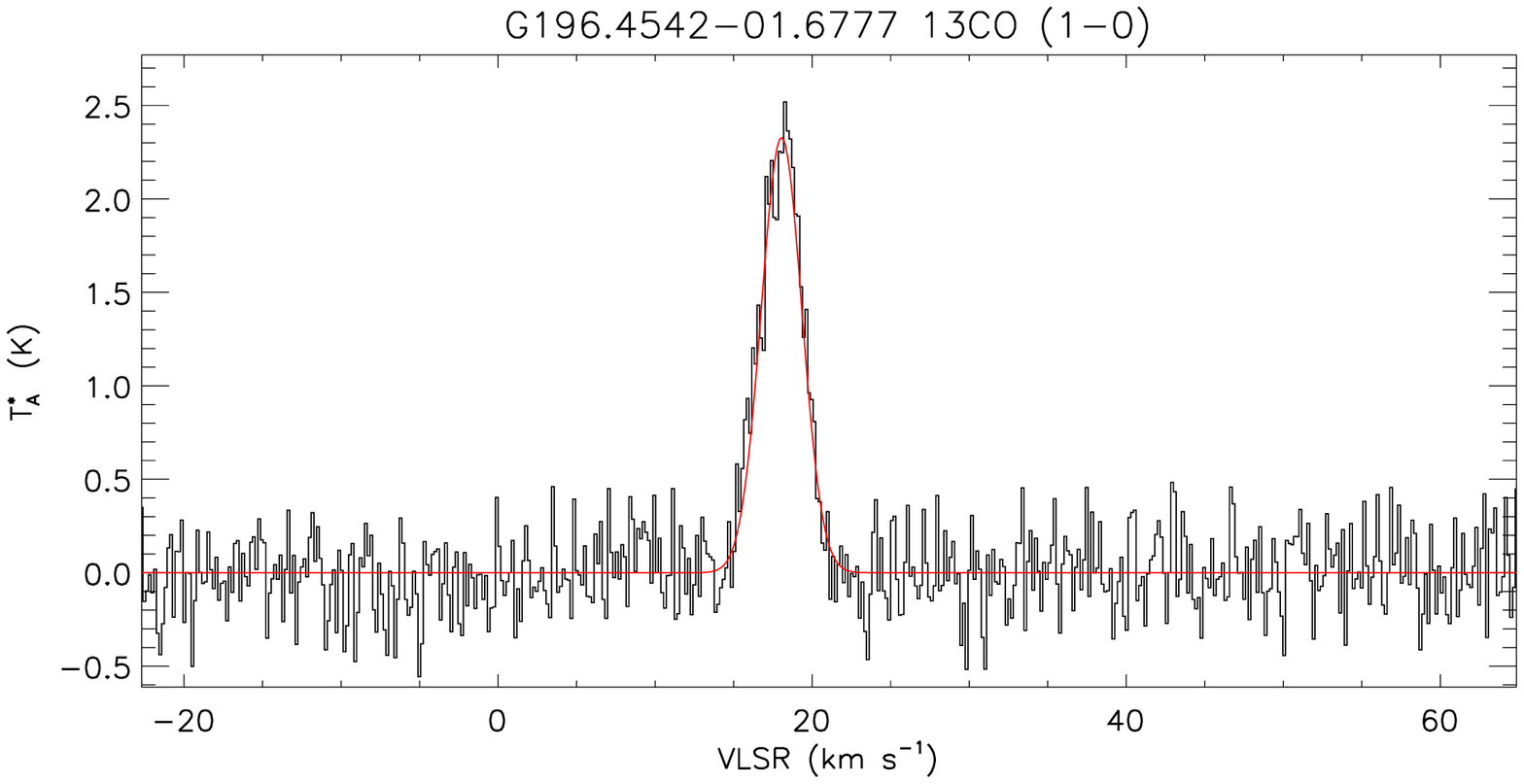} \\ 
\includegraphics[width=0.33\linewidth]{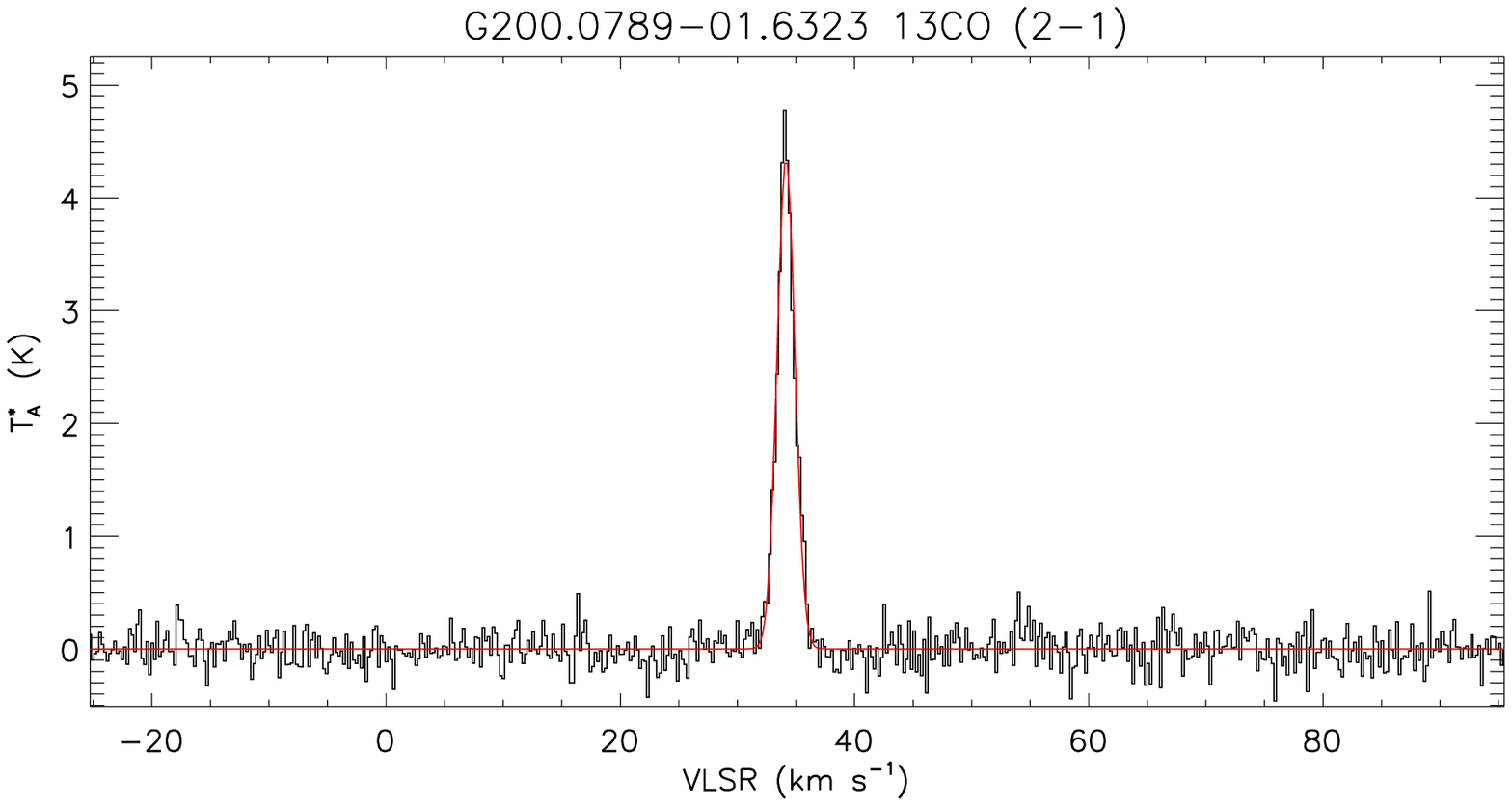} 
\includegraphics[width=0.33\linewidth]{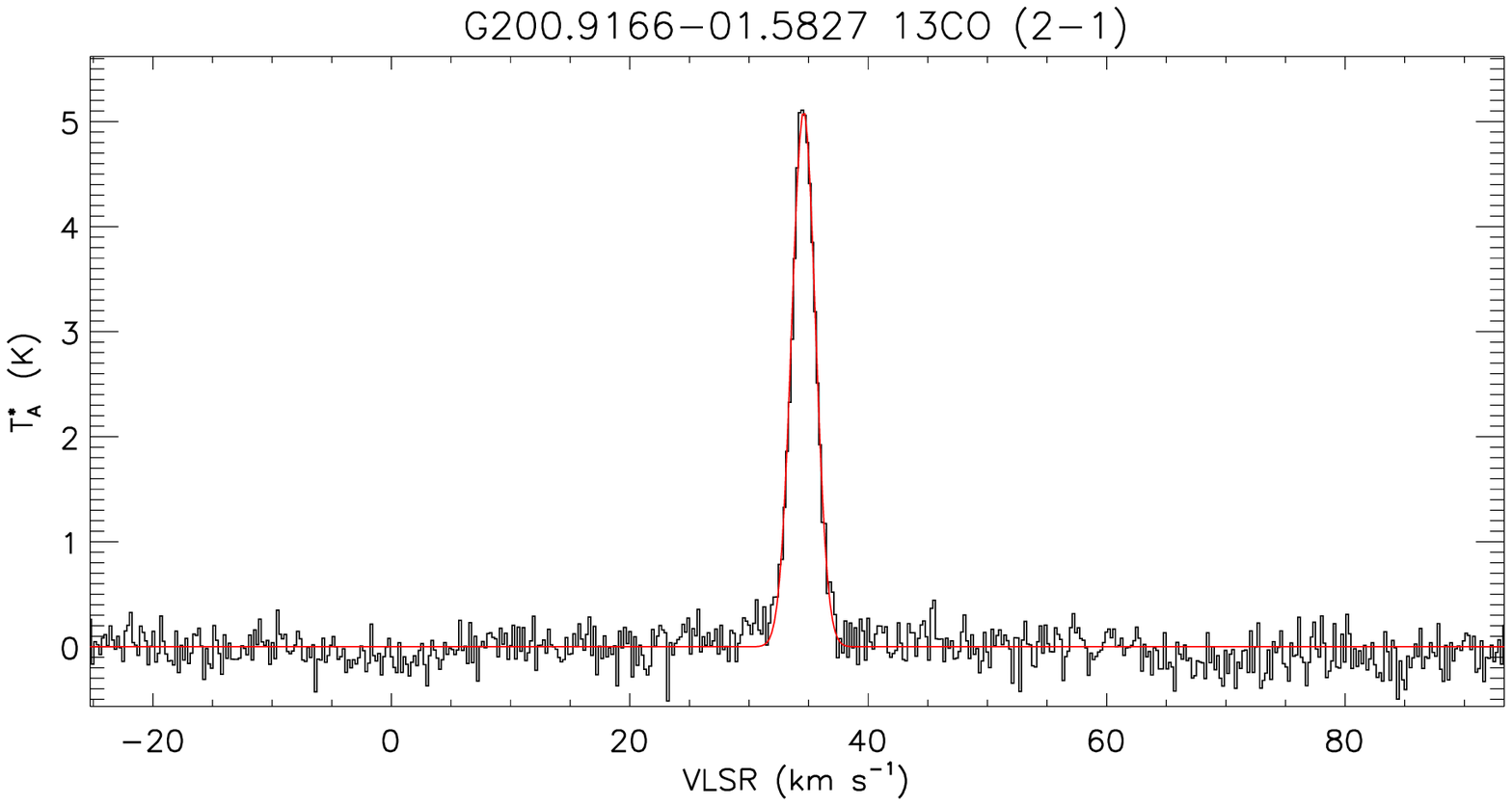} 
\includegraphics[width=0.33\linewidth]{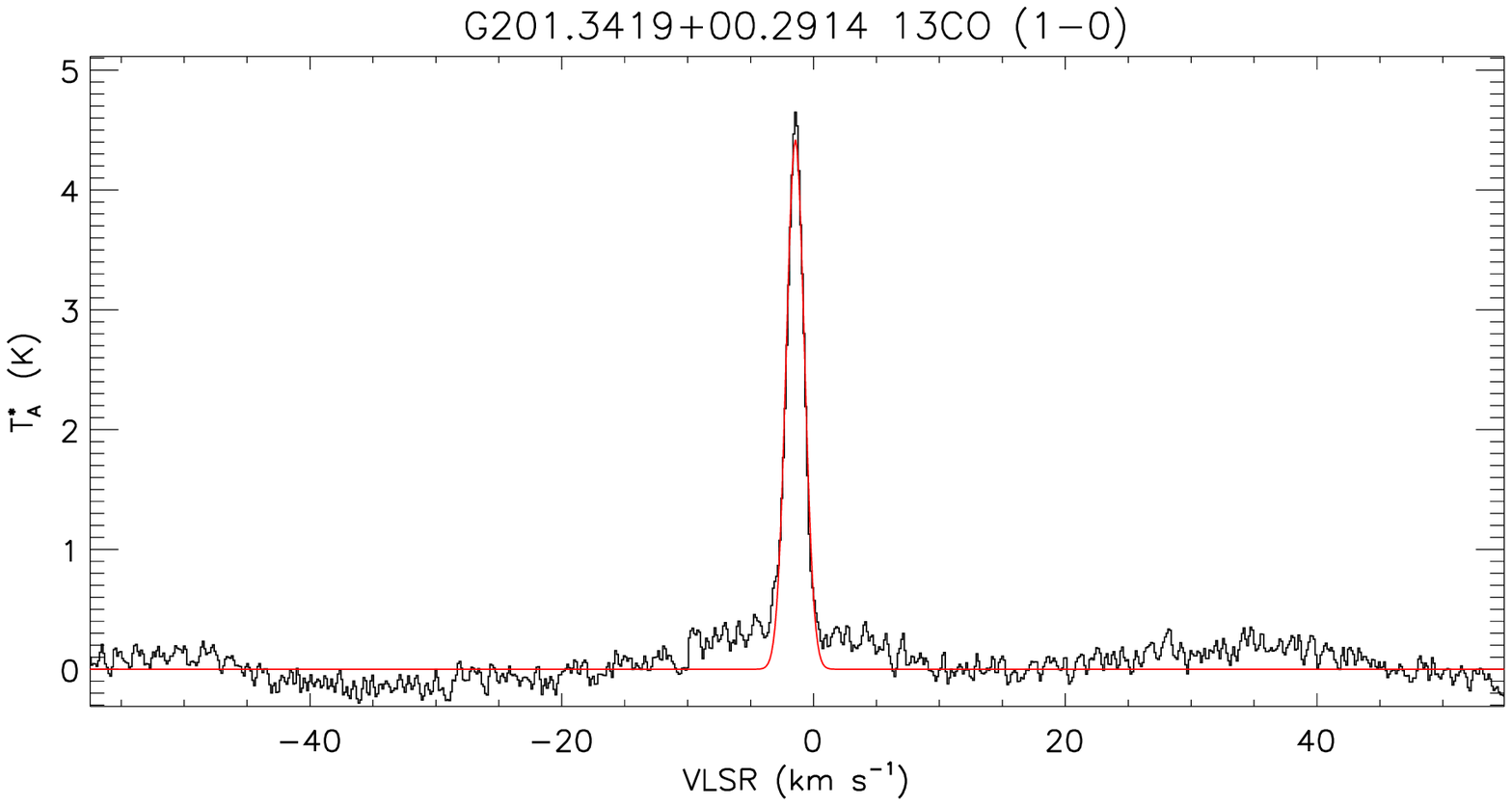} \\ 
\includegraphics[width=0.33\linewidth]{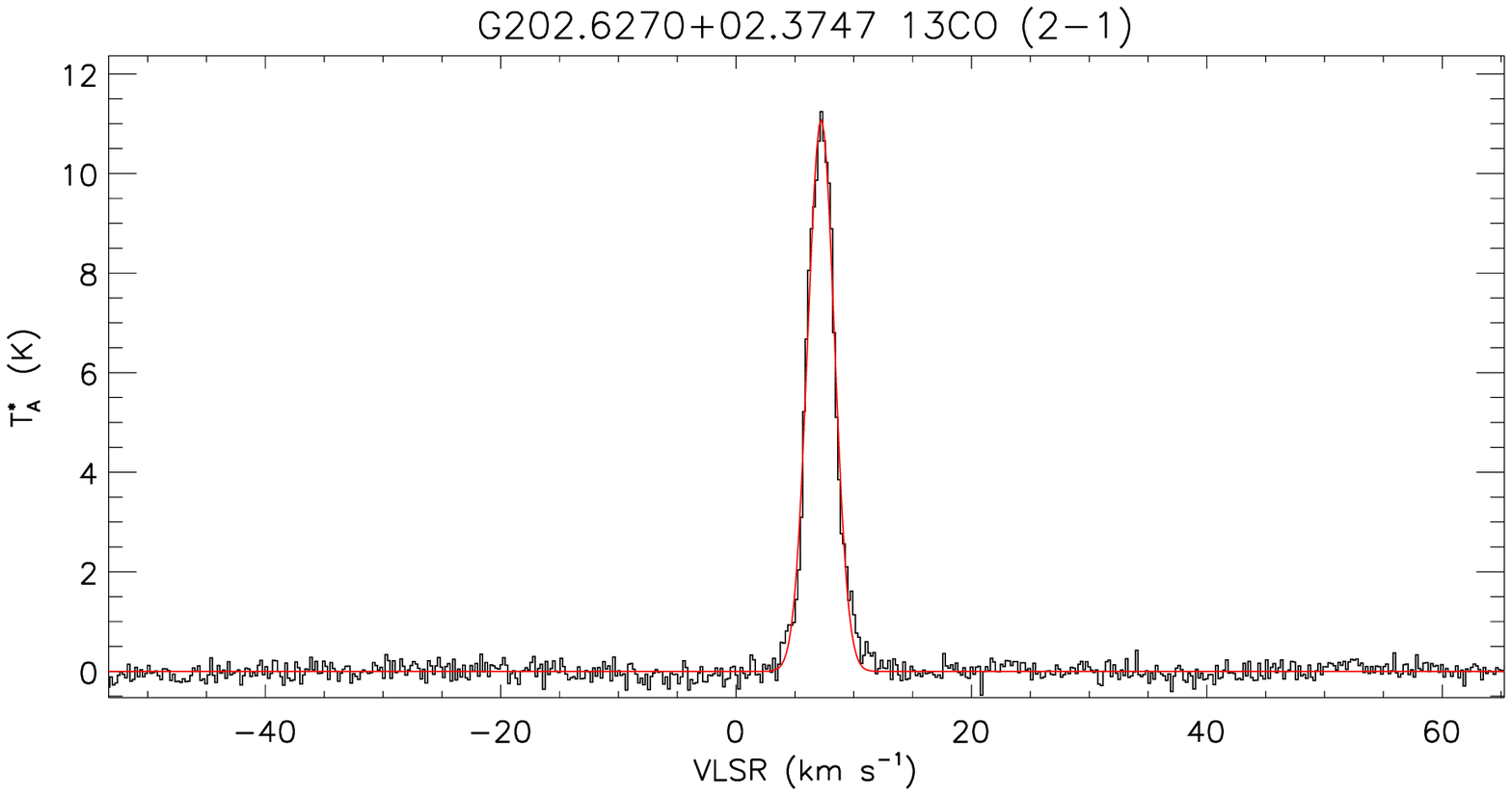} 
\includegraphics[width=0.33\linewidth]{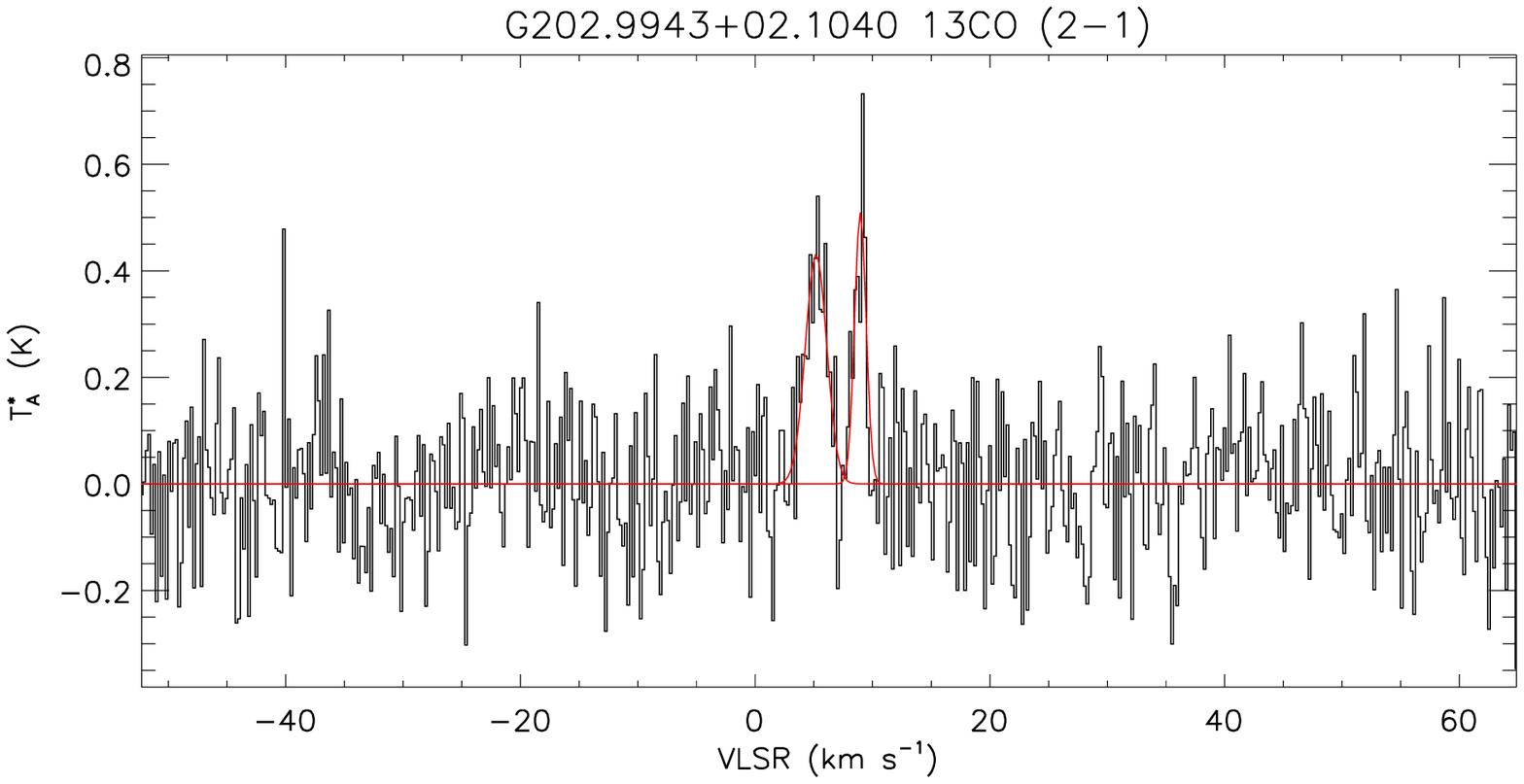} 
\includegraphics[width=0.33\linewidth]{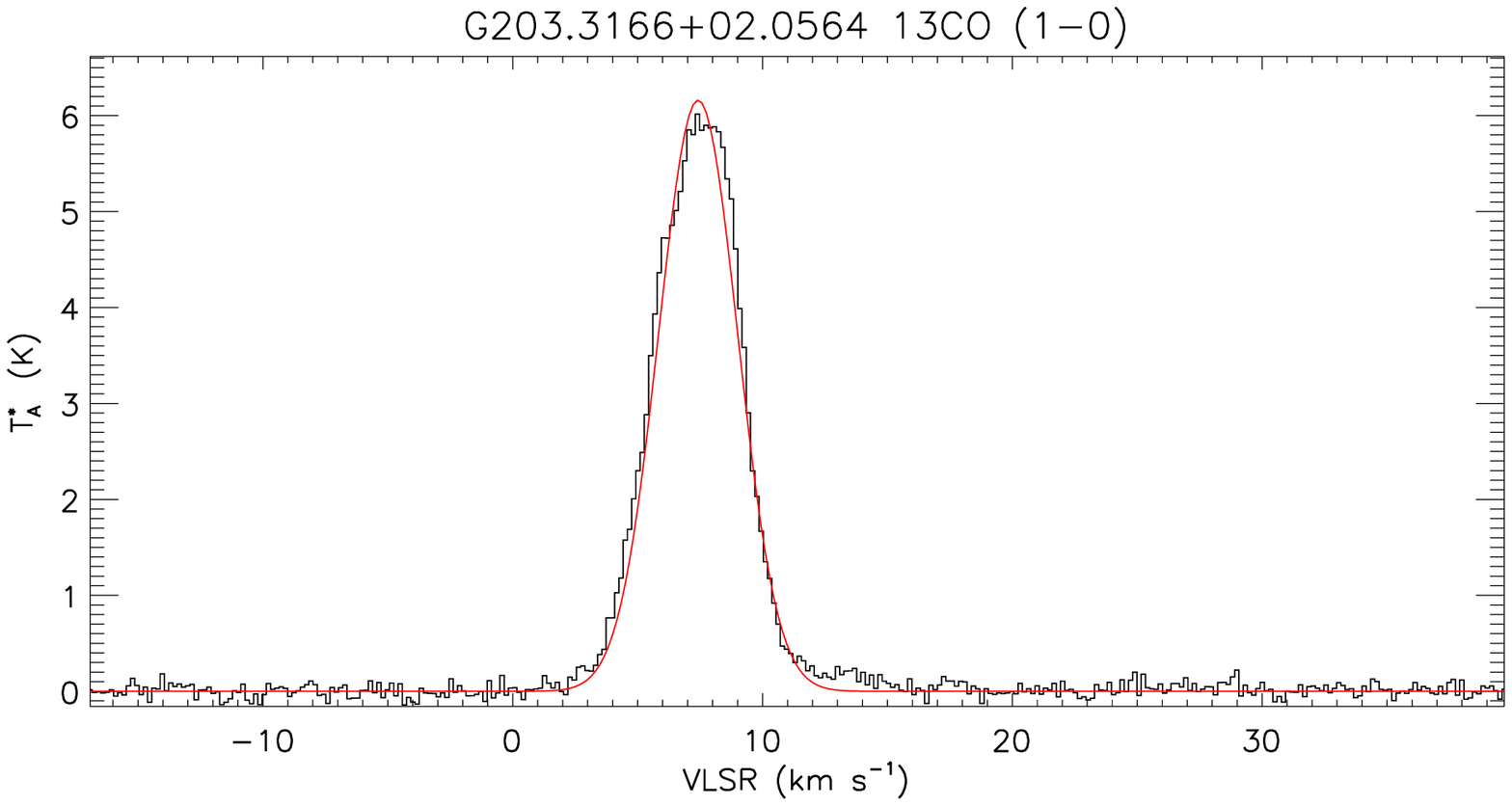} \\ 
\end{center}
\end{figure*}

\end{document}